\documentclass[authoryear,preprint,review]{elsarticle}
\usepackage[utf8]{inputenc}
\usepackage{amssymb}
\usepackage{amsmath}
\usepackage{subfigure}
\usepackage{array}
\usepackage{mathrsfs}
\usepackage{lineno,hyperref}
\usepackage{soul}
\hypersetup{colorlinks=true,linkcolor=blue,citecolor=blue,urlcolor=blue}
\renewcommand{\bar}{\overline}

\journal{Icarus}

\biboptions{authoryear}

\begin{document}

\begin{frontmatter}
\title{Global climate modeling of Saturn's atmosphere. \\Part IV: stratospheric equatorial oscillation.}
%% Authors
\author[mainaddress]{Deborah Bardet\corref{correspondingauthor}}
\cortext[correspondingauthor]{Corresponding author: deborah.bardet@lmd.jussieu.fr}
\author[mainaddress,secondaryaddress]{Aymeric Spiga}
\author[mainaddress]{Sandrine Guerlet}
\author[thirdaddress]{Simon Cabanes}
\author[mainaddress]{Ehouarn Millour}
\author[mainaddress]{Alexandre Boissinot}
%% Address
\address[mainaddress]{\emph{Laboratoire de M\'{e}t\'{e}orologie Dynamique / Institut Pierre-Simon Laplace (LMD/IPSL), Sorbonne Universit\'{e}, Centre National de la Recherche Scientifique (CNRS), \'{E}cole Polytechnique, \'{E}cole Normale Sup\'{e}rieure (ENS)}, address: Campus Pierre et Marie Curie BC99, 4 place Jussieu, 75005 Paris, France}
\address[secondaryaddress]{\emph{Institut Universitaire de France (IUF)}, address: 1 rue Descartes, 75005 Paris, France}
\address[thirdaddress]{\emph{DICEA. Sapienza Universt\`{a} di Roma}, address: Via Eudossiana 18, 00184 Rome, Italy}

%% Abstract
\begin{abstract}
The Composite InfraRed Spectrometer (CIRS) on board Cassini revealed an equatorial oscillation of stratospheric temperature, reminiscent of the Earth’s Quasi-Biennial Oscillation (QBO), as well as anomalously high temperatures under Saturn’s rings. To better understand these predominant features of Saturn’s atmospheric circulation in the stratosphere, we have extended to the upper stratosphere the DYNAMICO-Saturn global climate model (GCM), already used in a previous publication to study the tropospheric dynamics, jets formation and  planetary-scale waves activity. Firstly, we study the higher model top impact on the tropospheric zonal jets and kinetic energy distribution. Raising the model top prevents energy and enstrophy accumulation at tropopause levels. 
The reference GCM simulation with 1/2$^{\circ}$ latitude/longitude resolution and a raised model top exhibits a QBO-like oscillation produced by resolved planetary-scale waves. However, the period is more irregular and the downward propagation faster than observations. Furthermore, compared to the CIRS temperature retrievals, the modeled QBO-like oscillation underestimates by half both the amplitude of temperature anomalies at the equator and the vertical characteristic length of this equatorial oscillation. This QBO-like oscillation is mainly driven by westward-propagating waves; a significant lack of eastward wave-forcing explains a fluctuating eastward phase of the QBO-like oscillation. We also show that the seasonal cycle of Saturn is a key parameter of the establishment and the regularity of the equatorial oscillation. At 20$^{\circ}$N and 20$^{\circ}$S latitudes, the DYNAMICO-Saturn GCM exhibits several strong seasonal eastward jets, alternatively in the northern and southern hemisphere. These jets are correlated with the rings' shadowing. Using a GCM simulation without rings' shadowing, we show its impact on Saturn's stratospheric dynamics. Both residual-mean circulation and eddy forcing are impacted by rings' shadowing. In particular, the QBO-like oscillation is weakened by an increased drag caused by those two changes associated with rings' shadowing.
\end{abstract}

\end{frontmatter}

%%%%%%%%%%%%%%%%%%%%%%%%%%%%%%%%%%%%%%%%%%%%%%%%%%%%%%%%%%%%%%%%%%%%%%%%%%%%%%%%%%%%%%%%%%%%%%%%
%%%%%%%%%%%%%%%%%%%%%%%%%%%%%%%%%%%%%%%%%%%%%%%%%%%%%%%%%%%%%%%%%%%%%%%%%%%%%%%%%%%%%%%%%%%%%%%%
\section{Introduction}

The longevity of the Cassini mission permitted an unprecedented spatial and seasonal coverage of Saturn's stratosphere. In particular, the Composite InfraRed Spectrometer (CIRS) instrument on board Cassini revealed stratospheric phenomena analogous to ones occuring in Earth's and Jupiter's stratospheres \citep{dowling_2008}. 

%%%%%%%
Firstly, seasonal monitoring of hydrocarbons in Saturn's stratosphere suggests a conceivable inter-hemispheric transport of stratospheric hydrocarbons \citep{guerlet_2009,guerlet_2010,sinclair_2013,fletcher_2015,sylvestre_2015}, similar to the Earth's Brewer-Dobson circulation which affects the stratospheric ozone distribution \citep{murgatroyd_1961,dunkerton_1979,solomon_1986,butchart_2014}. The Cassini mission further revealed a lack of temperature minimum under the rings' shadow (that was expected by the radiative balance \cite{fletcher_2010,friedson_2012,guerlet_2009,guerlet_2010,guerlet_2014}), which is an additional hint of subsidence motion in the winter hemisphere.

%%%%%%%%
Secondly, a crucial discovery is that temperatures retrieved from thermal infrared spectra in Saturn's stratosphere exhibit an equatorial oscillation with semi-annual periodicity \citep{orton_2008,fouchet_2008}. During the 13-Earth-years cruise of Cassini around Saturn, CIRS and radio occultations measurements permitted to characterize this equatorial oscillation of temperature and its downward propagation over seasonal timescales \citep{guerlet_2011,li_2011,schinder_2011,guerlet_2018}. Alternatively eastward and westward stacked jets are associated with the temperature signatures detected by Cassini, according to the thermal wind equation. This oscillation is called Saturn ``Quasi-Periodic Oscillation'' or ``Semi-Annual Oscillation'', due to its half-Saturn-year periodicity. Equatorial oscillations appear to be common phenomena in planetary stratospheres. Observations of Jupiter's stratosphere revealed an equatorial oscillation of temperature \citep{leovy_1991,orton_1991} associated with a wind propagation reversal \citep{friedson_1999,simon-miller_2007}. This oscillation exhibits a period of 4.4 Earth years and has been designated as the jovian Quasi-Quadrennial Oscillation. Equatorial oscillations are also suspected on Mars \citep{kuroda_2008,ruan_2019}. Those equatorial oscillations in planetary atmospheres are reminiscent of the Earth's stratospheric Quasi-Biennial Oscillation (QBO) and Semi-Annual Oscillation (SAO) \citep{lindzen_1968,andrews_1983,baldwin_2001}.

%%%%%%%%
Earth's Quasi-Biennial Oscillation (QBO) results from wave-mean flow interactions \citep{reed_1961,andrews_1983,baldwin_2001}. The QBO is driven by the vertical propagation of tropospheric waves -- both planetary-scale and mesoscale waves -- to stratospheric altitudes where they break either by becoming convectively unstable or by encountering critical level (depending on vertical and horizontal wavelengths, wave phase speeds and vertical shear \citep{lindzen_1968,dunkerton_1997}). Critical levels play a central role in driving the alternate eastward/weastward QBO phases by zonal momentum transfer. Radiative and mechanical damping of the waves also induce wave absorption by the mean flow, as suggested by \cite{holton_1972}. Equatorially trapped waves carry eastward and westward momentum. During wave breaking, momentum is transferred to the mean flow, changing the large-scale zonal wind field and driving an axisymmetric meridional circulation. The observed Earth's QBO displays a vertical stack of warm and cold anomalies due to this meridional circulation: adiabatic warming in the subsidence region produces warm anomalies and adiabatic cooling in the upwelling region produces cold anomalies \citep{plumb_1982}. The induced large-scale thermal perturbations, and zonal wind reversal, oscillate with an average period of 28 months and a mean downward propagation rate of 1 km per month on Earth \citep{reed_1961,mayr_2016}. The eastward and westward phases of the Earth's equatorial oscillation are of similar amplitude and are created by a diversity of eastward- and westward-propagating wave sources \citep{dunkerton_1997,hamilton_2001,giorgetta_2002,ern_2009}. The terrestrial QBO's eastward phase is produced by the wave breaking of eastward-propagating gravity waves (from tropospheric moist convection) and Kelvin waves. Kelvin waves contribute to about 30\% of the eastward forcing \citep{ern_2009}, whereas mesoscale gravity waves contribute to 70\% of the eastward forcing \citep{dunkerton_1997}. The westward phase of the Earth's QBO is induced by Rossby and mixed Rossby-gravity waves' breaking, as they carry westward momentum. Furthermore, inertia-gravity waves transport eastward and westward zonal momentum, thus are involved in both the eastward and westward phases of the QBO \citep{baldwin_2001}.

%%%%%%%%
Despite differences in their periods, Jupiter's and Saturn's equatorial oscillations share similarities with the Earth's \citep{dowling_2008}, which raises questions about the driving mechanisms of the gas giants' equatorial oscillations.
Building on tools developed throughout the long history of Earth's atmospheric modeling, numerical models of global circulation on Jupiter and Saturn have been developed for almost 20 years. Stratospheric oscillations on gas giants have been previously addressed by a handful of those models. Three studies employed a quasi-two-dimensional modeling framework that only resolves meridional and vertical structure and parameterizes longitudinal forcing. On the one hand, \cite{li_read_2000} demonstrated that the major contribution to a QBO-like oscillation in Jupiter's atmosphere was planetary-scale waves (Rossby, mixed Rossby-gravity and Kelvin waves). On the other hand, \cite{friedson_1999} and \cite{cosentino_2017} showed that a parameterization of mesoscale gravity waves enabled to reproduce the observed jovian equatorial oscillation -- with no need to invoke planetary-wave forcing. 

%%%%%%%%%
Regardless of the nature of waves invoked in these studies, QBO-like oscillations are phenomena resulting from wave propagation in longitude, latitude and on the vertical that cannot be fully resolved by 2D models: the 3D propagation of waves, and their impact on global circulation, must be parameterized in 2D models. Thus, to better understand the wave-mean flow interactions leading to the observed gas giants' equatorial oscillations, atmospheric models must solve the three-dimensional dynamics. A fully three-dimensional model to study in detail the global troposphere-to-stratosphere circulation in gas giants has only been recently developed, because of the huge computational resources required to resolve eddies arising from hydrodynamical instabilities that putatively force equatorial oscillations. \cite{showman_2018} were the first to show the development of a QBO-like oscillation in an idealized 3D global primitive-equation model for gas giants and brown dwarfs. Their model uses a random wave forcing parameterization at the radiative-convective boundary to drive equatorial oscillations, and a Newtonian scheme to represent radiative heating/cooling in the thermodynamics energy equation. A stack of eastward and westward jets that migrate downward over time is created in the stratosphere of \cite{showman_2018}'s model, analogous to Earth's QBO, with a range of periods similar to Jupiter's and Saturn's equatorial oscillations. Nevertheless, the QBO-like oscillations depicted in their model occur at higher pressure (between 10$^{5}$ and 10$^{3}$ Pa) than the observations (Saturn's equatorial oscillation extends from 10$^{3}$ to 1 Pa). 

%%%%%%%%
To gain further insights on Saturn's equatorial oscillation, this work aims at adopting the three-dimensional approach of \citet{showman_2018} with a more realistic representation of radiative transfer and wave dynamics in Saturn's stratosphere. The present paper is part IV of a series of papers about global climate modeling of Saturn. In part I, \citet{guerlet_2014} built a complete seasonal radiative model for Saturn. In part II, \citet{spiga_2018} coupled this radiative model to a new hydrodynamical core \citep{dubos_2015} to obtain the DYNAMICO-Saturn Global Climate Model (GCM). Using this GCM tailored for Saturn, \cite{spiga_2018} simulated Saturn's atmosphere from the troposphere to the lower stratosphere for 15 Saturn years at fine horizontal resolution, without any prescribed sub-grid-scale wave parameterization. The DYNAMICO-Saturn GCM simulation described in \citet{spiga_2018} produced consistent thermal structure and seasonal variability compared to Cassini CIRS measurements, mid-latitude eddy-driven tropospheric eastward and westward jets commensurate to those observed (and following the zonostrophic regime as is argued in the part III paper by \cite{cabanes_2019}), and planetary-scale waves such as Rossby-gravity (Yanai), Rossby and Kelvin waves in the tropical waveguide. While the simulations in \citet{spiga_2018} exhibited stacked eastward and westward jets in the equatorial stratosphere, those jets were not propagating downwards, contrary to the observed equatorial oscillation. The likely reason for this is that the model top was too low and the vertical resolution too coarse to address the question of a QBO-like oscillation in the stratosphere.

%%%%%%%%
In this paper, we use the DYNAMICO-Saturn model as in \citet{spiga_2018} with a top of the model extended to the higher stratosphere. As in the part II paper, the simulations presented here include neither a gravity-wave drag nor the bottom thermal forcing invoked in \cite{showman_2018}. Planetary-scale waves involved in the equatorial dynamics are triggered by both baroclinic and barotropic instabilities associated to tropospheric jets in our DYNAMICO-Saturn GCM \citep{spiga_2018}. We describe here the benefits on the stratospheric dynamics of using a wider vertical extent in our GCM. In section \ref{GCM description}, we describe briefly our DYNAMICO-Saturn GCM. Section \ref{spectral_analysis} presents, through comparison between \cite{spiga_2018} and our results, the impact on dynamical spectral regime of raising the model top. Section \ref{QBO-like} focuses on the stratospheric dynamics and wave-mean flow interactions producing a QBO-like oscillation in our Saturn GCM. We also tested the Saturn annual cycle influences in its equatorial oscillation periodicity in section \ref{oscillation/season}. In addition, a strong seasonal wind reversal is obtained at 20$^{\circ}$N and 20$^{\circ}$S. We study this phenomenon and discuss the impact of the rings' shadowing on the stratospheric tropical dynamics in section \ref{oscillation/rings}. Summary, conclusions and perspectives for future improvements are explained in section \ref{conclusions}.
%%%%%%%%%%%%%%%%%%%%%%%%%%%%%%%%%%%%%%%%%%%%%%%%%%%%%%%%%%%%%%%%%%%%%%%%%%%%%%%%%%%%%%%%%%%%%%%%
%%%%%%%%%%%%%%%%%%%%%%%%%%%%%%%%%%%%%%%%%%%%%%%%%%%%%%%%%%%%%%%%%%%%%%%%%%%%%%%%%%%%%%%%%%%%%%%%
\newpage
\section{DYNAMICO-Saturn}
\label{GCM description}

A complete description of the DYNAMICO-Saturn is available in \cite{spiga_2018}. A Global Climate Model (GCM) is composed of two parts: a dynamical core to resolve the Navier-Stokes equations on planetary scales and a physical package which is an ensemble of parameterizations to describe sub-grid-scale processes. The Saturn~GCM developed at Laboratoire de M\'{e}t\'{e}orologie Dynamique employs DYNAMICO, a dynamical core using an icosahedral grid that ensures conservation and scalability properties in massively parallel resources \citep{dubos_2015}. It solves the primitive hydrostatic equations assuming a shallow atmosphere. Our DYNAMICO-Saturn features an optional absorbing ("sponge") layer with a Rayleigh drag acting on the topmost model layers. As in \cite{spiga_2018}, we did not use it in simulations presented in this paper since previous studies \citep{shaw_2007,schneider_2009,liu_2010} show that using such an absorbing layer poses problems of angular momentum conservation.

The physical package used in our DYNAMICO-Saturn is tailored for Saturn, particularly regarding radiative transfer (this is fully detailed in the part I paper by \cite{guerlet_2014}). Radiative transfer computations use correlated-k tables, tabulated offline from detailed line-by-line computations. Methane, ethane and acetylene are included in radiative contributions, as well as H$_2$-H$_2$ and H$_2$-He collision-induced absorption and aerosol layers in the troposphere and the stratosphere. Rings'~shadowing is taken into account in radiative computations. The internal heat flux prescribed at the bottom of the model is used as a boundary condition for radiative transfer calculations. Possible unstable temperature lapse rates, induced after radiative transfer calculations, are modified by a convective adjustment scheme: the vertical profile of temperature is instantaneously brought back to the dry adiabat. 

Most simulation settings in this part IV paper are analogous to those adopted in the part II paper by \cite{spiga_2018}. For our \emph{reference simulation}, we employ DYNAMICO-Saturn with an approximate horizontal resolution of 1/2$^{\circ}$ in longitude/latitude. The time step of calculations is 118.9125 seconds, with physical packages called every half a Saturn day and radiative computations done every 20 Saturn days, because of the long radiative timescales in troposphere and stratosphere of gas giants. The values of all parameters detailed in Table 1 of \citet{spiga_2018} are set similarly in the simulations of this paper. The major difference between \cite{spiga_2018} (Part II) and the present Part IV reference simulation is the inclusion of 61 levels in the vertical dimension, extending from the troposphere p$_{bottom}$ = 3.10$^{5}$ Pa (or 3~bars) to the upper stratosphere p$_{top}$ = 10$^{-1}$ Pa (1~$\mu$bar). 

In this study, we opt for a reasonable yet not particularly high vertical discretization. Unlike \cite{showman_2018}, we do not attempt to test the sensitivity of the modeled equatorial oscillation to vertical resolution. We choose, instead, to make a simulation long enough to reach a steady state in the troposphere (at least 8 simulated Saturn years, see \citet{spiga_2018}) and to simulate several cycles of the equatorial oscillation in the stratosphere. This is crucial to obtain interactions between the troposphere and the stratosphere, and create a conceivable and stable equatorial oscillation. This choice of modeling strategy does not allow us to attest that the results presented here are vertically convergent, which is deferred to another study. It is possible that a simulation with twice as many levels over the same pressure range would give quantitatively different results.

Simulations using our DYNAMICO-Saturn are initialized at every horizontal grid point with a vertical profile of temperature computed by a 1D radiative-convective equilibrium model of Saturn's atmosphere  \citep{guerlet_2014}, using the same physical parameterizations than our DYNAMICO-Saturn GCM. The single-column simulation starts with an isothermal profile and runs for twenty Saturn years to reach the annual-mean steady-state radiative-convective equilibrium. The initial zonal and meridional winds are set to zero in our simulations. DYNAMICO-Saturn simulations requires radiative and dynamical spin-up of about five simulated Saturn years for the tropopause levels to ensure a dynamical steady-state \citep{spiga_2018}. In what follows, we present 13-Saturn-year-long simulations to study Saturn's stratospheric dynamics. 

%%%%%%%%%%%%%%%%%%%%%%%%%%%%%%%%%%%%%%%%%%%%%%%%%%%%%%%%%%%%%%%%%%%%%%%%%%%%%%%%%%%%%%%%%%%%%%%%
%%%%%%%%%%%%%%%%%%%%%%%%%%%%%%%%%%%%%%%%%%%%%%%%%%%%%%%%%%%%%%%%%%%%%%%%%%%%%%%%%%%%%%%%%%%%%%%%
\section{Impact of the extended model top on Saturn's atmospheric dynamics}
\label{spectral_analysis}
Saturn's atmosphere is dominated by zonal jets, invariant in the longitudinal direction (i.e. flow axisymmetry) and alternatively prograde (eastward) and retrograde (westward) in latitude. 
The spherical curvature of the planetary fluid layer acts to channel kinetic energy into the zonal direction via the so-called $\beta$-effect, where the parameter $\beta = 2 \Omega cos \varphi/a$ is the variation of the Coriolis force with Saturn's radius $a$, rotation rate $\Omega$ and latitude $\varphi$. This dynamical regime is referred to as zonostrophic macroturbulence \citep{sukoriansky02, galperin14}. Further discussions on this topic can be found in the Part III paper \cite{cabanes_2019}.

\begin{figure}
    \centering
    \subfigure[][]{
        \label{subfig:compare_simu_32}
        \centering
        \includegraphics[scale=0.05]{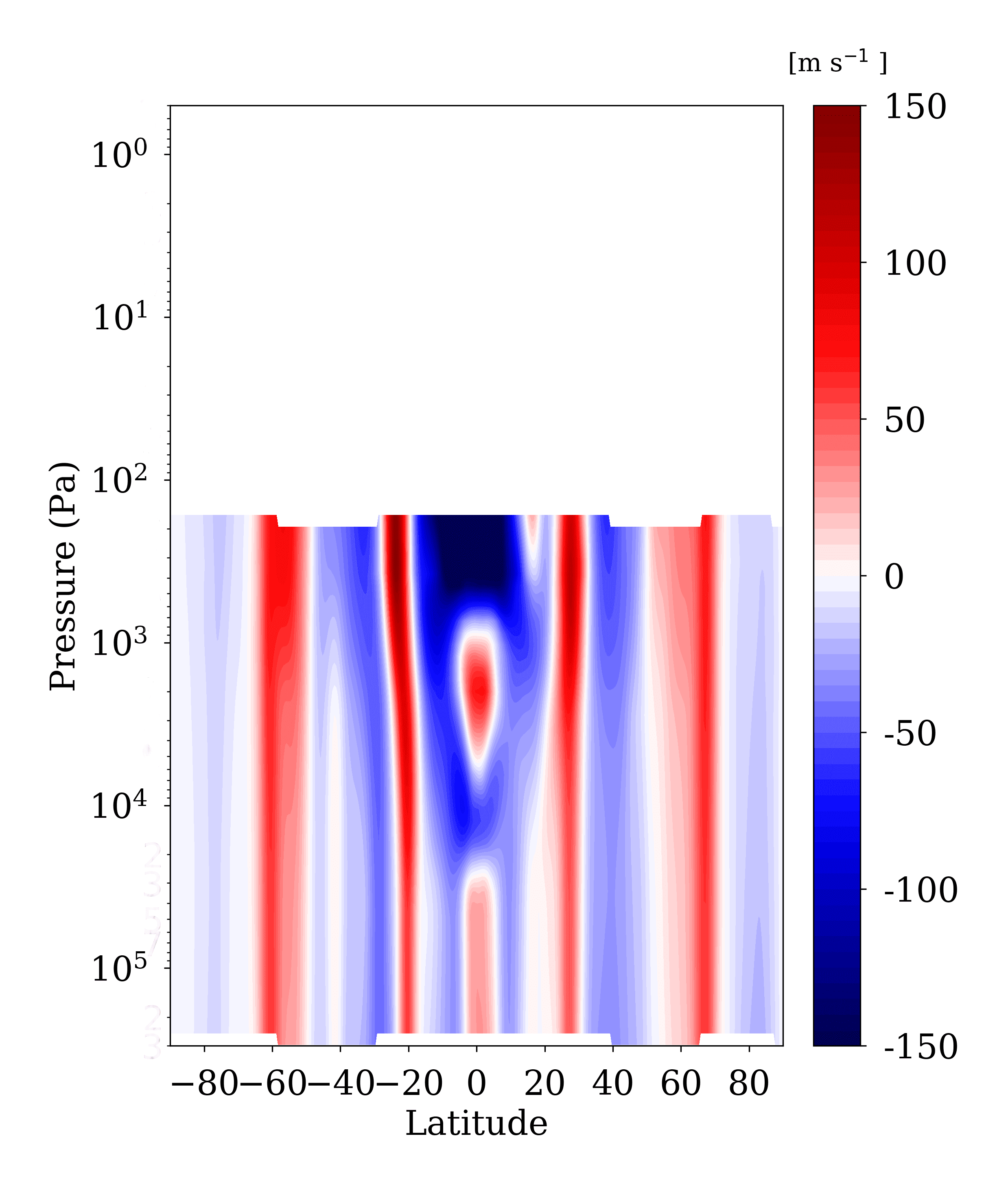}}
    \subfigure[][]{
        \label{subfig:compare_simu_61}
        \centering
        \includegraphics[scale=0.05]{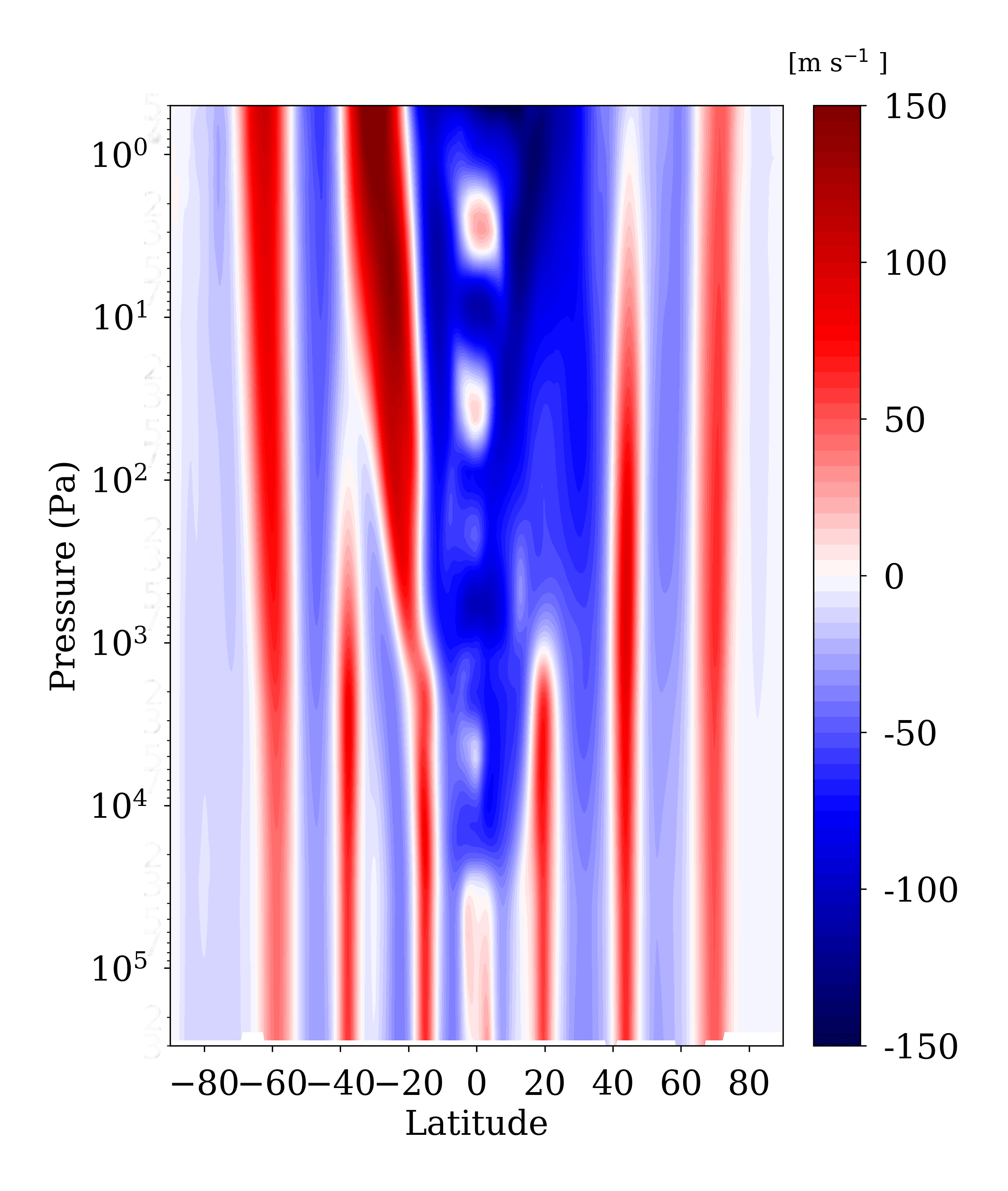}}
    \caption{Altitude/latitude cross sections of the zonal-mean zonal wind, at the northern spring equinox, comparing \cite{spiga_2018} results (32 vertical levels, \ref{subfig:compare_simu_32}) and our results using 61 vertical levels (\ref{subfig:compare_simu_61}).}
    \label{fig:compare_simu}
\end{figure}

Here, we present a comparison of tropospheric and stratospheric jets as simulated by \cite{spiga_2018} versus the present work, to check that the dynamical regime is conserved with the new vertical grid.
Figure \ref{fig:compare_simu} displays altitude/latitude sections of zonal-mean zonal wind after 11 years of simulation for each vertical grid. Both simulations present similar results in the troposphere, although with some notable differences. The 32-vertical-level simulation (\cite{spiga_2018}, Figure \ref{subfig:compare_simu_32}) exhibits about 10 jets, over latitude, each one extending from 3$\times$10$^{5}$ Pa (or 3 bars) to the model top (10$^{2}$ Pa or 1 mbar) except the equatorial one. At the equator, there is a stacking of eastward and westward jets between 2$\times$10$^{4}$ Pa and the model top. With an extended vertical range (Figure \ref{subfig:compare_simu_61}, model top at 10$^{-1}$ Pa), we now distinguish 14 jets into the troposphere and 9 jets in the stratosphere. The equatorial stacking of wind seen in 32-level simulation is now clearly localized between 2$\times$10$^{4}$ Pa (tropopause) and the model top. Contrary to previous GCM works \citep{lian_2008, lian_2010, schneider_2009}, we remark that some of eastward jets of the 61-level simulation, particularly ones closer to the equator, decay with altitude, which is not the case in the 32-level one (except for the equatorial jet). In both simulations, the jets beyond 60$^{\circ}$ latitudes do not exhibit this decay with altitude. Thus the behavior of our 61-level troposphere-to-stratosphere simulation is more consistent with observations for the lower latitudes \citep{fletcher_2019}. We also note that the equatorial tropospheric eastward jet is less intense in our reference (61-level) simulation than in the 32-level simulation of \citet{spiga_2018}.

\begin{figure}
    \centering
    \includegraphics[scale=0.1]{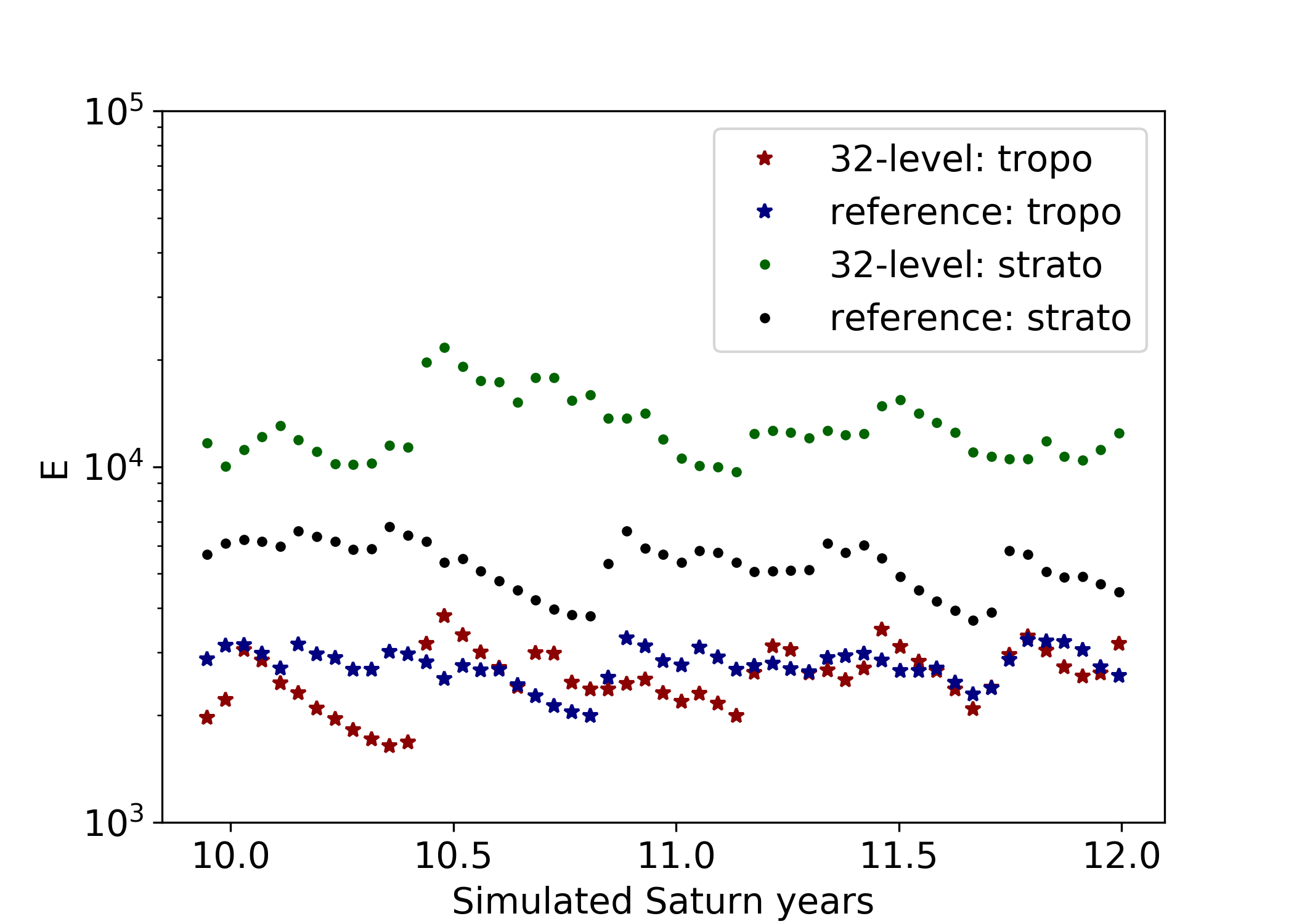}
    \caption{Time evolution of the total energy (m$^{2}$ s$^{-2}$) in the two last simulated Saturn years for each simulations. Red and green curves correspond to the 32-level simulation at $p \sim$ 1.9$\times$10$^{4}$ Pa and $p \sim$ 4.9$\times$10$^{2}$ Pa respectively. Blue and black curves correspond to the reference simulation at $p \sim$ 1.9$\times$10$^{4}$ Pa and $p \sim$ 4.9$\times$10$^{2}$ Pa respectively.}
    \label{fig:Ek_all_simu}
\end{figure}

%%%%%%%%
We now extend our comparison by computing the spherical harmonic decomposition of horizontal velocity maps from our 61-level simulation, using the same tools as in the Part III paper \citep{cabanes_2019} for the 32-level simulation. When spherical harmonic functions are invoked, energy spectra depend on non-dimensional total and zonal indices $n$ and $m$ respectively \citep{boer83}. The axisymmetric flow component defines the jets and  corresponds to the zonal index $m=0$. The non-axisymmetric flow component, or residual, defines waves and eddies and corresponds to all other indices $m \neq 0$. We report on Figure~\ref{fig:Ek_all_simu} the time evolution of the total energy integrated over all indices, for the upper troposphere and the lower stratosphere. The general tendency shows that the total energy in the troposphere is similar in our 61-level simulation and in the 32-level simulation of \citet{spiga_2018}. However, the stratosphere is more energetic than the troposphere and the energetic intensity reduces by a factor of two when we extend the model top in the 61-level simulation.

\begin{figure}
    \centering
    \begin{subfigure}
        \centering
        \includegraphics[scale=0.08]{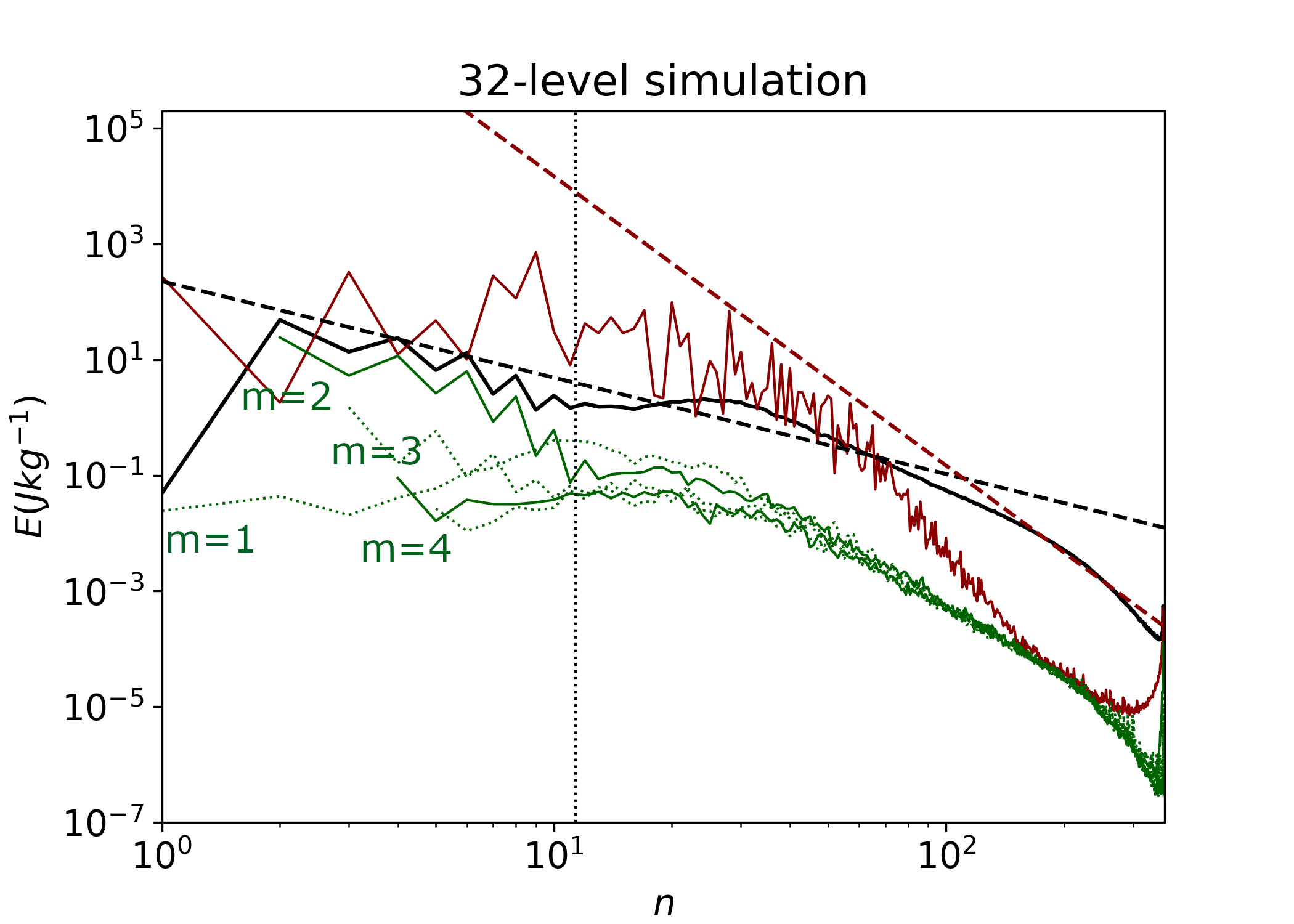}
    \end{subfigure}
    \begin{subfigure}
        \centering
        \includegraphics[scale=0.08]{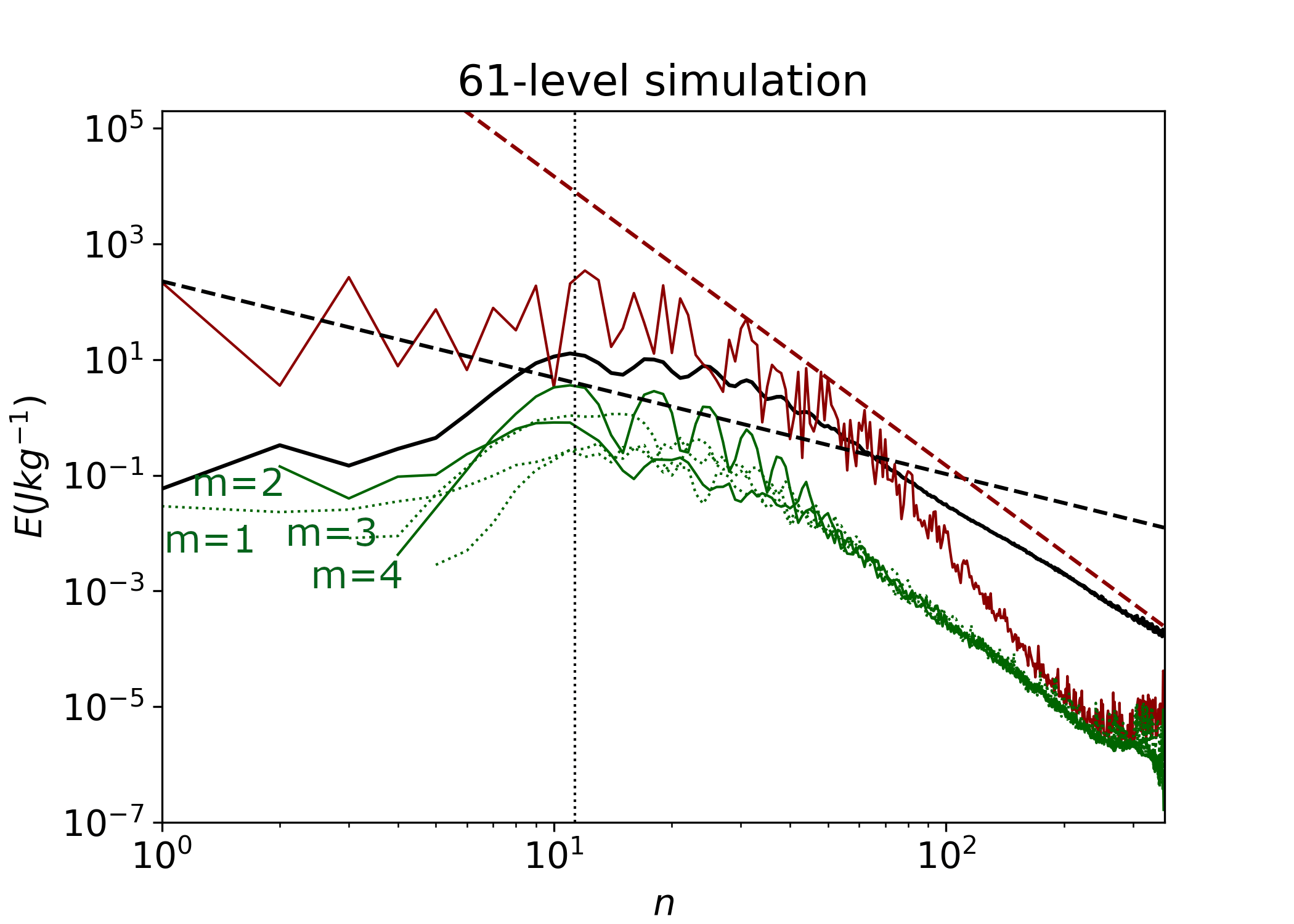}
    \end{subfigure}
    \begin{subfigure}
        \centering
        \includegraphics[scale=0.08]{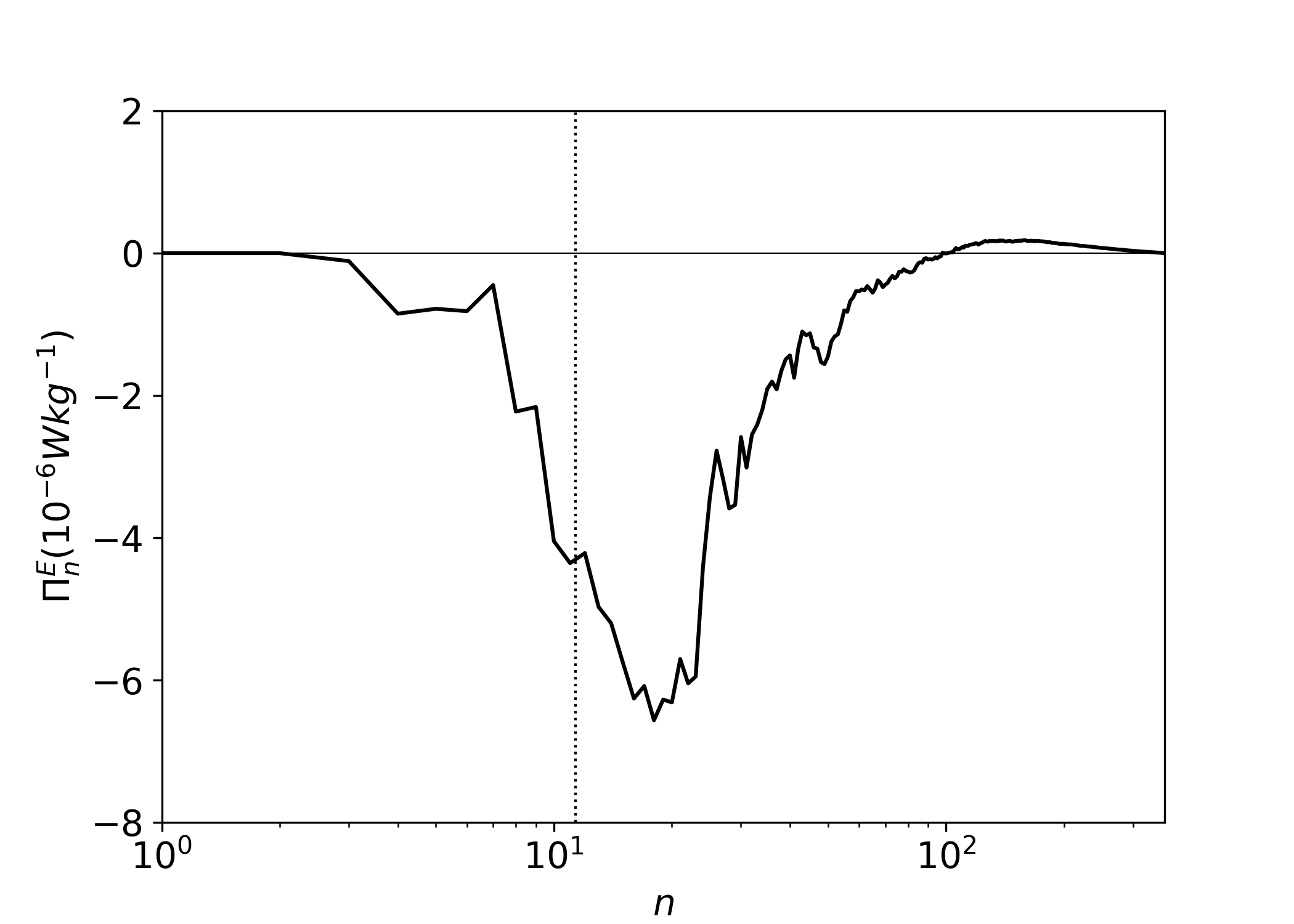}
    \end{subfigure}
    \begin{subfigure}
        \centering
        \includegraphics[scale=0.08]{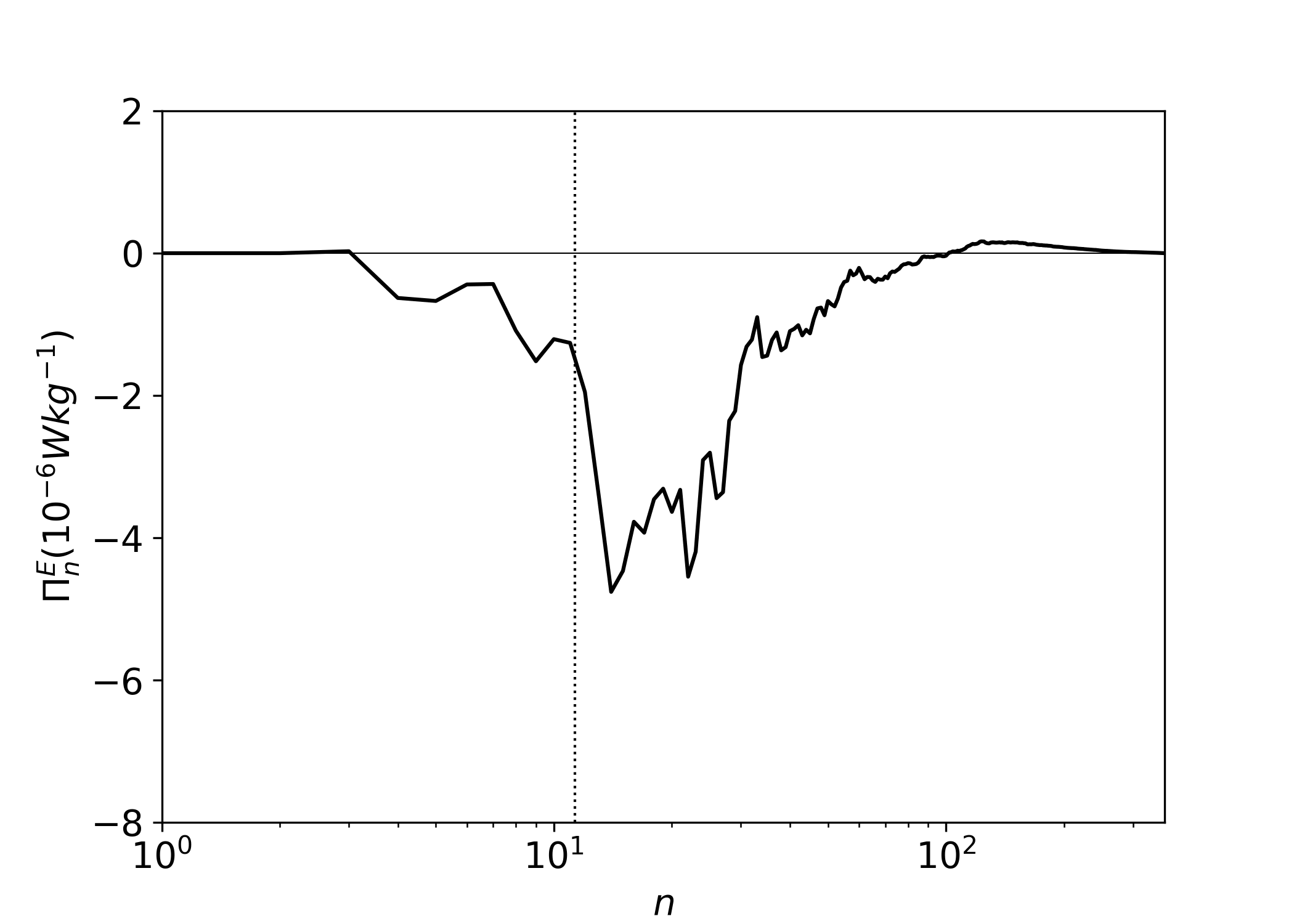}
    \end{subfigure}
    \begin{subfigure}
        \centering
        \includegraphics[scale=0.08]{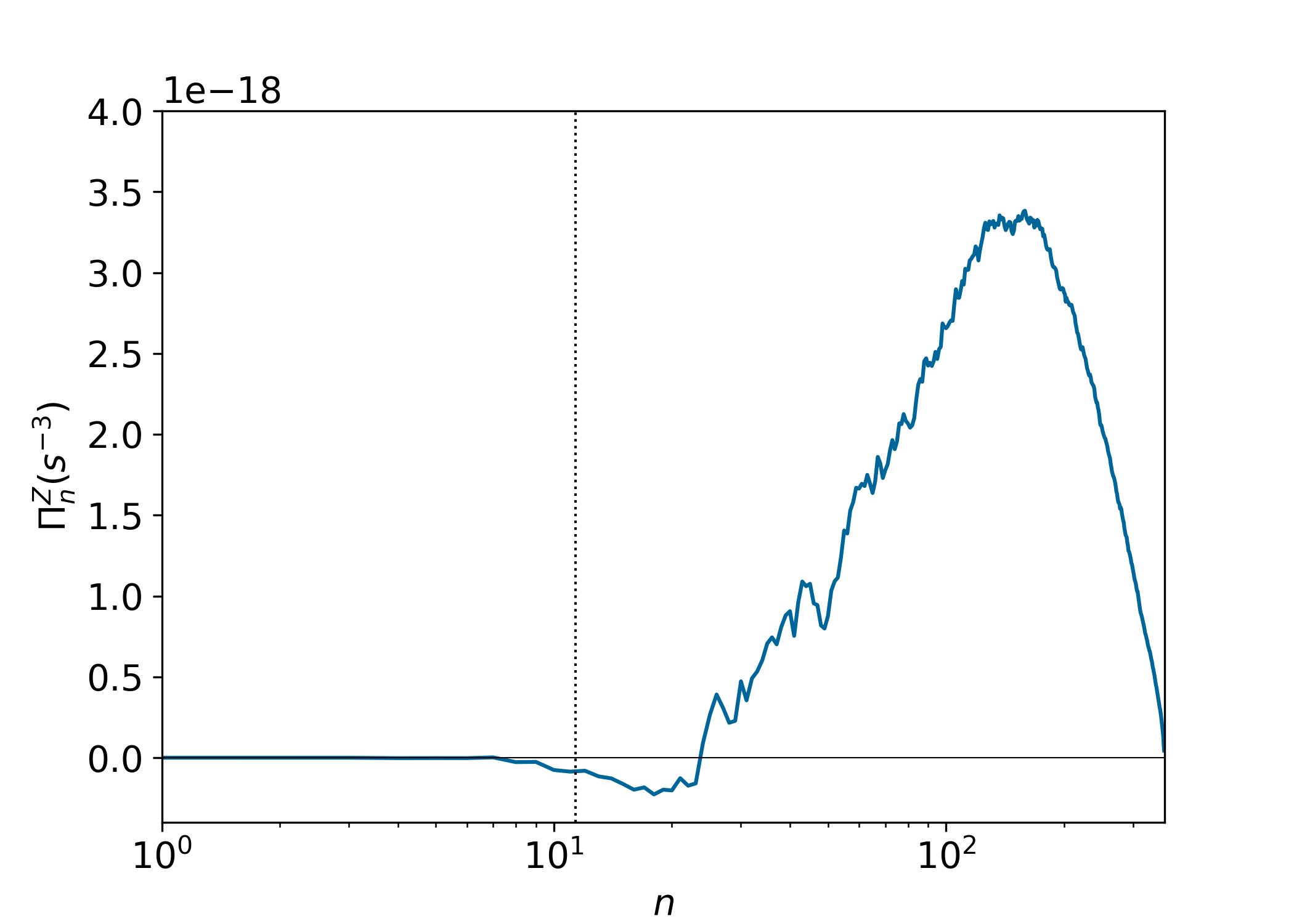}
    \end{subfigure}
    \begin{subfigure}
        \centering
        \includegraphics[scale=0.08]{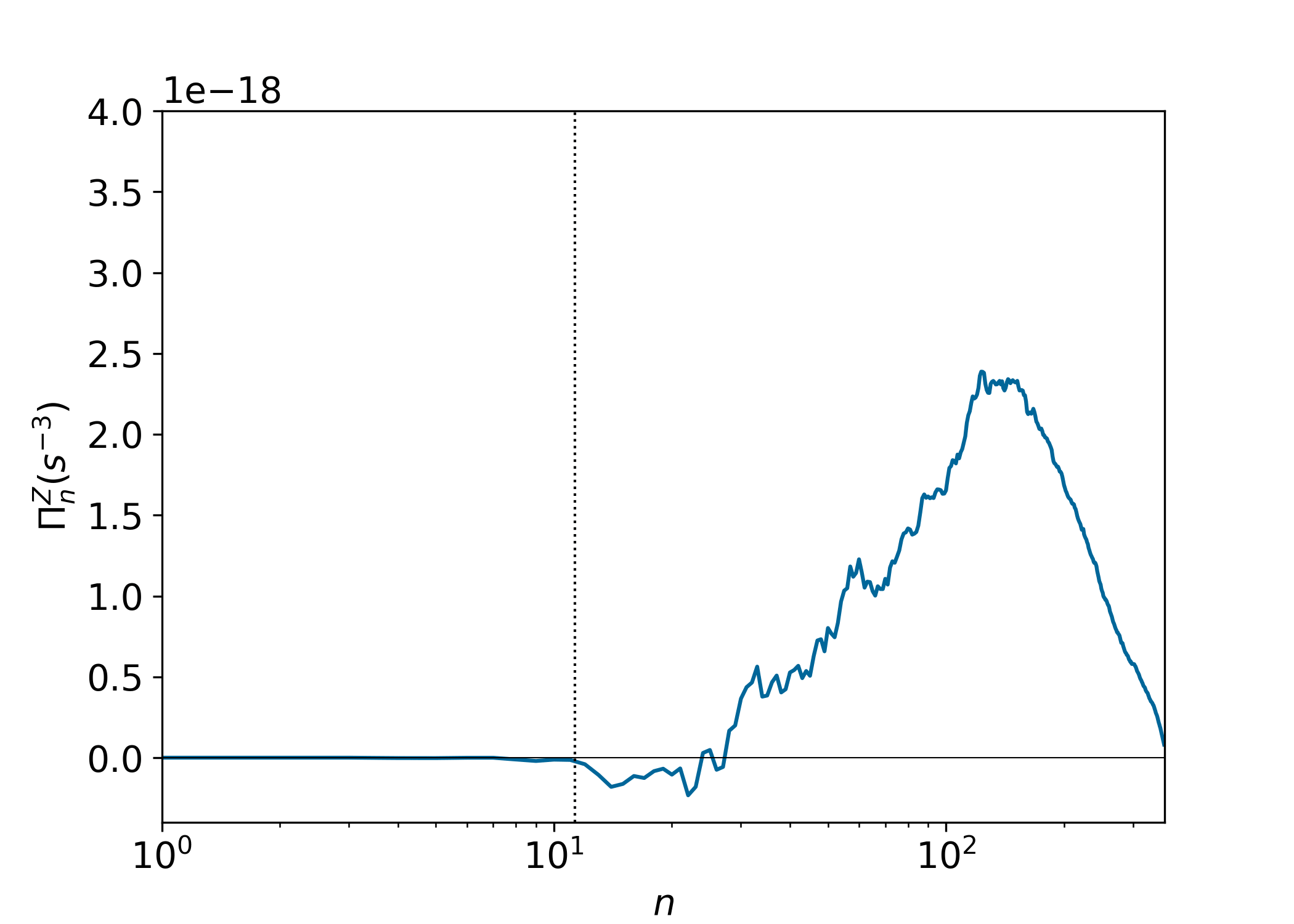}
    \end{subfigure}
    \caption{Comparative spectral analysis of the 32-level simulation (left) and the reference simulation (right) in the upper tropospheric level $p \sim$ 1.9$\times$10$^{4}$ Pa. Spectral quantities are averaged in time, over the tenth to twelfth simulated Saturn years, and in altitude, over pressure levels  (1.7 $< p <$ 2)$\times$10$^{4}$ Pa. Top panels show kinetic energy spectra: the zonal spectra in red (i.e. axisymmetric $m=0$ mode), the residual spectra in black (i.e. sum over all non-axisymmetric $m \neq 0$ modes shown separately), the modal spectra are shown for zonal modes $m$ = 1 to 5 in green. Middle and bottom panels represent the spectral energy and enstrophy fluxes respectively. Positive fluxes corresponds to downscale energy/enstrophy transfers (a ``direct''  cascade) and negative fluxes are upscale energy/enstrophy transfers (an ``inverse''  cascade). All spectral quantities are function of the total indices $n$. At any typical length scale in latitude L, one can attribute a typical index by the relation $n=2 \pi a /L$. Dashed lines are (black) the theoretical Kolmogorov-Kraichnan law $6 \epsilon^{2/3} n^{-5/3}$ with the energy transfer rate $\epsilon = 2 \times 10^{-6}$ W kg$^{-1}$ \citep{kraichnan_1967} and (red) the theoretical zonostrophic law $0.5 \beta^2 n^{-5}$. The vertical dotted lines are the Rhines indices $n_R = R(\beta/2U)^{1/2}$, with $U$ the square root of the total energy presented in Figure~\ref{fig:Ek_all_simu}. It corresponds to the typical length scale $\sim 32,000$ km for both simulations. All our calculation consider a parameter $\beta = 4\times10^{-12}$ m$^{-1}$ s$^{-1}$ estimated at mid-latitude $\varphi = 45^{\circ}$.}
    \label{fig:compare_simu_spectre}
\end{figure}

\begin{figure}
    \centering
    \begin{subfigure}
        \centering
        \includegraphics[scale=0.08]{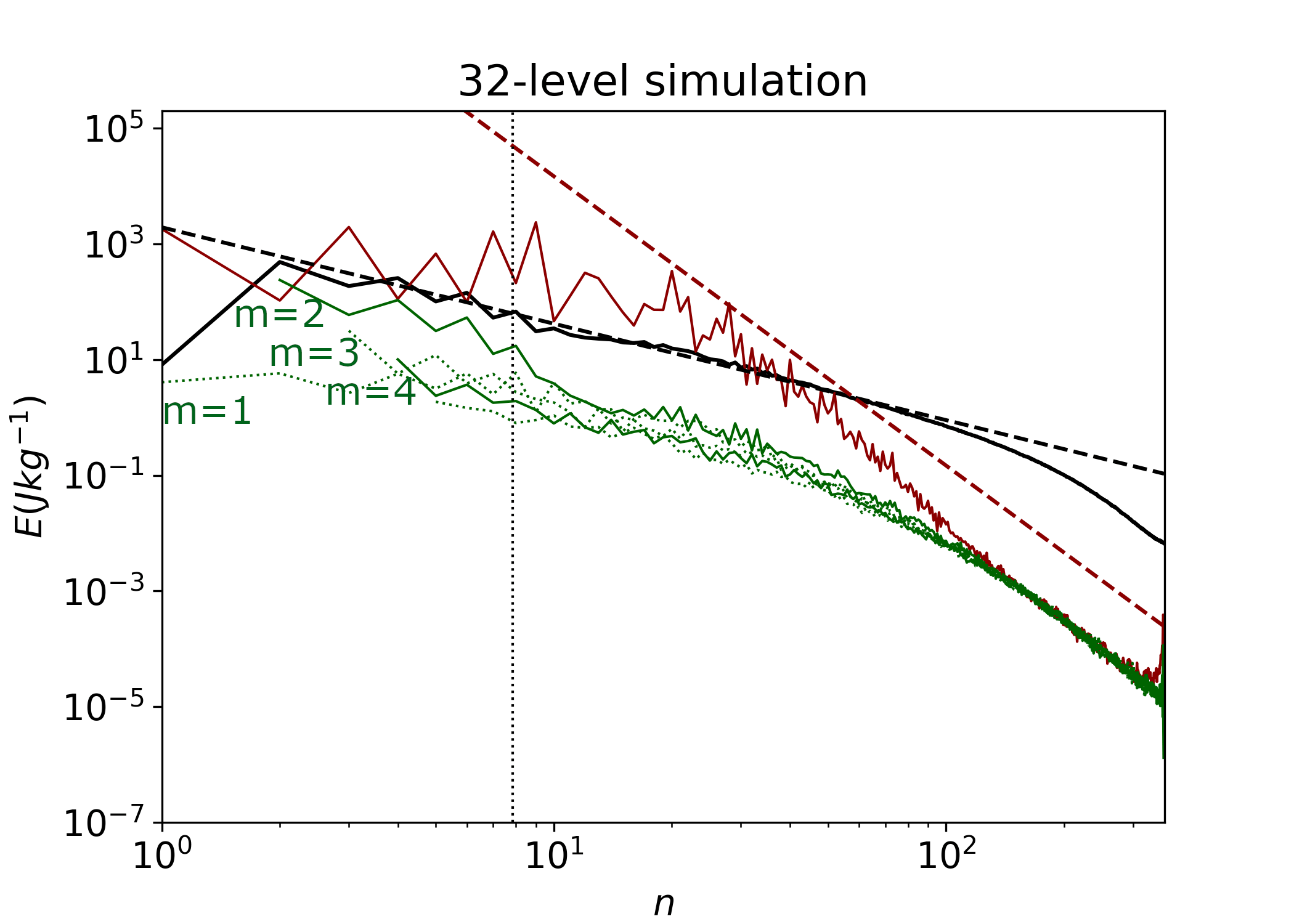}
    \end{subfigure}
    \begin{subfigure}
        \centering
        \includegraphics[scale=0.08]{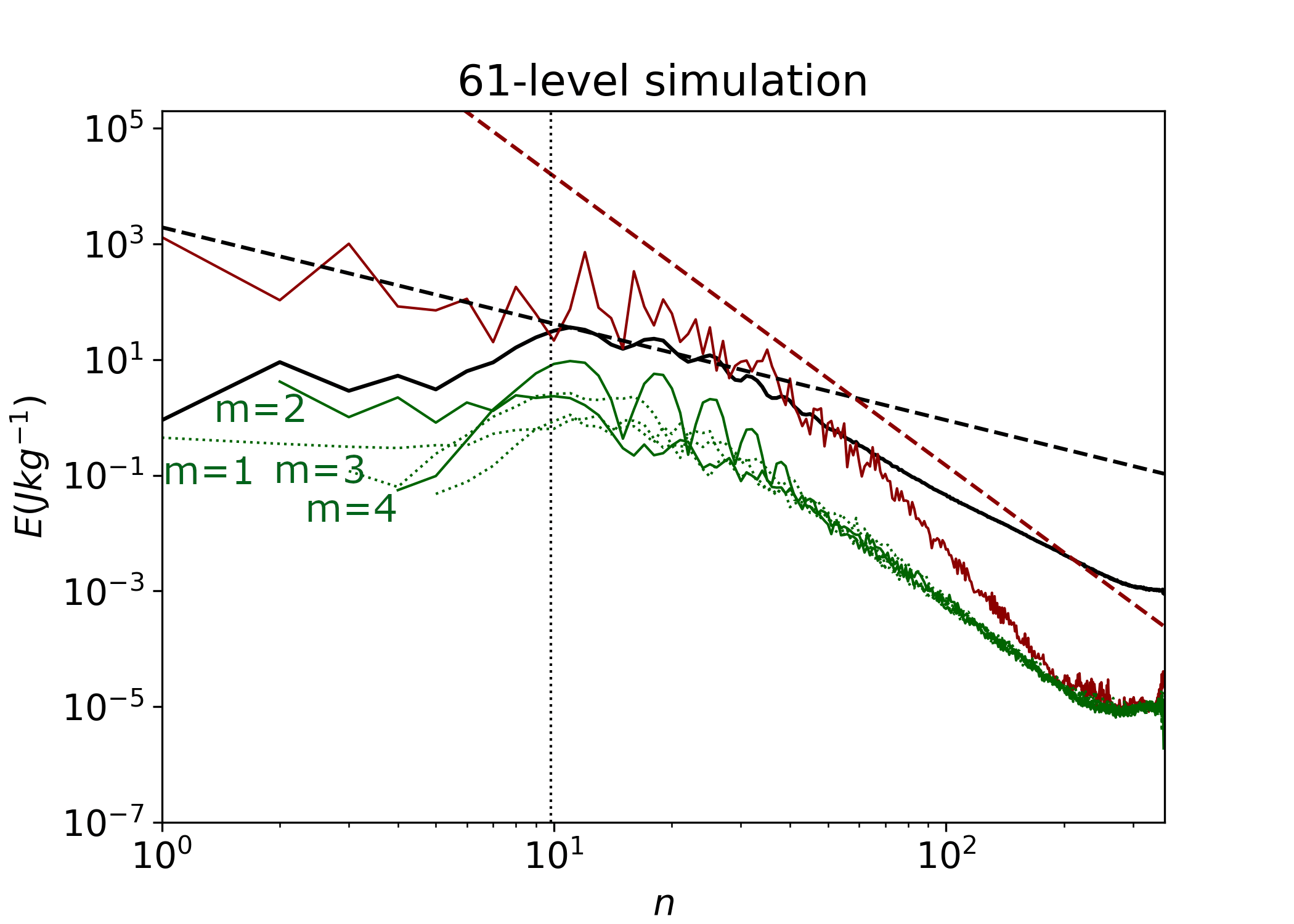}
    \end{subfigure}
    \begin{subfigure}
        \centering
        \includegraphics[scale=0.08]{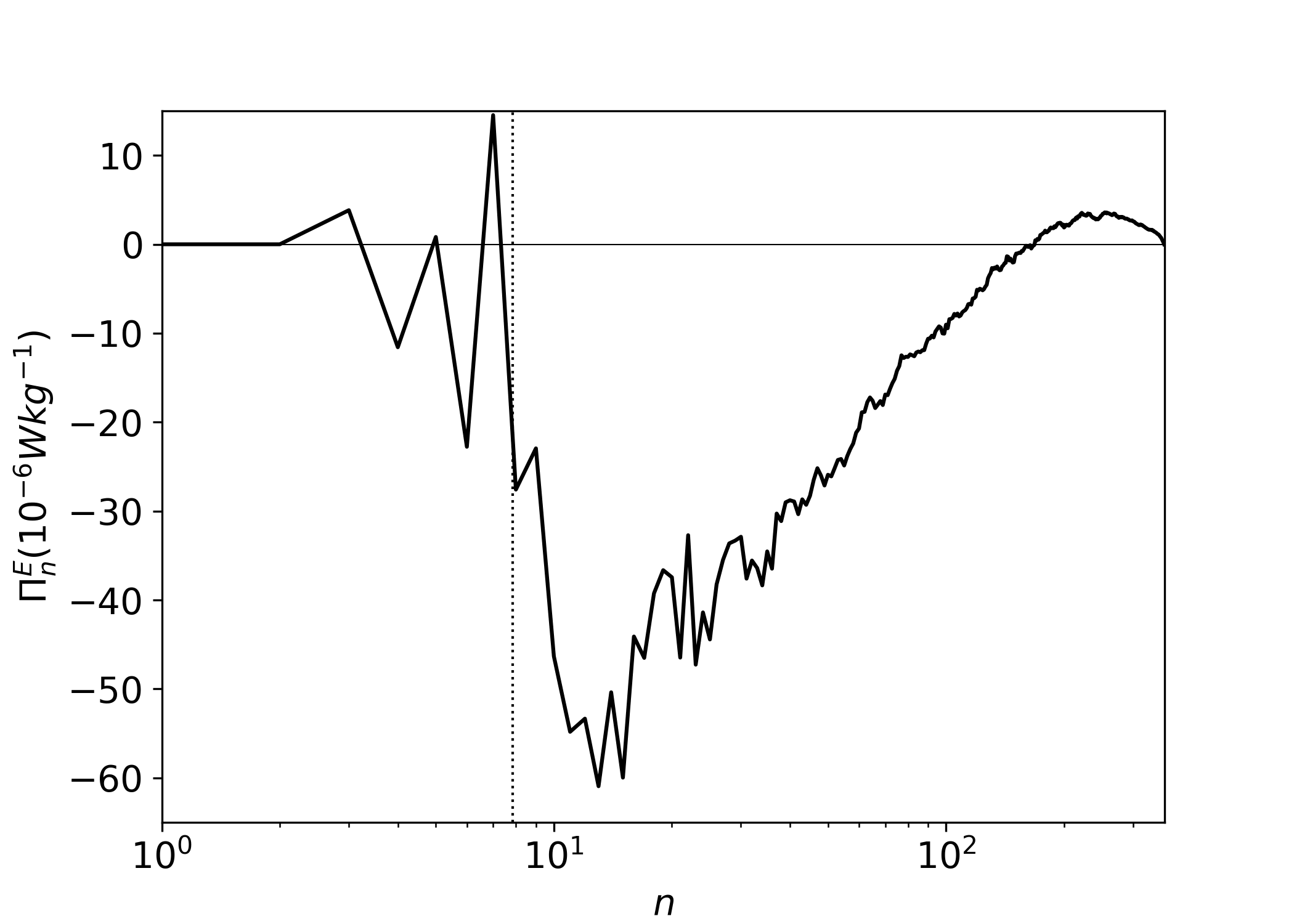}
    \end{subfigure}
    \begin{subfigure}
        \centering
        \includegraphics[scale=0.08]{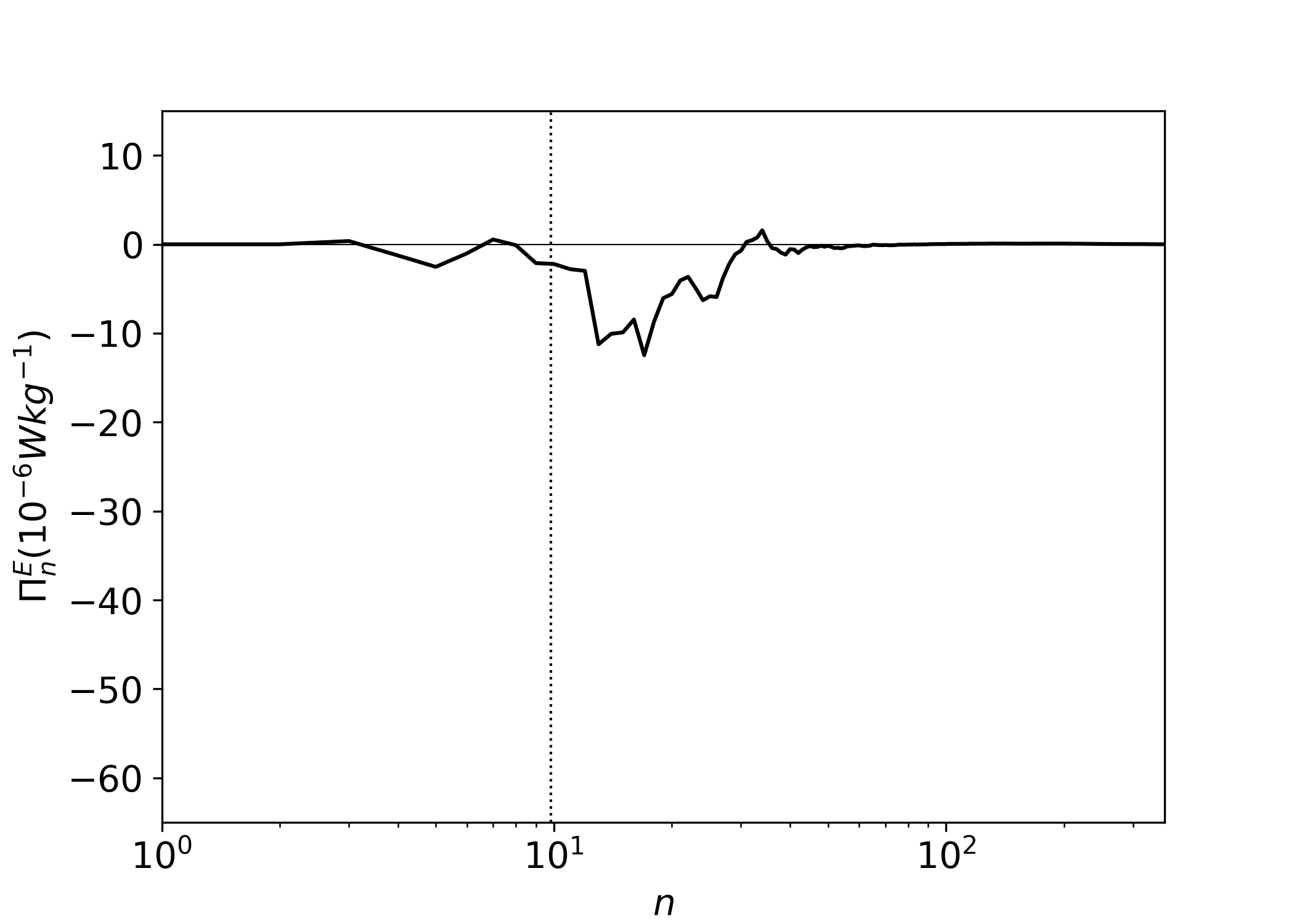}
    \end{subfigure}
    \begin{subfigure}
        \centering
        \includegraphics[scale=0.08]{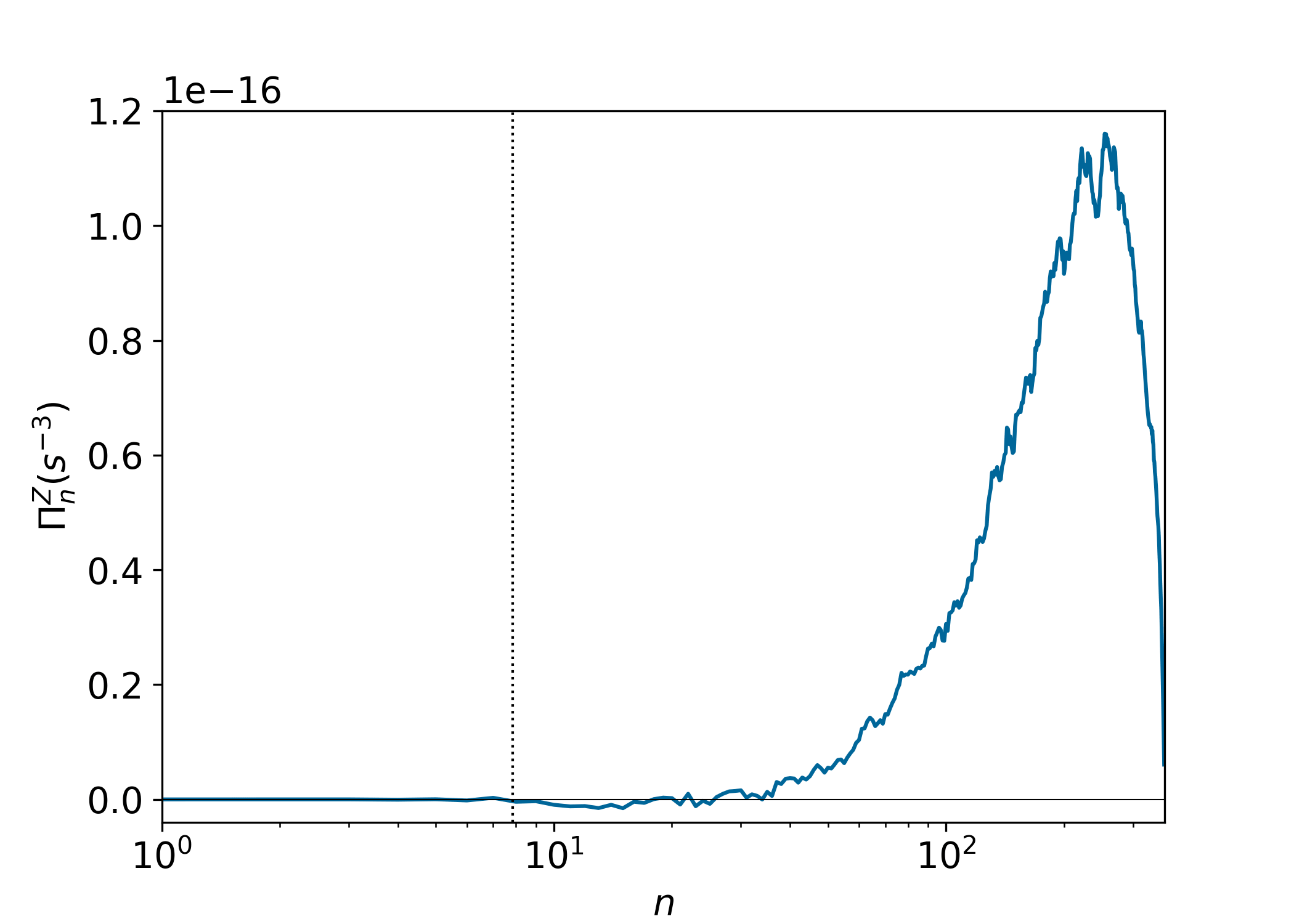}
    \end{subfigure}
    \begin{subfigure}
        \centering
        \includegraphics[scale=0.08]{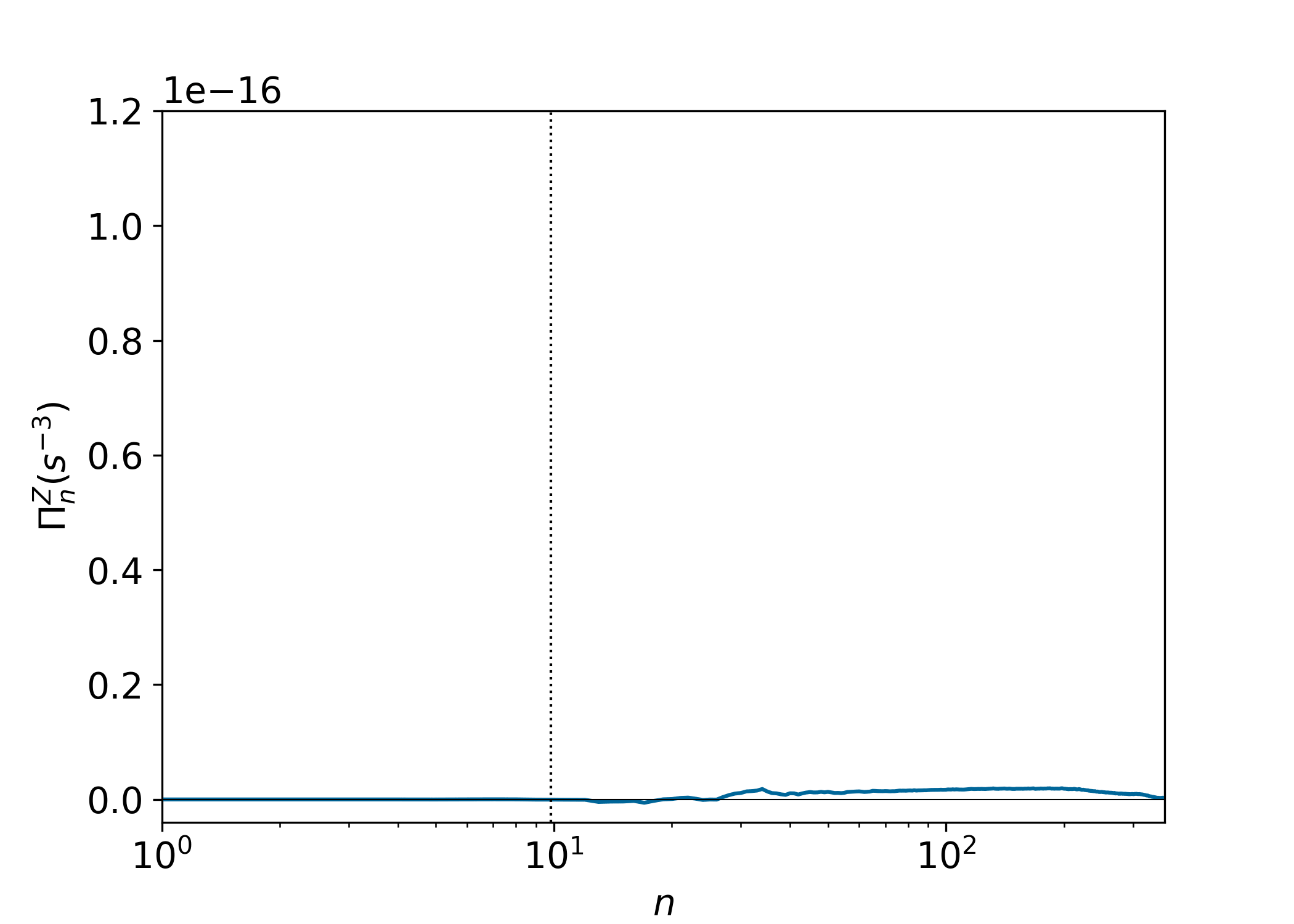}
    \end{subfigure}
    \caption{Comparative spectral analysis of the 32-level simulation (left) and the reference 61-level simulation (right) in the stratospheric level p $\sim$ 4.9$\times$10$^{2}$ Pa. Spectral quantities are averaged in time, over the tenth to twelfth simulated Saturn years, and in altitude, over pressure levels (4.7 $<$ p $<$ 6)$\times$10$^{2}$ Pa. The dashed black lines are the KK-law with the energy transfer rate $\epsilon = 5 \times 10^{-5}$ W kg$^{-1}$  \citep{kraichnan_1967}. The Rhines typical indices correspond to a typical length scale of $\sim 47,000$ and $\sim 37,000$ km for the 32-level and 61-level simulations respectively.  All other spectral quantities are similar to the caption of Figure~\ref{fig:compare_simu_spectre}}
    \label{fig:compare_simu_spectre_toit}
\end{figure}

We show the temporal average of the spectral quantities at steady state in the upper troposphere in Figure~\ref{fig:compare_simu_spectre} and the lower stratosphere in Figure~\ref{fig:compare_simu_spectre_toit}.  A common feature of the 32-level and 61-level simulations is that the energetic magnitude of the zonal jets (i.e. axisymmetric spectra in red) is bounded by the $0.5 \beta^2 n^{-5}$ spectral law of the zonostrophic theory coined by \cite{sukoriansky02}. The energetic maximum is well estimated by the Rhines typical index $n_R = a(\beta/2U)^{1/2}$ \citep{rhines77}, which sets typical jets' size to $\sim 38,000$ km in the troposphere and $\sim 44,000$ km in the stratosphere (see details in Figure~\ref{fig:compare_simu_spectre}). At Rhines scale, the energy of the jets is two orders of magnitude higher than the residual energy spectra, namely the energy contained in waves and eddies. To disclose the dynamical transfers underlying this jet-dominated flow, we also report in Figures~\ref{fig:compare_simu_spectre} and \ref{fig:compare_simu_spectre_toit} the spectral fluxes of energy and enstrophy. In all simulations, spectral fluxes clearly show the predominance of an inverse cascade of energy at large scale and a direct cascade of enstrophy at small scale. This energy-enstrophy double cascade scenario is a well-known feature of quasi-two-dimensional (2D) turbulence, when a turbulent flow is made quasi-2D by confinement to a shallow atmosphere and under rapid planetary rotation \citep{pouquet13,cabanes_2019}. For the 32-level simulation, \cite{cabanes_2019} showed that the inverse cascade of energy results from the barotropization of baroclinic eddies at small scales and drives the large scale atmospheric flow.

What Figures~\ref{fig:compare_simu_spectre} and \ref{fig:compare_simu_spectre_toit} also show is that the eddy spectra is different in our 61-level simulation compared to the 32-level simulation analyzed in \citet{spiga_2018} and \citet{cabanes_2019}. In the 32-level simulation, the tropospheric residual spectrum barely fits the well-known Kolmogorov-Kraichnan (KK) law $ \epsilon^{2/3} n^{-5/3}$ \citep{Kraichnan67} evocative of the inverse turbulent cascade with the energy transfer rate $\epsilon = 2 \times 10^{-6}$ W Kg$^{-1}$. Conversely, in the 61-level reference simulation, the $-5/3$ KK slope does not apply to the residual spectrum that shows an energetic bump at intermediate indices (i.e. $ 10< n < 100$).
The 32-level simulation also features very energetic non-axisymmetric modes $m=1,2,3$ (see modal spectra in green) that dominate in the residual spectrum beyond the Rhines index ($n<n_R$) and extend the $-5/3$ slope to small indices (i.e. down to $n \sim 2$). Such modes or waves are absent in the 61-level simulation and energy of the residual spectrum decreases at indices smaller than the Rhines index. It is likely that the energetic $m=1,2,3$ modes enforced in the 32-level simulation are artifacts of the shallow geometry. By extending the model top, we also release part of the flow confinement in the vertical that promotes a quasi-2D flow. This release might explain the reduction of the energy and enstrophy fluxes in the 61-level simulation compared to the 32-level simulation. The raise of model top towards stratospheric levels appears to significantly impact the statistical properties of the tropospheric flow. Saturn's direct observations, akin to those used for Jupiter in \citet{young17}, are required to further validate our model.

Statistical properties are mostly similar in the upper troposphere and in the lower stratosphere (Figure~\ref{fig:compare_simu_spectre_toit}). The 32-level simulation preserves a $-5/3$ residual spectrum that prevails at all indices in the lower stratosphere. In the 61-level simulation, the residual spectrum is substantially steepened and spectral energy piles up in the range $n>n_R$, to decrease beyond Rhines index. 
In both simulations, the integrated total energy in the stratosphere is more energetic than in the troposphere (Figure~\ref{fig:Ek_all_simu}), with energy fluxes one order of magnitude higher in the stratosphere. 
This might result from the growth of baroclinic instabilities that preferentially form in the stratosphere \citep{cabanes_2019}. 
The integrated total energy also show that the 32-level simulation predicts a flow at stratospheric levels more energetic by a factor of 2 than in the 61-level simulation. In the former simulation, the top of the model is close to the lower stratosphere, therefore an energy accumulation likely produces these high values of total energy -- and might also create wave reflection leading to energetic non-axisymmetric modes $m=1,2,3$.

\begin{figure*}
    \centering
    \begin{subfigure}
        \centering
        \includegraphics[scale=0.08]{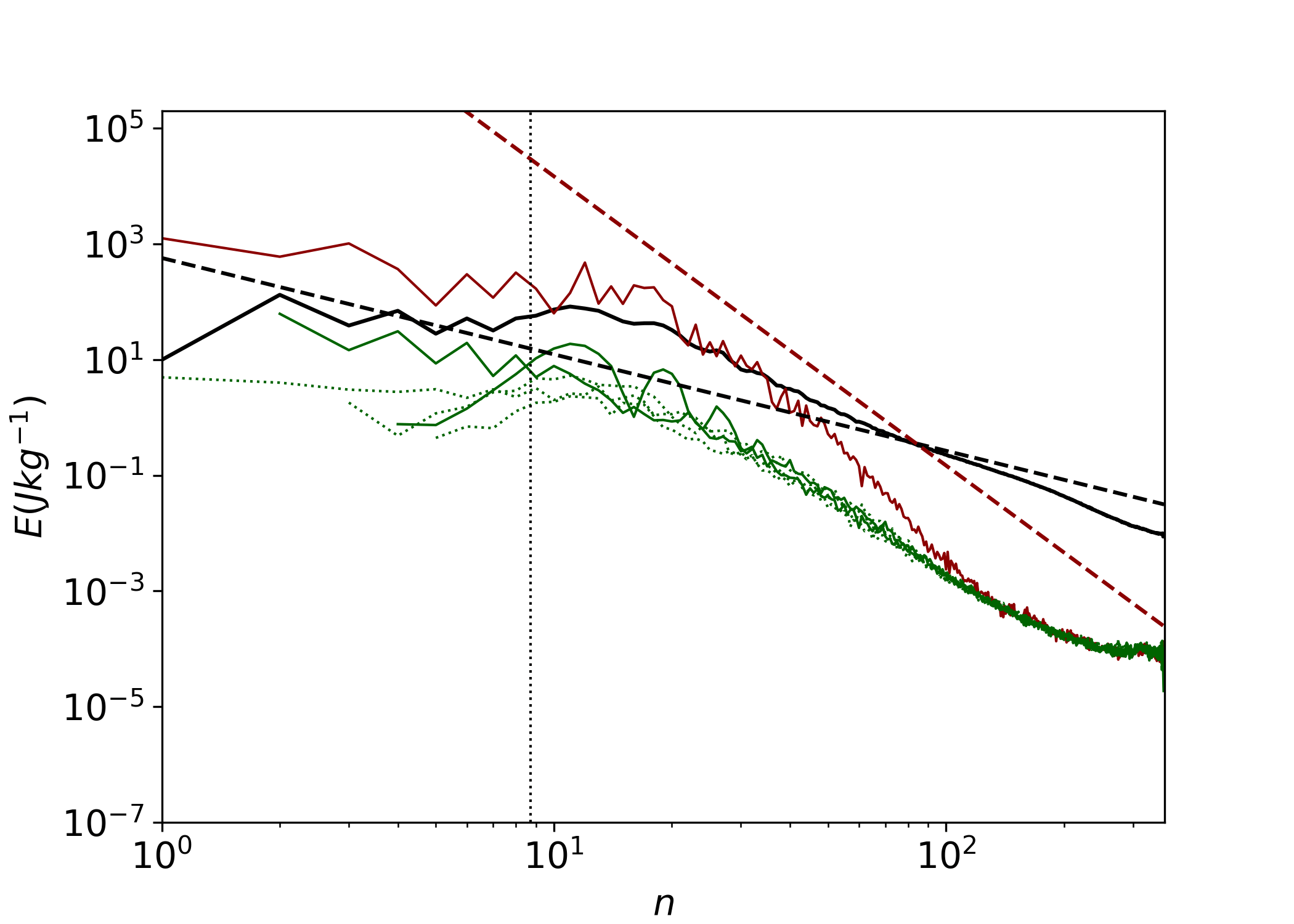}
    \end{subfigure}
    \begin{subfigure}
        \centering
        \includegraphics[scale=0.08]{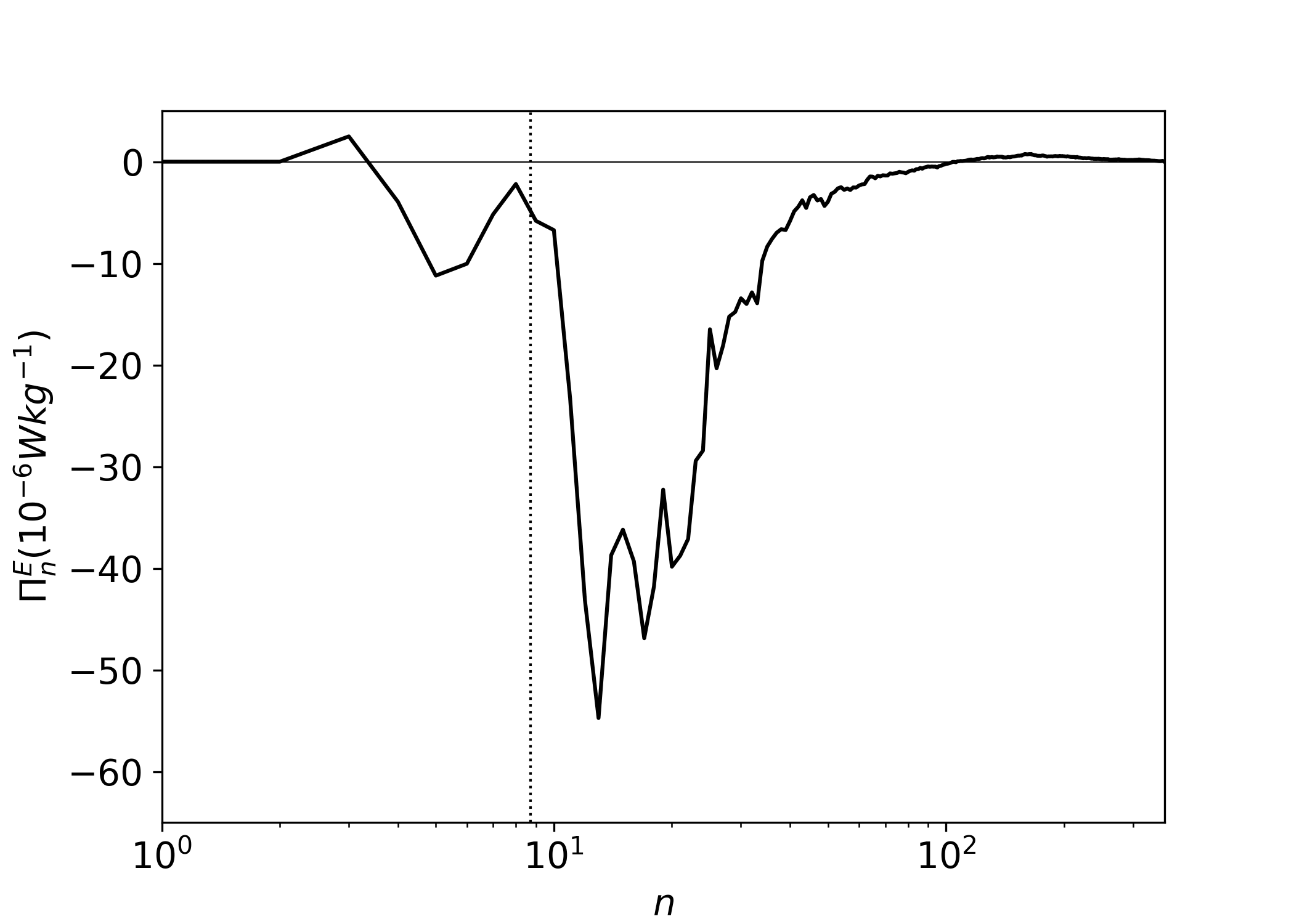}
    \end{subfigure}
    \begin{subfigure}
        \centering
        \includegraphics[scale=0.08]{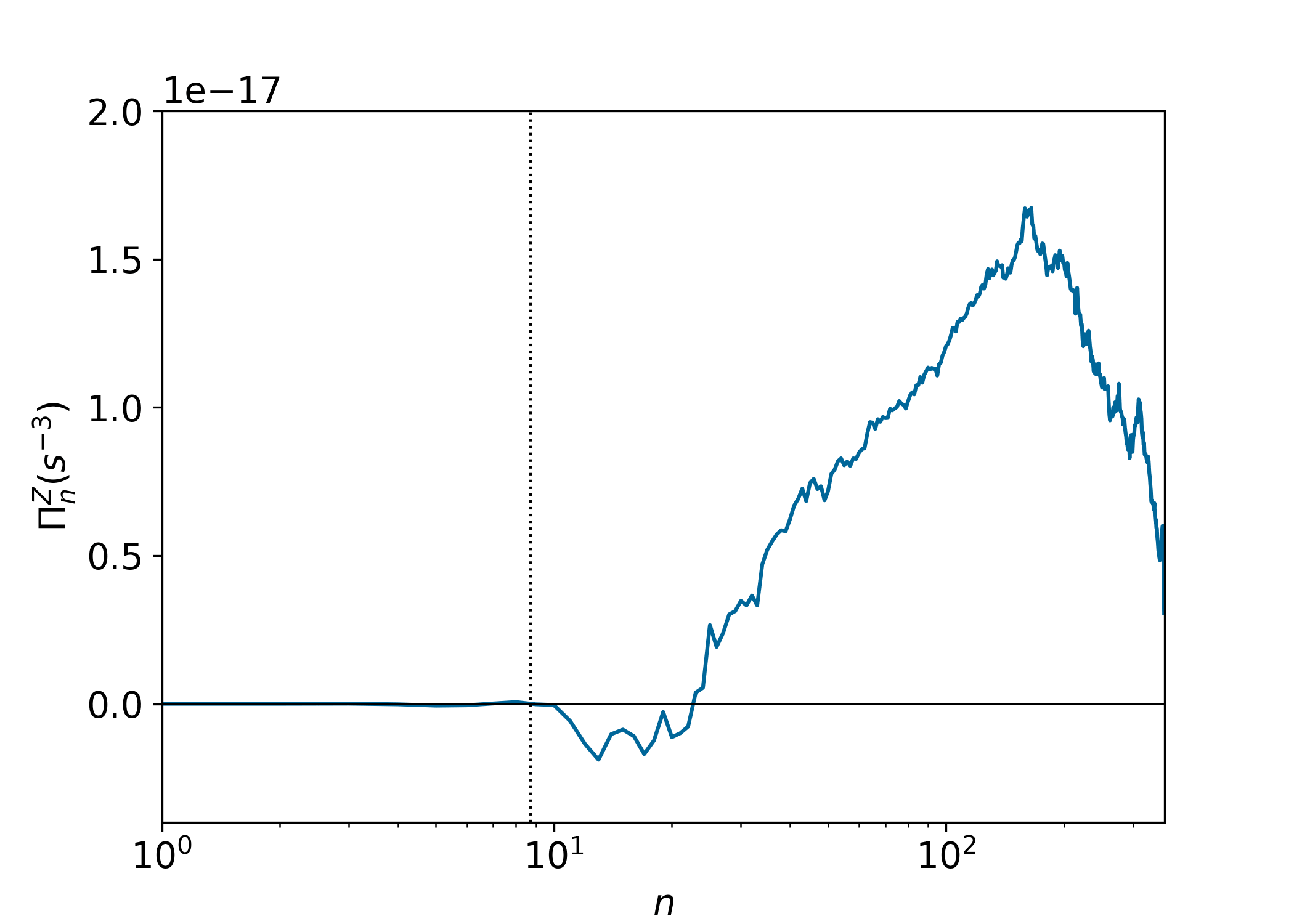}
    \end{subfigure}
    \caption{Spectral analysis of the reference simulation in the stratospheric level p $\sim$ 25 Pa. Spectral quantities are averaged in time, over the tenth to twelfth simulated Saturn years, and in altitude, over pressure levels 21 $<$ p $<$ 27 Pa. The dashed black lines are the KK-law with the energy transfer rate $\epsilon = 8 \times 10^{-6}$ W kg$^{-1}$ \citep{kraichnan_1967}. The Rhines typical indices correspond to a typical length scale of $\sim 42,000$ km.  All other spectral quantities are similar to the caption of Figure~\ref{fig:compare_simu_spectre}}
    \label{fig:strato_spectre}
\end{figure*}

%%%%%%%%
Finally, we propose in Figure \ref{fig:strato_spectre} a spectral analysis similar to those of Figure~\ref{fig:compare_simu_spectre} and \ref{fig:compare_simu_spectre_toit}
for the stratospheric level 25 Pa of our 61-level simulation, left unresolved in the previous Part II and III papers.
At the smallest resolved scales, the zonal kinetic spectrum depicts a -5 slope with estimated values of $\epsilon$=8$\times$10$^{-6}$ W kg$^{-1}$, two orders of magnitude less than the residual kinetic spectrum. As in the troposphere (Figure \ref{fig:compare_simu_spectre}, right top row), the large scales are driven by a strong zonal anisotropy linked to zonostrophic turbulence. In the stratosphere, our simulation indicates that the modes m = 2 and 4 are dominant in the residual spectrum, suggesting that strongly-energetic waves propagate in the zonal direction. The spectral flux of energy depicts huge negative values for moderate $n$, contrary to tropospheric results in Figure \ref{fig:compare_simu_spectre}. The inverse energy cascade, from the small scales to the large scales, is operating in the stratosphere for most intermediate scales between $n = 3$ and $n = 10^{2}$. At both the largest and the smallest resolved scales, the energy flux becomes positive denoting a direct energy cascade (energy is transferred from the large to the small scales). Aside from the positive peak around $n = 3$, the global trend of energy flux is equivalent in the troposphere and the stratosphere. The enstrophy flux confirms the dynamical regime of the stratosphere (bottom row of Figure \ref{fig:strato_spectre}): it is positive for $n$ higher than 10, which suggests the existence of an inverse cascade.

%%%%%%%%%%%%%%%%%%%%%%%%%%%%%%%%%%%%%%%%%%%%%%%%%%%%%%%%%%%%%%%%%%%%%%%%%%%%%%%%%%%%%%%%%%%%%%%%
%%%%%%%%%%%%%%%%%%%%%%%%%%%%%%%%%%%%%%%%%%%%%%%%%%%%%%%%%%%%%%%%%%%%%%%%%%%%%%%%%%%%%%%%%%%%%%%%
\section{Equatorial stratospheric dynamics with the DYNAMICO-Saturn}
\label{QBO-like}

%%%%%%%%
\subsection{Global stratospheric zonal wind}
\label{QBO-like_2_features}

\begin{figure}
    \centering
    \includegraphics[scale=0.1]{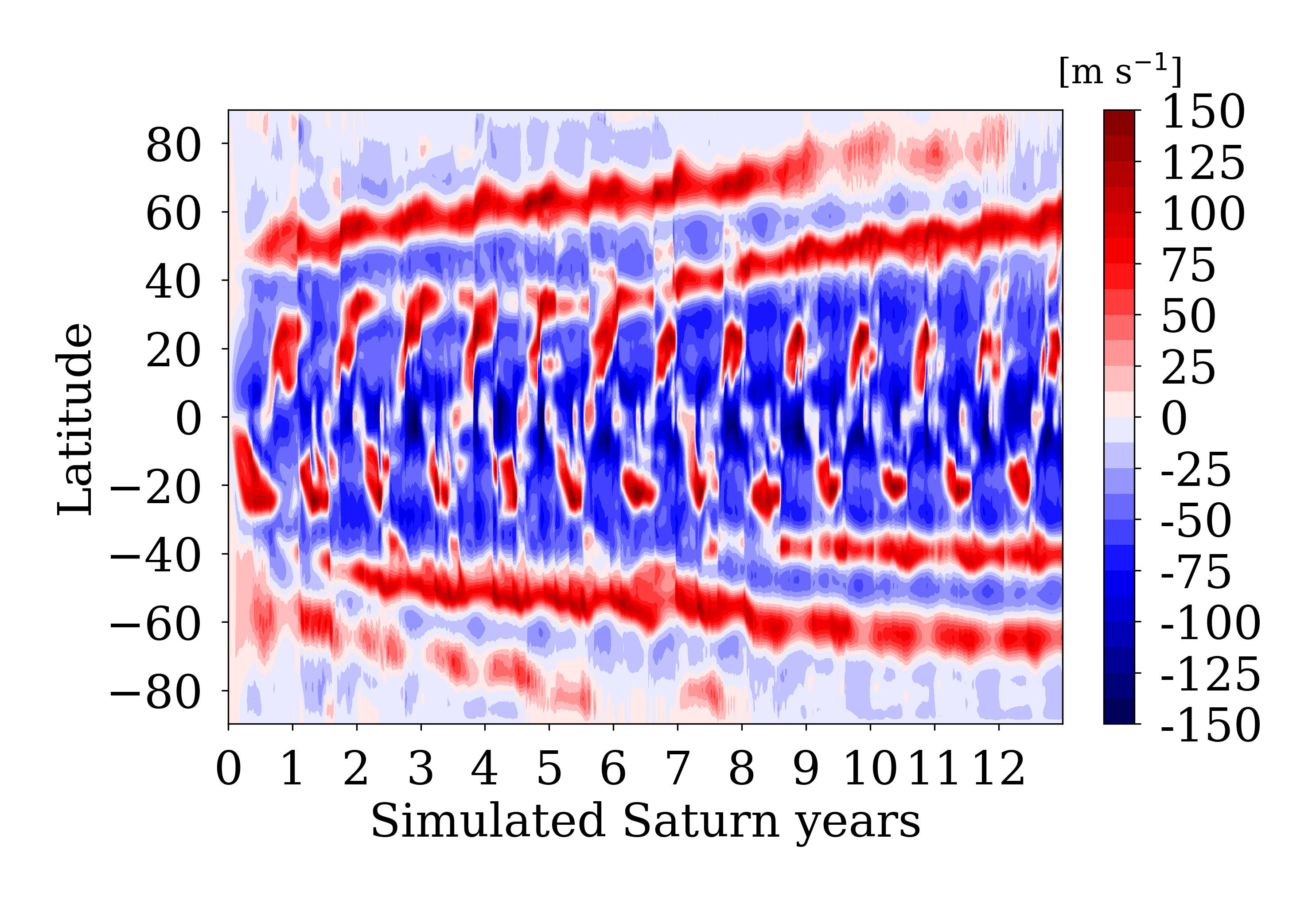}
    \caption{Time evolution of the zonal-mean zonal wind in Saturn's stratosphere (40 Pa) within the whole 13-years duration of our DYNAMICO-Saturn simulation. The wind structure at this pressure level is typical of conditions between the tropopause and the model top in the upper stratosphere.}
    \label{fig:u_time_evo_04mb}
\end{figure}

The temporal evolution of the zonal-mean stratospheric zonal wind over the whole 13-year duration of our DYNAMICO-Saturn simulation is shown in Figure \ref{fig:u_time_evo_04mb} at the pressure level of 40 Pa. 
Before turning to low latitudes, we note that at high and mid-latitudes, zonal jets undergo a poleward migration due to bursts of eddies induced by baroclinic instabilities. This migration was already witnessed by \cite{spiga_2018} in the troposphere and is thought to be partly caused by baroclinicity at the bottom of the model (3$\times$10$^{5}$ Pa), where the meridional gradient of temperature is slightly overestimated compared to the observations. Furthermore, \cite{chemke_2015} show that the poleward migration of baroclinic eddy-driven jets is consistent with an asymmetry in the baroclinic growth around the jet's core: the poleward flank of the jet can be slightly more baroclinically unstable than the equatorward flank. The baroclinic growth precedes the increase of the eddy momentum flux convergence, causes the eddy forcing to be shifted slightly poleward of the jet peak, and thus drives the poleward migration of the jets.

Now, turning to the low latitudes simulated in our reference 61-level Saturn simulation, we notice in Figure \ref{fig:u_time_evo_04mb} two key features in the temporal evolution of the stratospheric zonal-mean zonal wind:
\begin{itemize}
    \item an alternatively eastward and westward wind direction at the equator, with a sub-annual periodicity, appearing in the second simulated year and maintained for the rest of the 13 simulated years;  
    \item an alternatively eastward and westward wind direction at 20$^{\circ}$N and 20$^{\circ}$S, with an annual periodicity and a phase opposition between the northern and the southern hemisphere.
\end{itemize}
These two features are discussed in detail in the following sections.

%%%%%%%%
\subsection{Equatorial stratospheric zonal jets}
\label{QBO-like_zonal_wind}

\begin{figure*}
    \centering
    \includegraphics[scale=0.15]{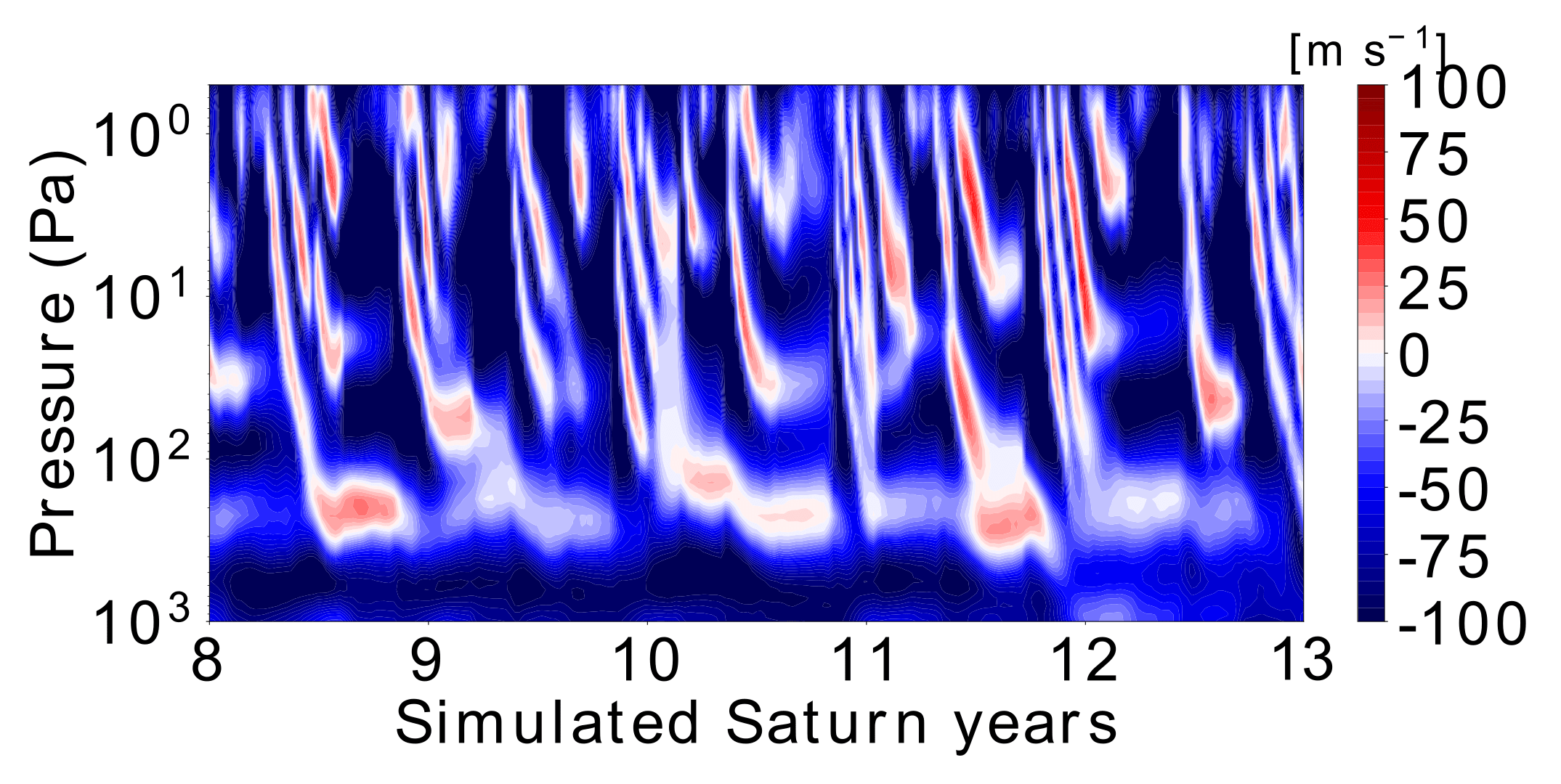}
    \caption{Altitude/time section at the equator of the zonal-mean zonal wind in Saturn's stratosphere for the 5 last years of our 13-years DYNAMICO-Saturn simulation.}
    \label{fig:u_y00_strato}
\end{figure*}

By making an altitude/time section of the zonal-mean zonal wind at the equator (Figure \ref{fig:u_y00_strato}), for the four last years of our 13-years simulation, the downward propagation with time of the stacked stratospheric eastward-westward jets is well visible. At a given pressure level, the zonal wind alternates between eastward and westward direction, with a westward phase more intense -- about -100 m s$^{-1}$ (Figure 9 in \cite{showman_2018} shows a similar behaviour) -- than the eastward one (only 60 m s$^{-1}$). In addition, at a given pressure level, the westward phase lasts longer than the eastward one. The equatorial eastward phases of the oscillation seem to be unstable and disturbed compared to westward phases. For this reason, the period of the resulting equatorial oscillation is irregular in time. 

As is reminded in the introduction, a pattern of alternatively eastward and westward jets stacked on the vertical, with a downward propagation, reminds both the Quasi-Biennial Oscillation (QBO) on Earth and the QBO-like oscillation evidenced in Cassini observations (\cite{fouchet_2008}, \cite{guerlet_2011}, \cite{li_2011}, \cite{orton_2008}, \cite{guerlet_2018}). 
We tracked the equatorial jets to calculate the periodicity and the downward propagation rate of the modeled QBO-like oscillation. Our DYNAMICO-Saturn produces a QBO-like oscillation with a period in the range 0.3-0.7 Saturn year. The corresponding stacked jets' downward propagation rate is respectively between 73 and 60 km per Saturn month. In \cite{fletcher_2017}, Cassini observations were used to estimate a period of 0.50$\pm$0.03 Saturn year and a downward propagation speed of (42$\pm$2) km per Saturn month. \cite{guerlet_2018} estimated a downward propagation of 24.5 km per Saturn month at 500 Pa and 49 km per Saturn month at 10 Pa. Hence, the DYNAMICO-Saturn simulation exhibit an equatorial oscillation periodicity of the right order of magnitude compared to the observations, but more irregular. Moreover, the stacked zonal jets in our simulation propagate downward 1.5 times too fast compared to the observations. The irregularity in both the oscillating period and the downward rate propagation could be due to the absence of sub-grid-scale waves parameterization, which, as is explained in introduction, contribute to 70\% of the eastward phase forcing in the Earth's QBO.

\begin{figure}
    \centering
    \includegraphics[scale=0.1]{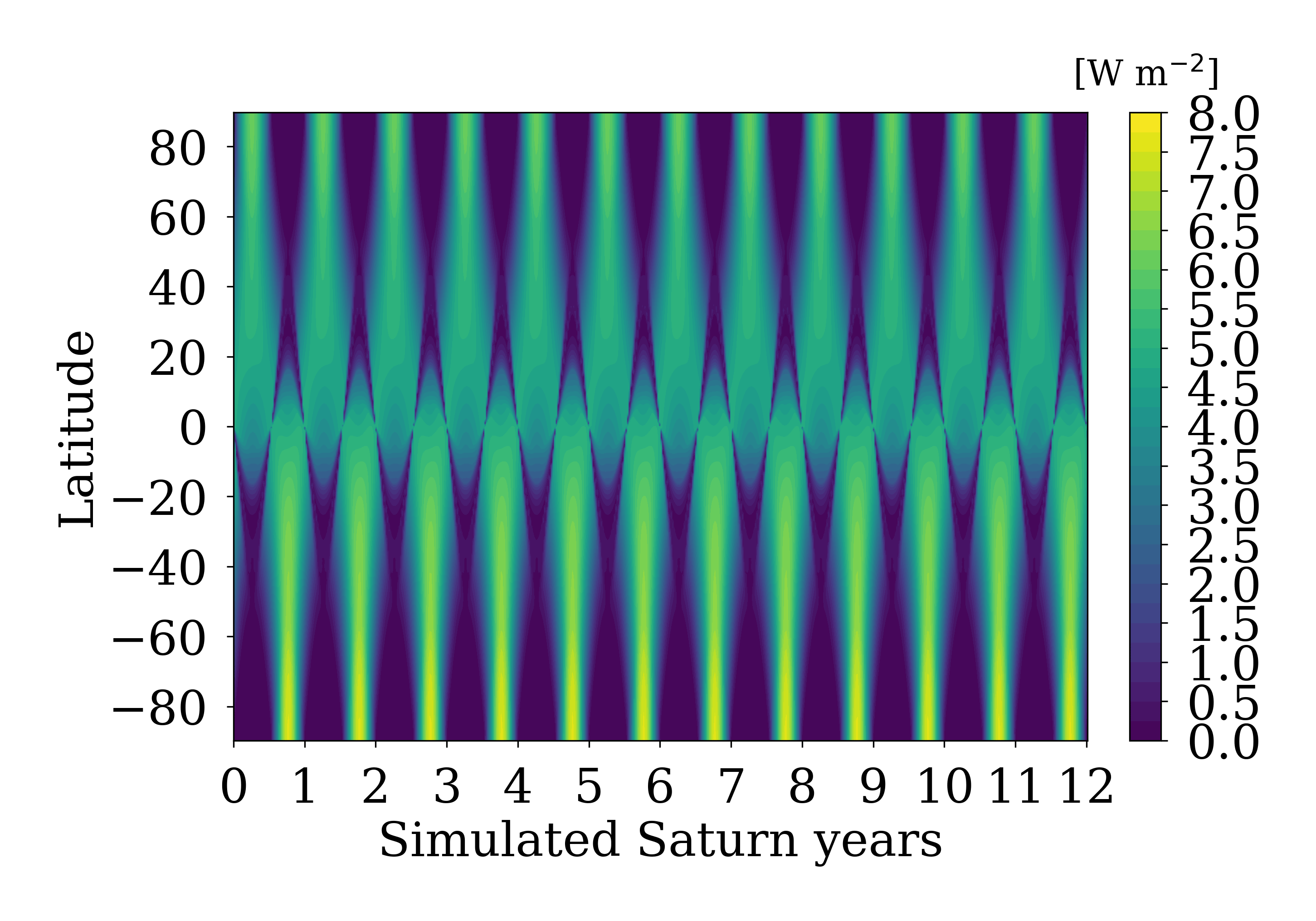}
    \caption{Temporal evolution of the incoming solar radiation in Saturn's atmosphere on the whole 12-year DYNAMICO-Saturn simulation}
    \label{fig:isr}
\end{figure}

\begin{figure}
    \centering
    \subfigure[][]{
        \label{subfig:u_y20N_strato}
        \centering
        \includegraphics[scale=0.15]{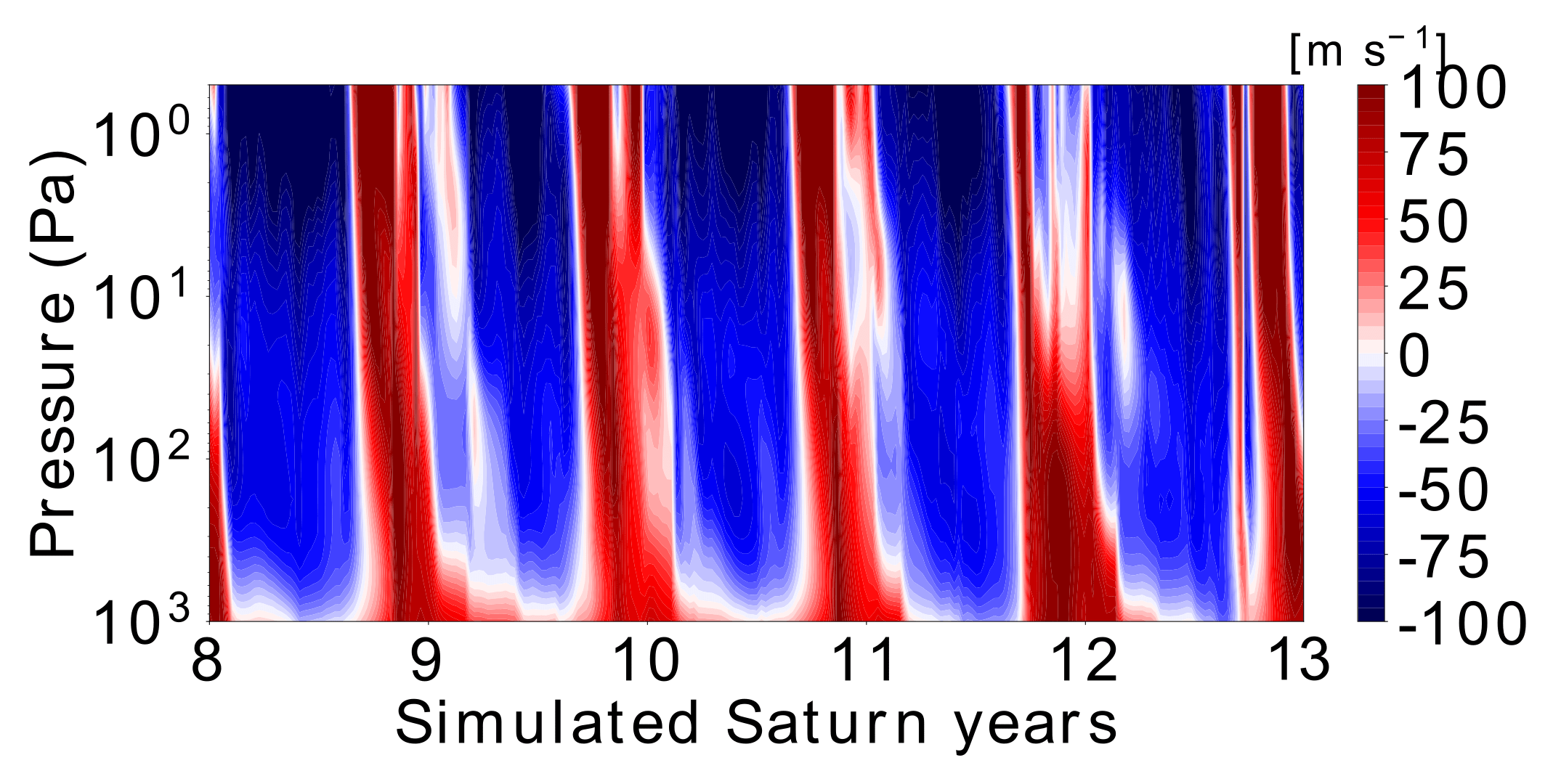}}
    \subfigure[][]{
        \label{subfig:u_y20S_strato}
        \centering
        \includegraphics[scale=0.15]{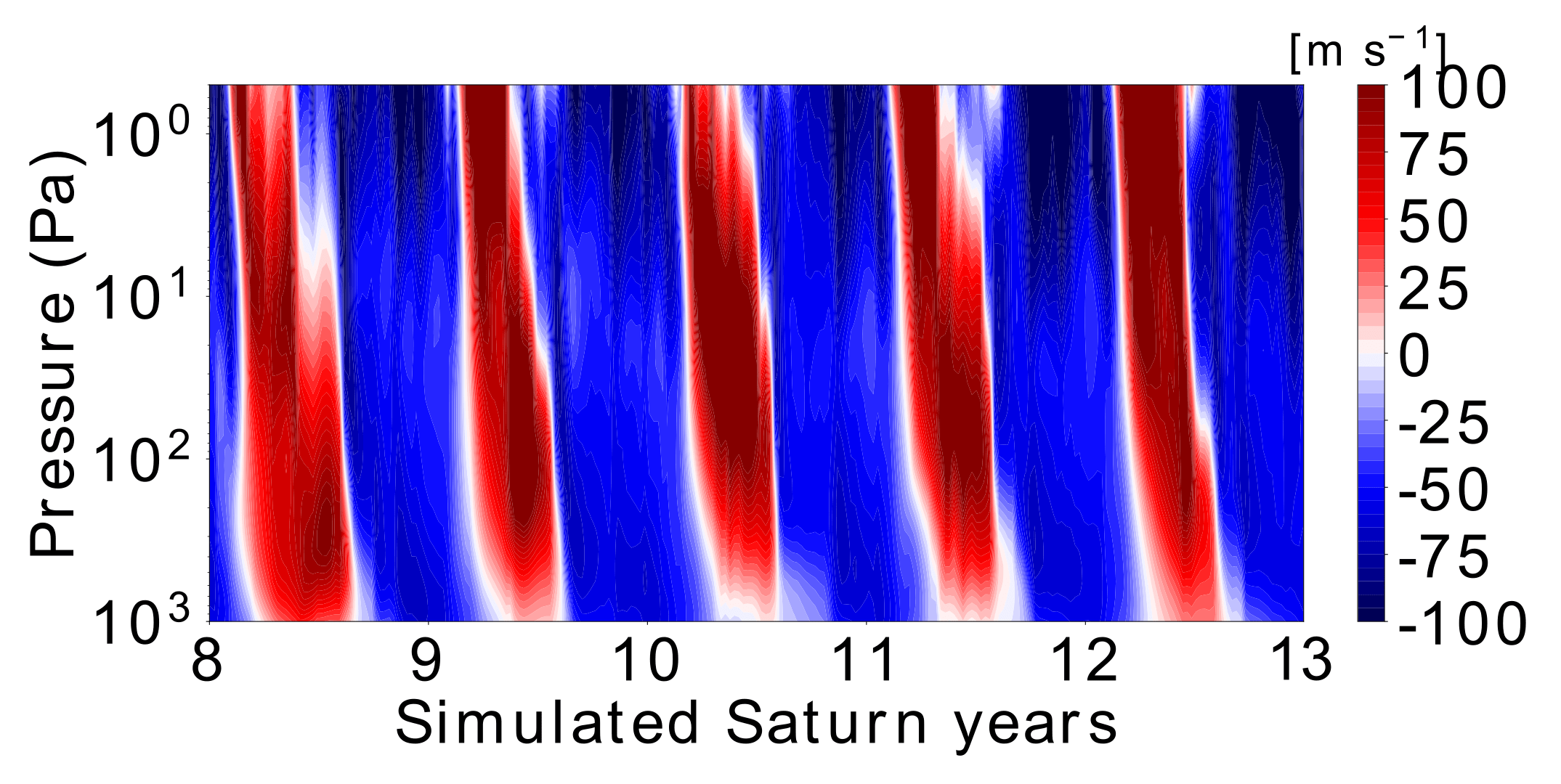}}
    \caption{Altitude/time section at 20$^{\circ}$N (\ref{subfig:u_y20N_strato}) and 20$^{\circ}$S (\ref{subfig:u_y20S_strato}) of the zonal-mean zonal wind in Saturn's stratosphere for the 5 last years of our 13-years DYNAMICO-Saturn simulation.}
    \label{fig:u_y20_strato}
\end{figure}

The second notable feature in Figure \ref{fig:u_time_evo_04mb} is the strong alternating eastward and westward jets at around 20$^{\circ}$N and 20$^{\circ}$S, with a seasonal phase opposition between the northern and southern hemispheres. Compared to a time evolution of the incoming solar radiation (Figure \ref{fig:isr}), we remark that each eastward jets emerge during the winter and seem to be correlated to the rings' shadowing. Those tropical eastward and westward jets both have an amplitude of about 100 m~s$^{-1}$, contrary to the QBO-like equatorial oscillation which exhibits an eastward phase weaker than the westward phase. This pattern exhibits a regular one-Saturn-year periodicity without downward propagation: the jets shown in Figure \ref{fig:u_time_evo_04mb} at 40 Pa extend from 10$^{3}$ Pa to the model top (Figure \ref{fig:u_y20_strato}).

%%%%%%%%
\subsection{Equatorial stratospheric thermal structure}
\label{QBO-like_temperature}

\begin{figure}
    \centering
    \begin{subfigure}
        \centering
        \includegraphics[scale=0.05]{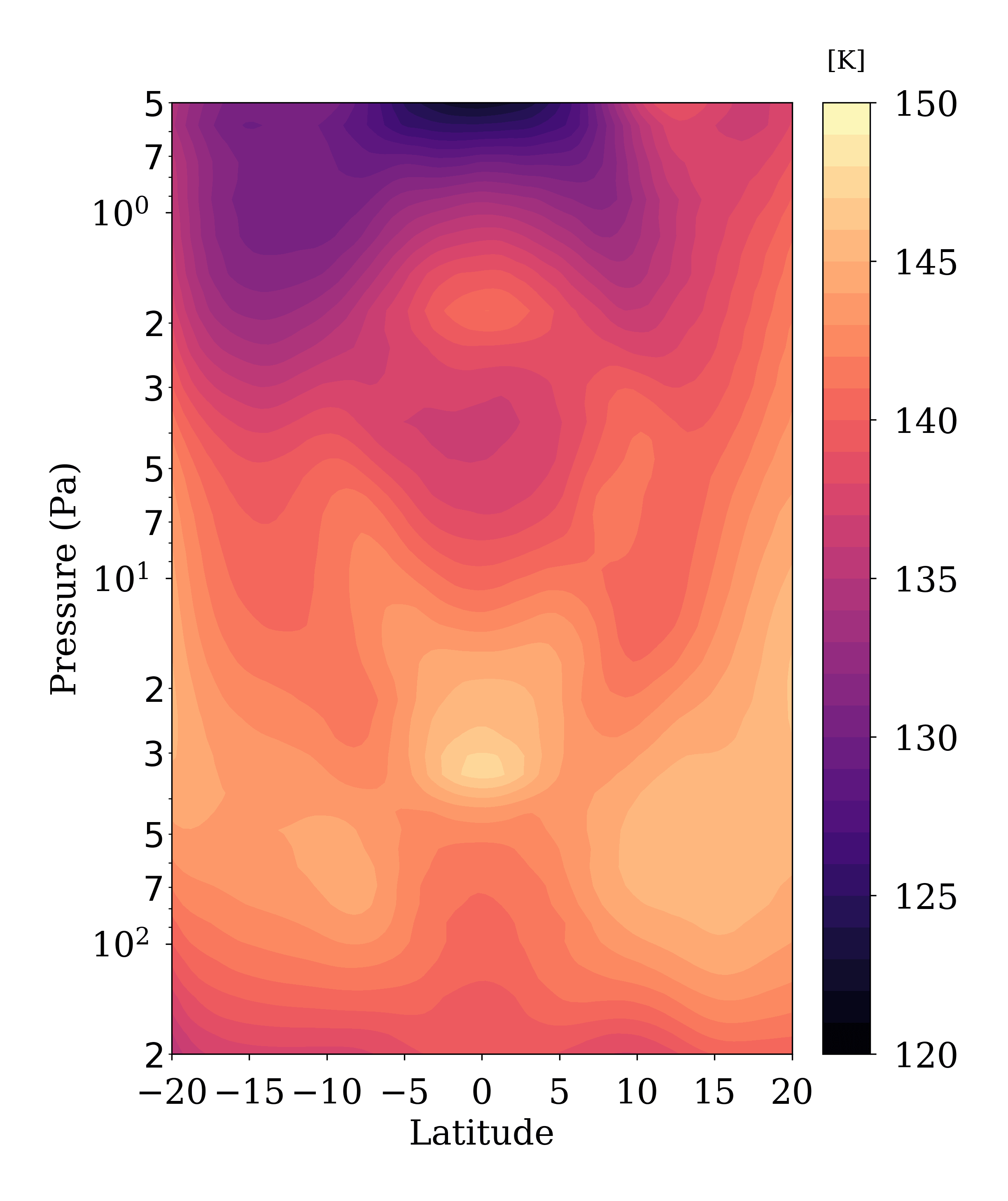}
    \end{subfigure}
    \begin{subfigure}
        \centering
        \includegraphics[scale=0.05]{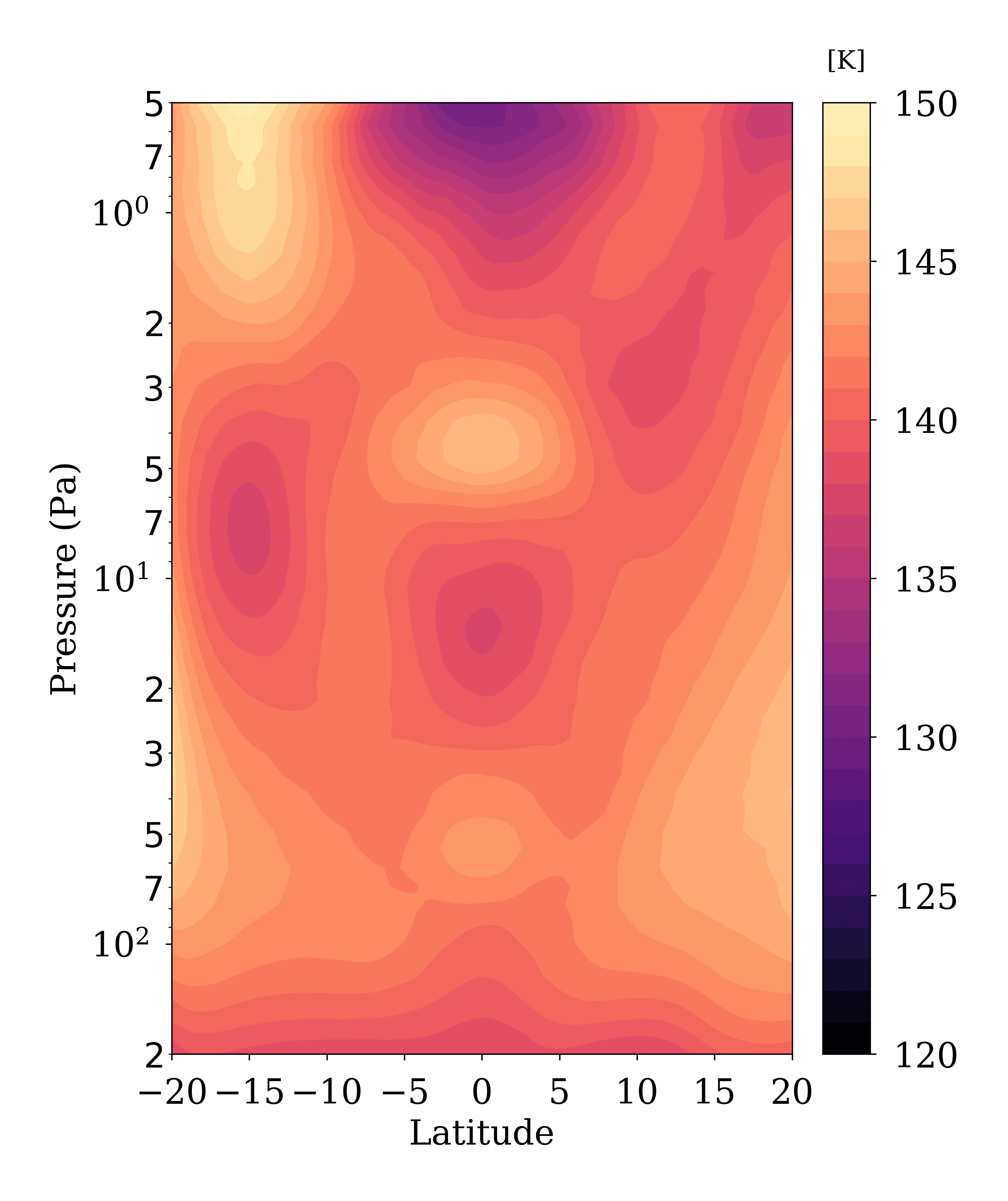}
    \end{subfigure}
    \begin{subfigure}
        \centering
        \includegraphics[scale=0.05]{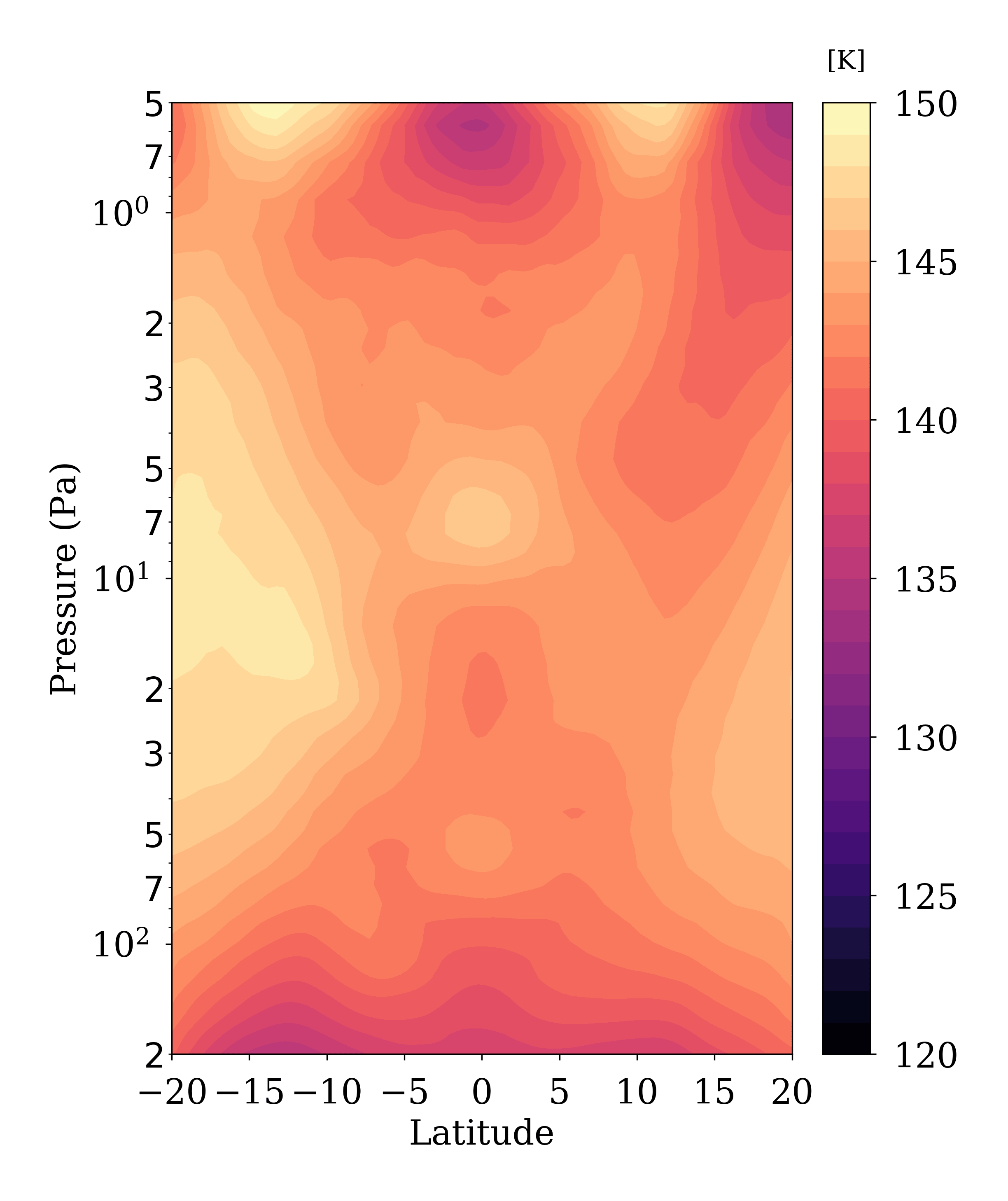}
    \end{subfigure}
    \caption{Altitude/latitude sections of zonal-mean temperature in Saturn's stratosphere at three dates of simulation: top panel at 10.45 years, middle panel at 10.55 years and bottom panel at 10.60 years.}
    \label{fig:temperature_osc}
\end{figure}

In this section, we compare our DYNAMICO-Saturn simulations with the observations of Saturn's equatorial stratosphere by Cassini/CIRS. Stratospheric winds have never been measured; rather, the CIRS instrument on-board the Cassini spacecraft allowed the retrieval of stratospheric temperature vertical profiles and stratospheric winds were calculated using thermal wind balance. It is therefore relevant to compare the observed and simulated stratospheric temperatures. Figure \ref{fig:temperature_osc} represents altitude/latitude section of temperature simulated by DYNAMICO-Saturn in the stratosphere. Between 8$^{\circ}$N and 8$^{\circ}$S, this meridional section exhibits a stack of local maxima (around 148~K) and minima (130~K) of temperature from 10$^{2}$ Pa to the top of the model. The pattern of alternating temperature extrema propagate downward with time. Temperature oscillates with a 10~K-magnitude around 140~K with a period and a descent rate equivalent to the wind's. Moreover, temperature extrema at the equator are anti-correlated to those occurring in the tropics. At a given pressure level, the cold equator regions are flanked by warmer tropical regions at 10 to 15$^{\circ}$N and at 10 to 15$^{\circ}$S. Using the thermal-wind equation applied assuming a geostrophically balanced flow, this significant meridional gradient of temperature confirms the presence of a significant vertical shear of the zonal wind at the equator. 

\begin{figure*}
    \centering
    \begin{subfigure}
        \centering
        \includegraphics[scale=0.05]{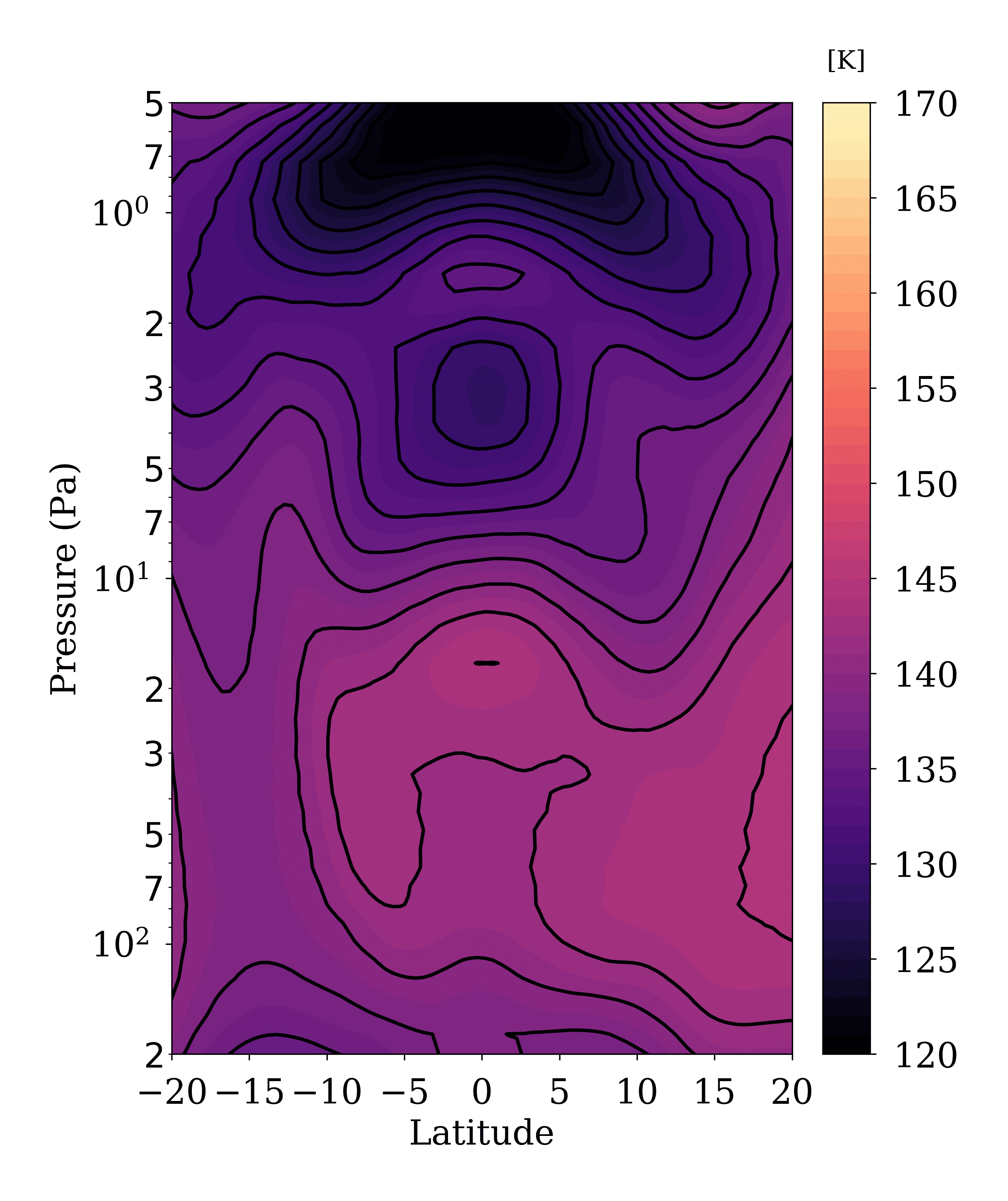}
    \end{subfigure}
    \begin{subfigure}
        \centering
        \includegraphics[scale=0.05]{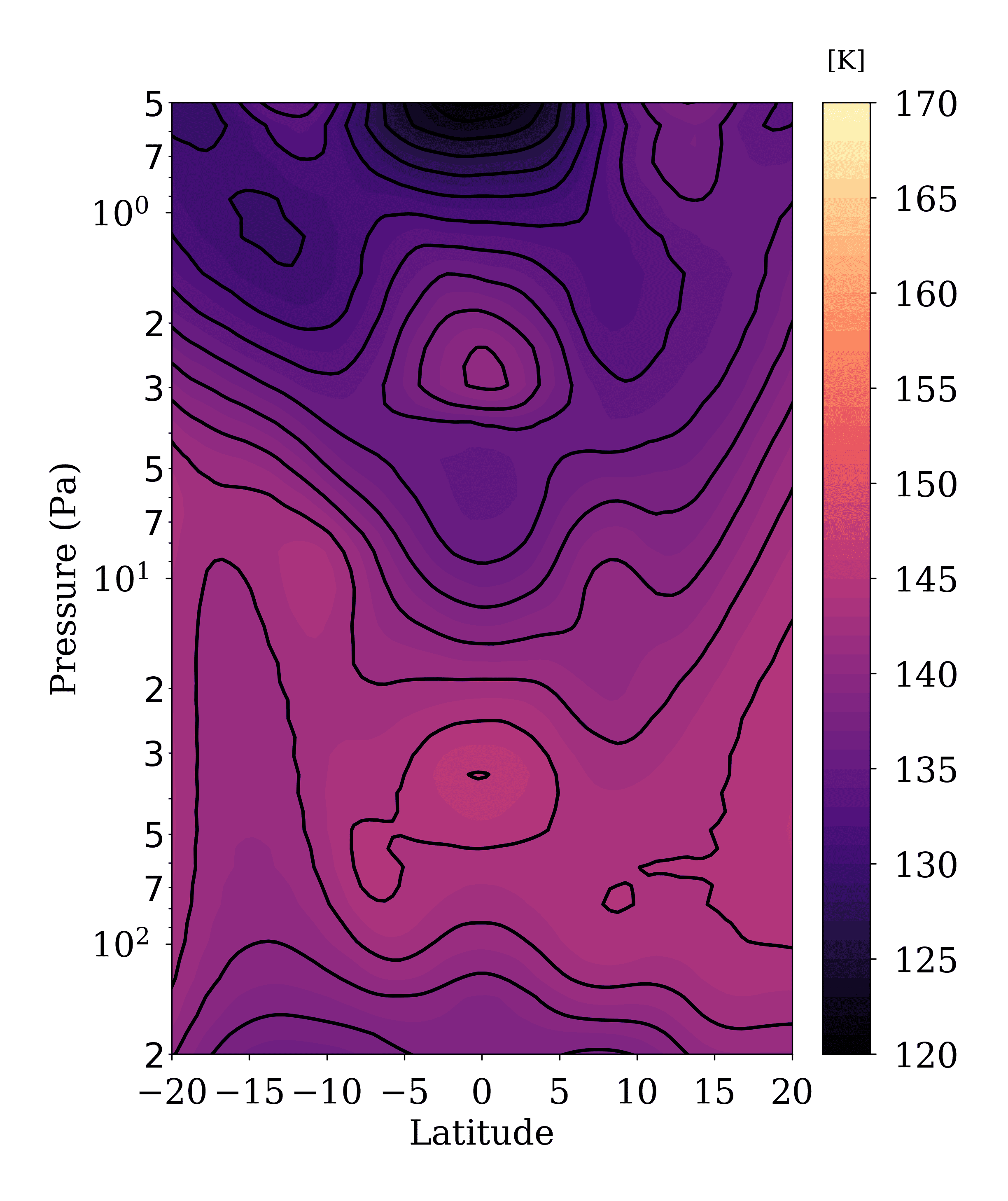}
    \end{subfigure}
    \begin{subfigure}
        \centering
        \includegraphics[scale=0.05]{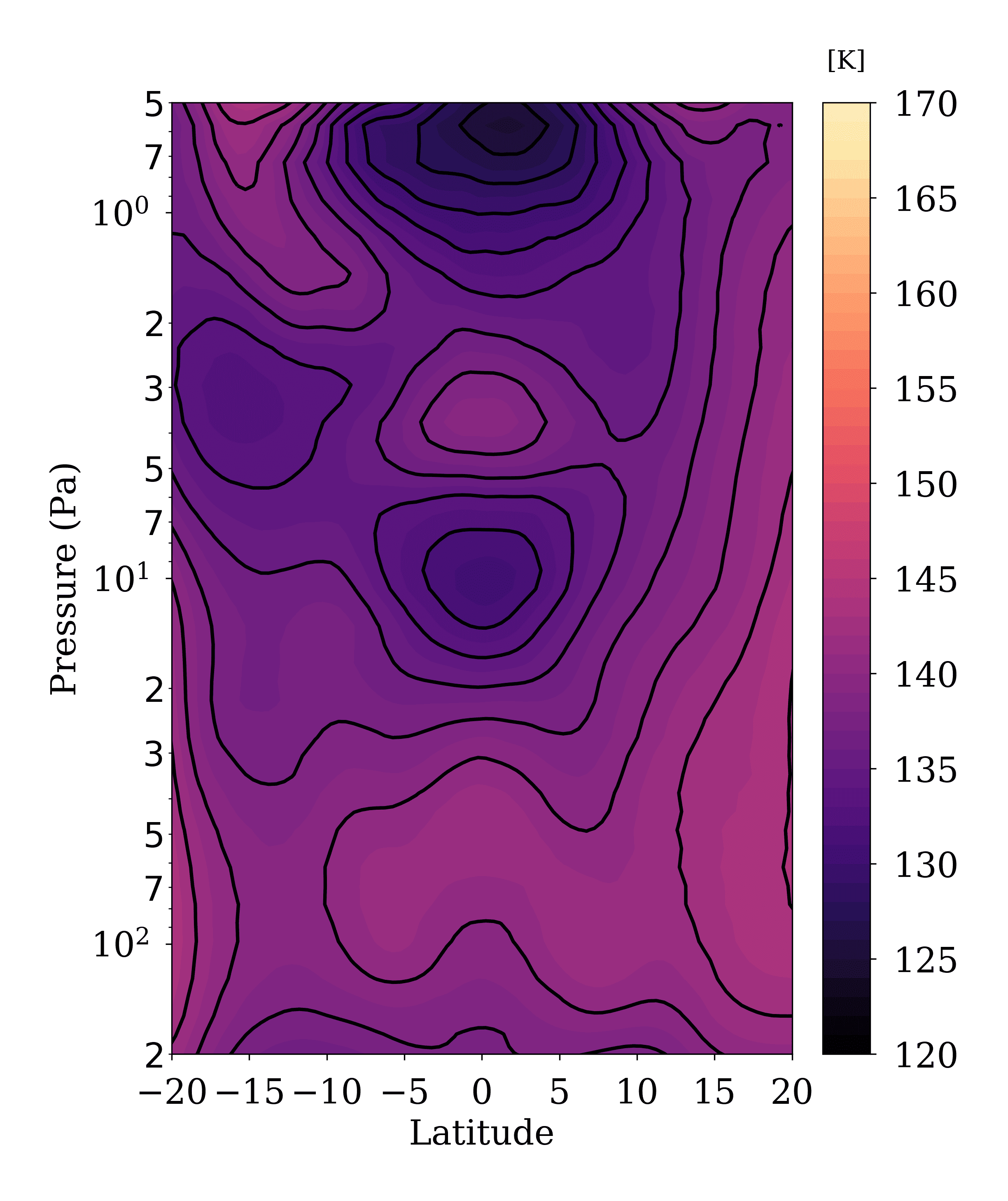}
    \end{subfigure}
    \caption{Altitude/latitude sections of zonal-mean temperature in Saturn's stratosphere at the 8.5, 8.53 and 8.6 years to compare with CIRS measurements (Figure \ref{fig:temperature_CIRS_guerlet2018}). Contour lines are separated by 2 Kelvin.}
    \label{fig:temperature_comparison_guerlet2018}
\end{figure*}

\begin{figure*}
    \centering
    \begin{subfigure}
        \centering
        \includegraphics[scale=0.05]{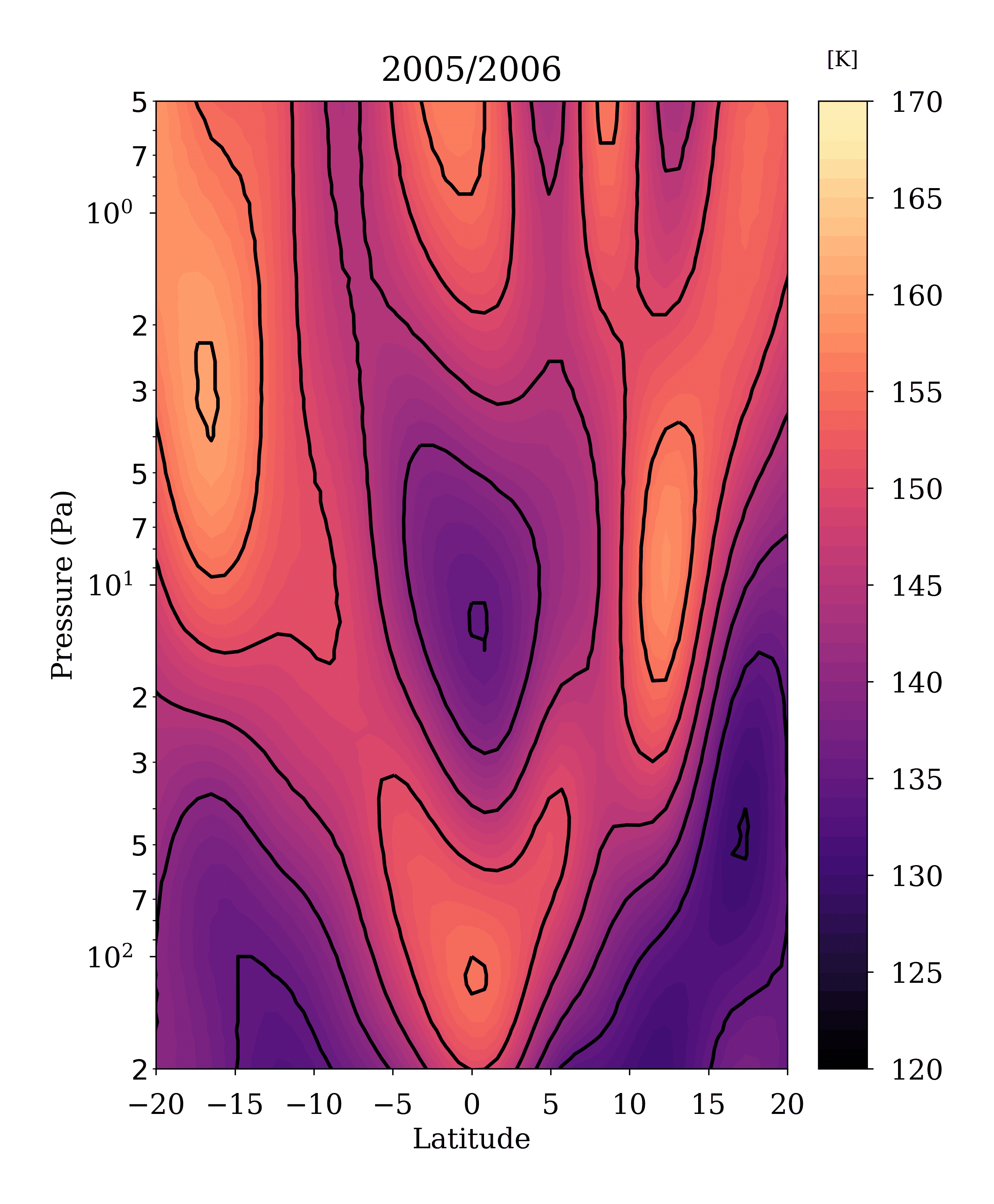}
    \end{subfigure}
    \begin{subfigure}
        \centering
        \includegraphics[scale=0.05]{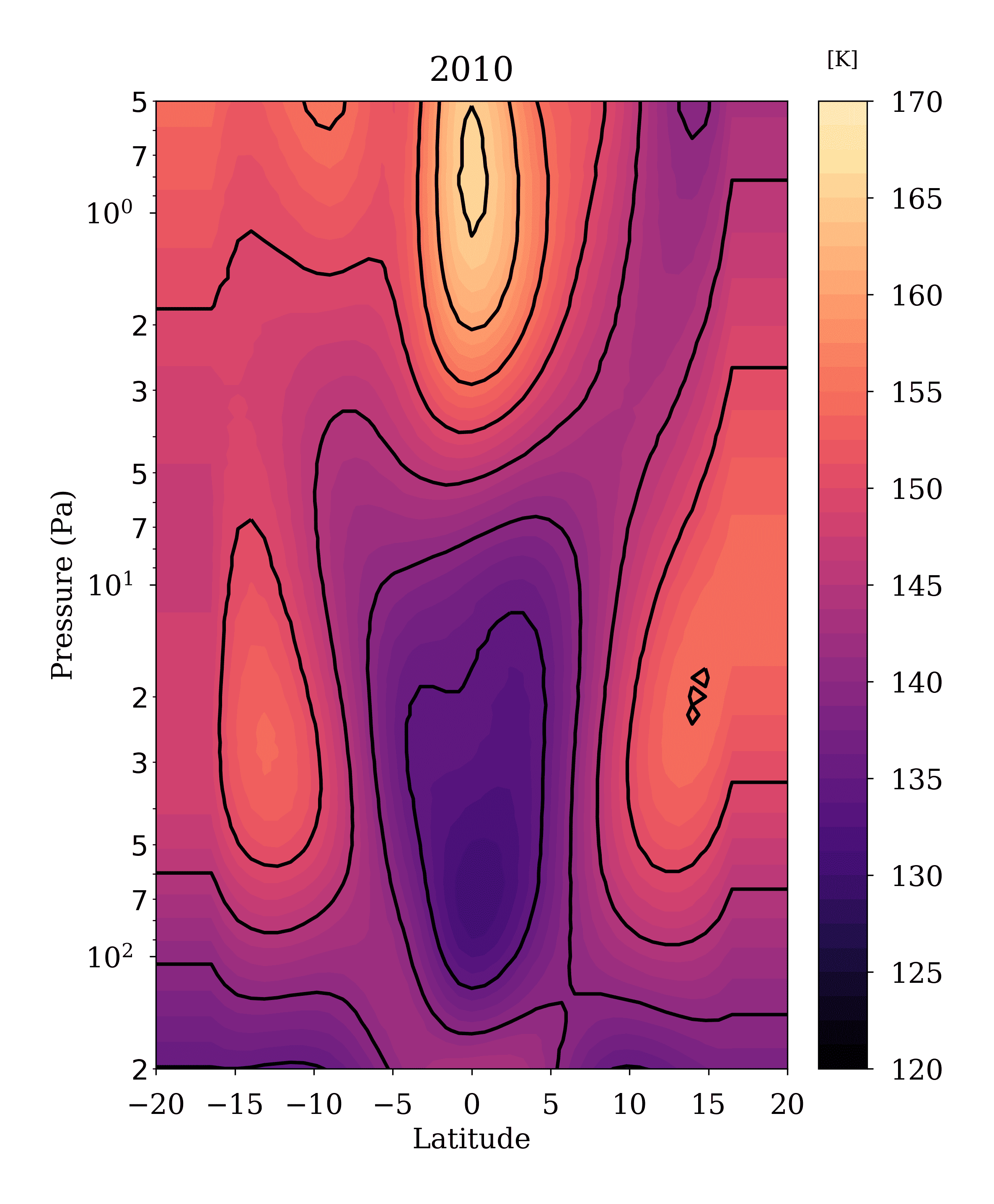}
    \end{subfigure}
    \begin{subfigure}
        \centering
        \includegraphics[scale=0.05]{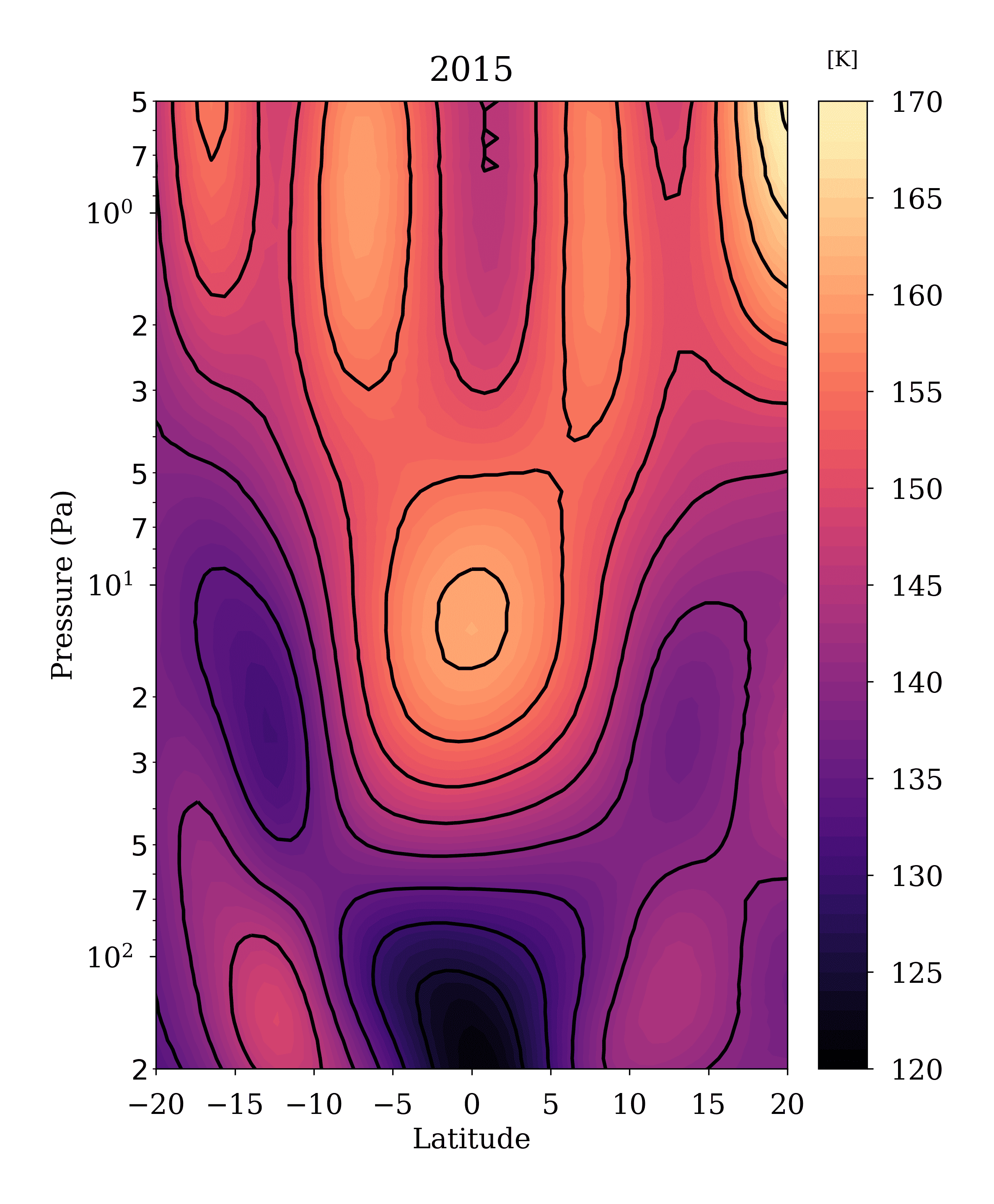}
    \end{subfigure}
    \caption{Altitude/latitude sections temperature measured by Cassini CIRS instrument at 2005/2006, 2010 and 2015 (analogous to \cite{guerlet_2018} Figure 3 top row but focused here on high stratospheric levels). Contour lines are separated by 5 Kelvin.}
    \label{fig:temperature_CIRS_guerlet2018}
\end{figure*}

To compare our simulation to CIRS observations, we re-mapped altitude vs latitude sections of modeled and observed temperatures with the same temperature range. GCM results are mapped in Figure \ref{fig:temperature_comparison_guerlet2018} and CIRS limb observations (acquired in 2005, 2010 and 2015) in Figure \ref{fig:temperature_CIRS_guerlet2018}, focused on high stratospheric levels. To carry out this comparison, we used CIRS data already processed in \cite{guerlet_2018}. Our DYNAMICO-Saturn simulation produces a temperature field qualitatively consistent with observations, with alternative maxima and minima of temperature stacked on the vertical in the stratosphere. There are between 3 and 4 temperature extrema in our DYNAMICO-Saturn model whereas there are only 2 (2010) or 3 (2005 and 2015) temperature extrema in CIRS temperature retrievals. Moreover, observations show a temperature oscillation from 170~K to 120~K, with a 20~K-magnitude, hence twice as much as in our Saturn GCM simulations. 

\begin{figure*}
    \centering
    \begin{subfigure}
        \centering
        \includegraphics[scale=0.05]{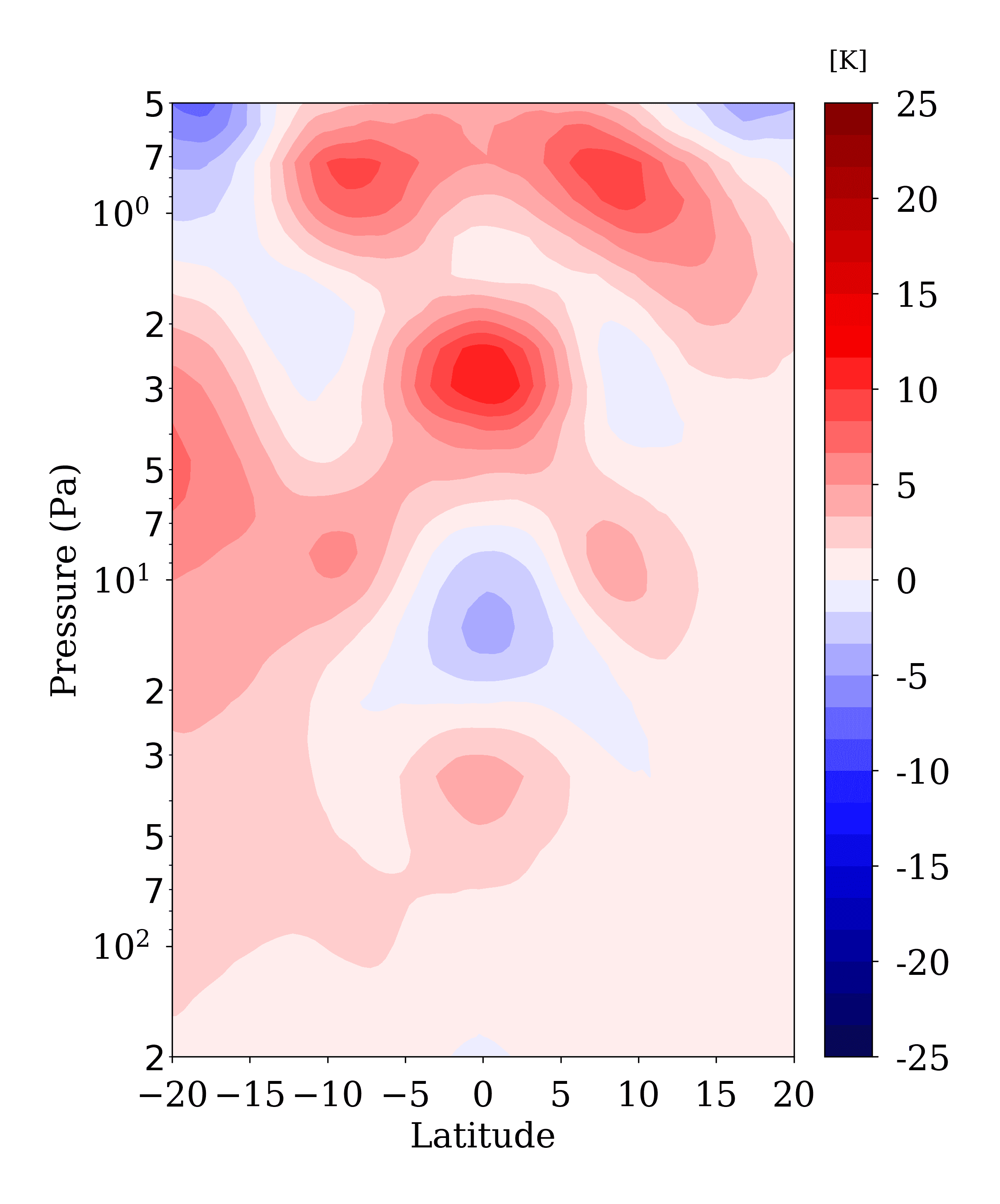}
    \end{subfigure}
    \begin{subfigure}
        \centering
        \includegraphics[scale=0.05]{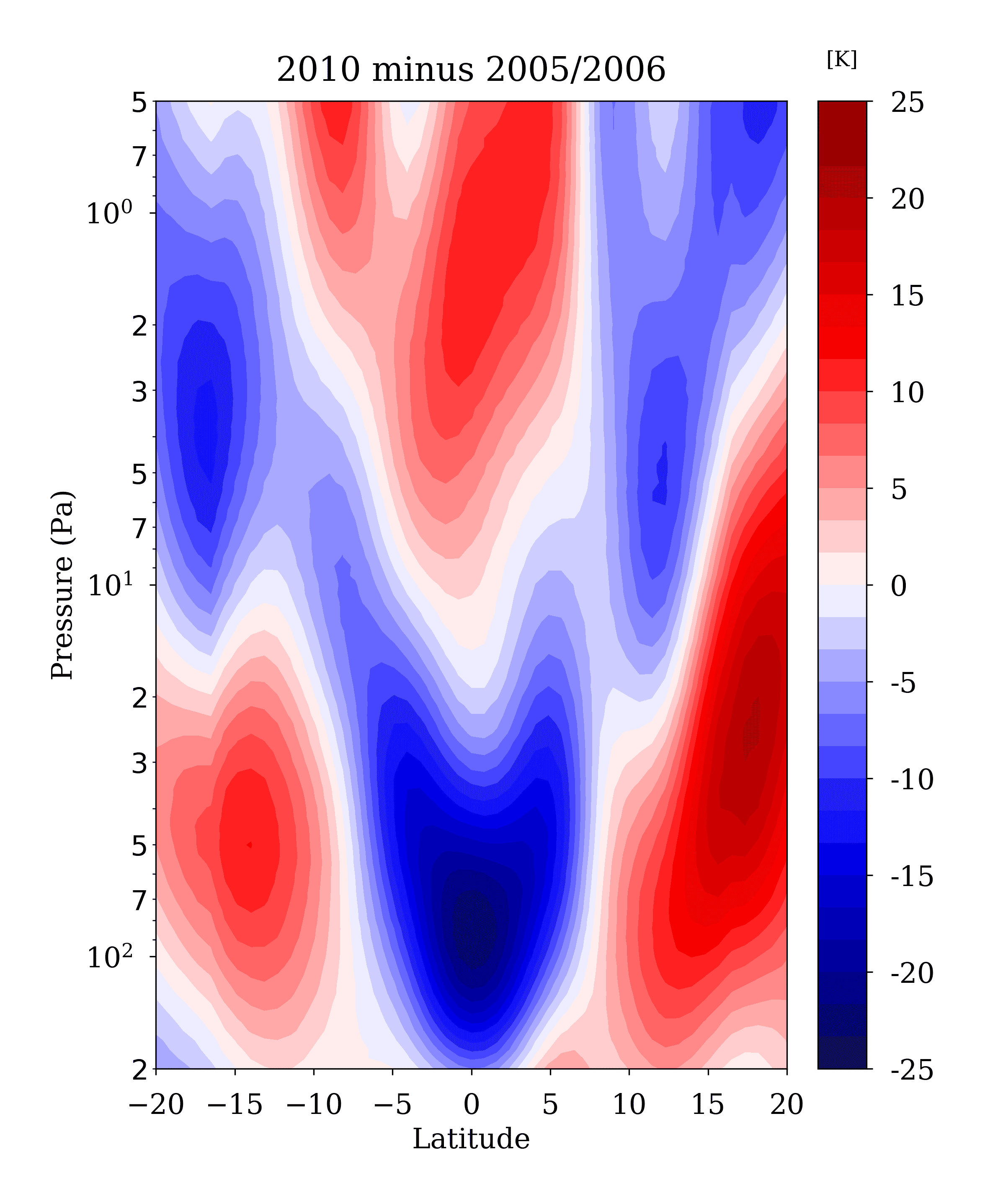}
    \end{subfigure}
    \begin{subfigure}
        \centering
        \includegraphics[scale=0.05]{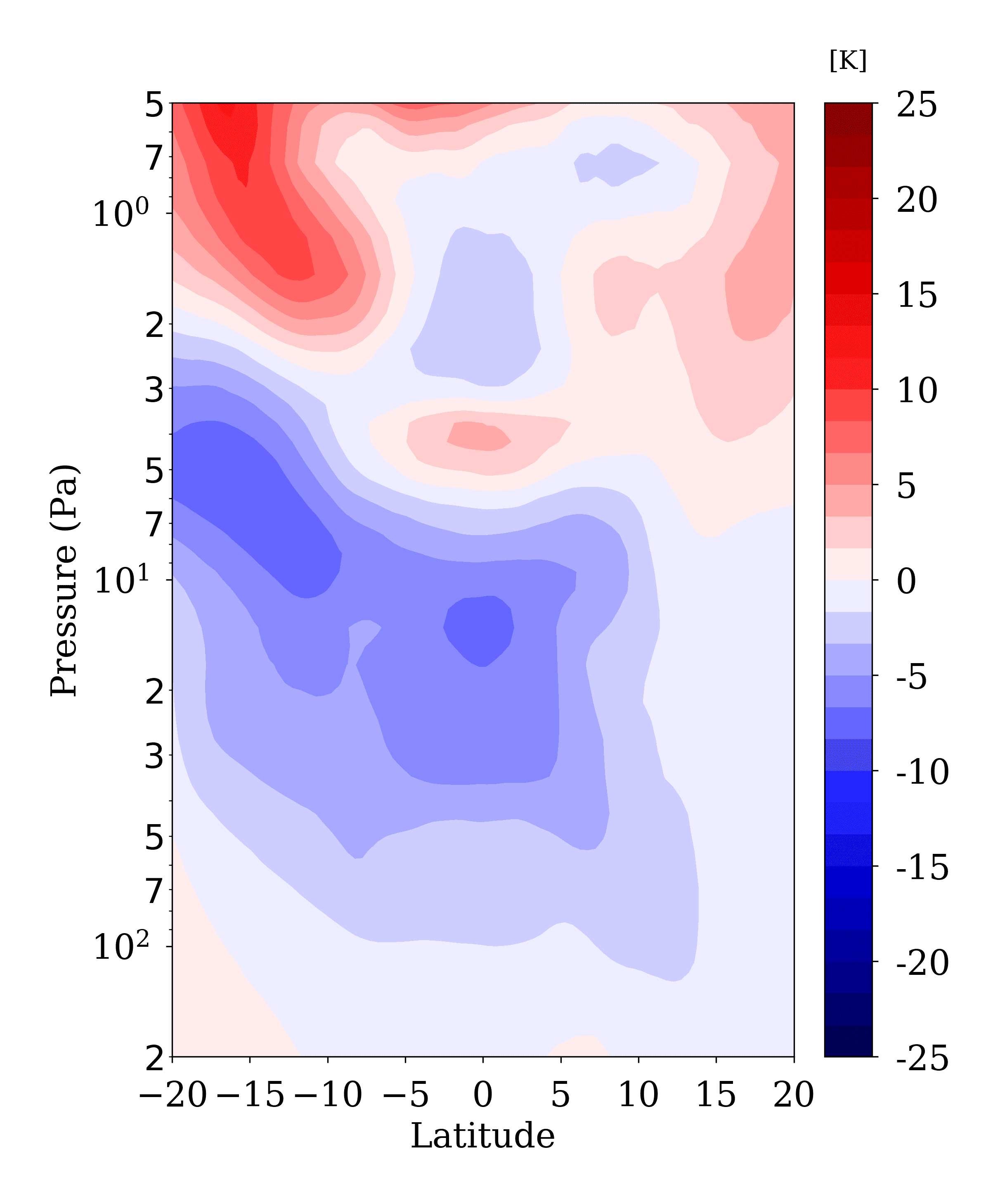}
    \end{subfigure}
    \begin{subfigure}
        \centering
        \includegraphics[scale=0.05]{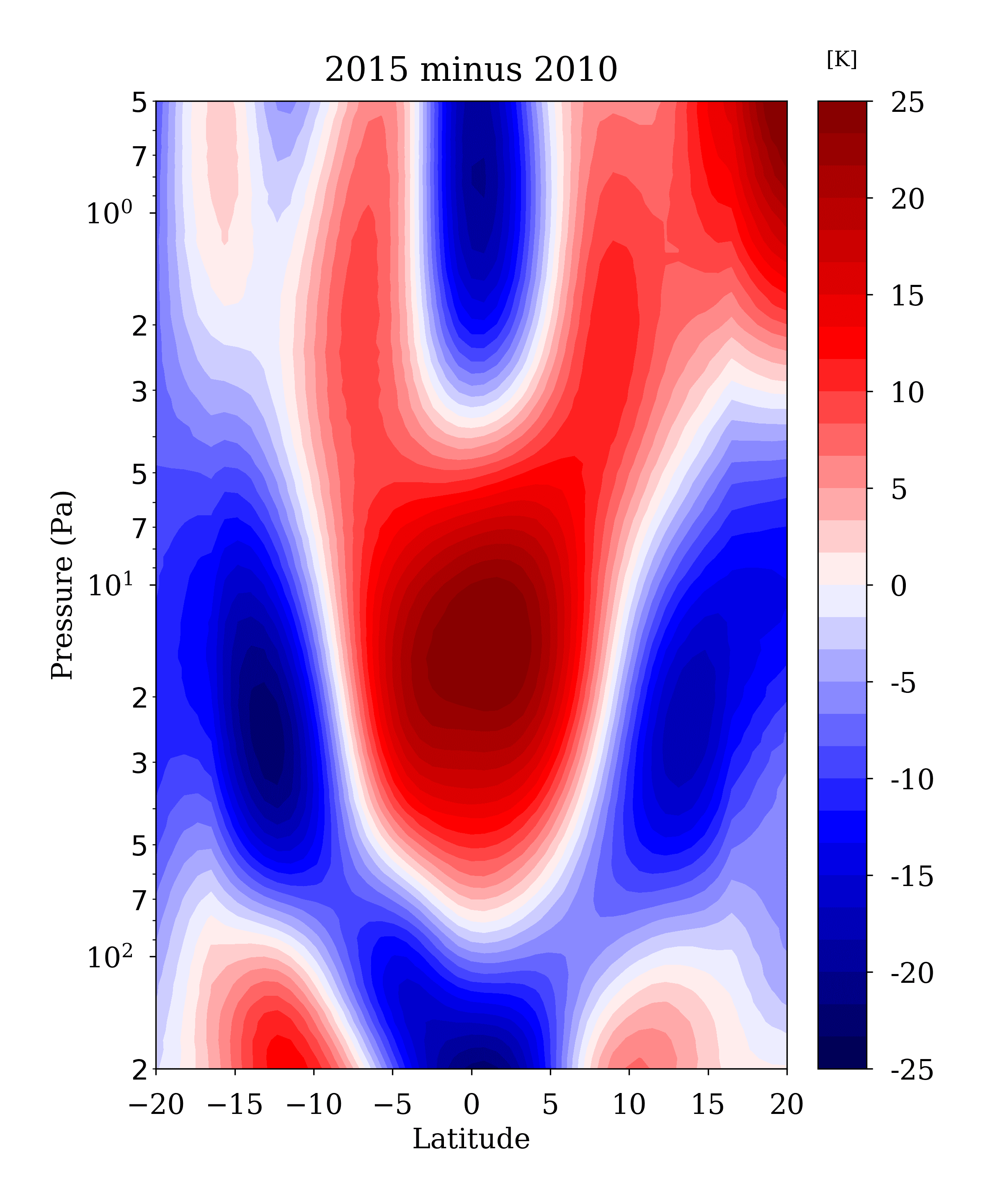}
    \end{subfigure}
    \caption{Altitude/latitude sections of temperature differences between two half-period-separated dates (left: DYNAMICO-Saturn results and right: as derived from CIRS measurements, analogous to \cite{guerlet_2018} Figure 3 bottom row) in Saturn's stratosphere. The modeled temperature oscillation underlines a vertical characteristic length twice as small as the observed one and a amplitude differences weaker than derived from CIRS measurements.}
    \label{fig:diff_temperature_comparison_guerlet2018}
\end{figure*}

To highlight the anomalies temperature resulting of the modeled and observed Saturn equatorial oscillation, we map temperature differences between two half-period-separated dates. Figure \ref{fig:diff_temperature_comparison_guerlet2018} displays altitude/latitude sections of temperature differences between the dates 8.53 and 8.5 years (top left) and between the dates 8.6 and 8.53 years (bottom left) of Figure \ref{fig:temperature_comparison_guerlet2018}. Comparison with CIRS measurements (right rows of Figure \ref{fig:diff_temperature_comparison_guerlet2018}) shows the similarities and differences in the modeled and observed QBO-like oscillations. Temperature differences underline a stack of positive and negative anomalies of temperature, with Saturn-DYNAMICO underestimating the amplitude by a factor of 2 compared to the observations. The characteristic vertical size of the DYNAMICO-Saturn QBO-like oscillations is two times smaller than the observed one. Our DYNAMICO-Saturn underestimates the temperature maximum by around 20~K and underestimates the magnitude of the temperature anomalies by 10~K. Another difference between the Cassini observations and the GCM-modeled equatorial oscillation is the vertical extent of the temperature anomalies at the equator. Cassini radio occultation data showed that the descending temperature minima and maxima extend down to the tropopause level, around 100 hPa \citep{schinder_2011} whereas in the Saturn-DYNAMICO simulations there is no more equatorial-oscillation signal below the 1-2 hPa level.

We can also compare our simulations to the idealized work of \cite{showman_2018}. In their idealized simulation, they obtain a temperature oscillation of about $\sim$ 10 K that migrate downward over time (Figure 8 of \cite{showman_2018}), as in our simulation. Moreover, we obtain a stack of temperature extremes spatially close to each other on the vertical, contrary to their simulation, where the extremes of temperature are half a decade (in pressure) apart from each other by a constant temperature zone. In our and their simulations, opposite extremes of temperature (at a given pressure level), centered between 10 and 15 latitude degrees (both north and south), are generated. There is, in the two simulations, an anti-correlation between temperature extremes at the equator and those occurring off the equator. With DYNAMICO-Saturn, compared to the CIRS observations, we obtain a more realistic patterns of temperature anomalies at the equator than the idealized work of \cite{showman_2018}, but there are still some notable differences with the CIRS observations.

Another feature of Saturn's stratosphere observed by Cassini/CIRS is the anomalously high temperatures under the rings' shadows \citep{fletcher_2010}. Previous radiative simulations of Saturn's atmosphere failed to reproduce these temperature anomalies \citep{guerlet_2014}, hence dynamical heating was considered to explain this feature. 
 
\begin{figure}
    \centering
    \includegraphics[scale=0.3]{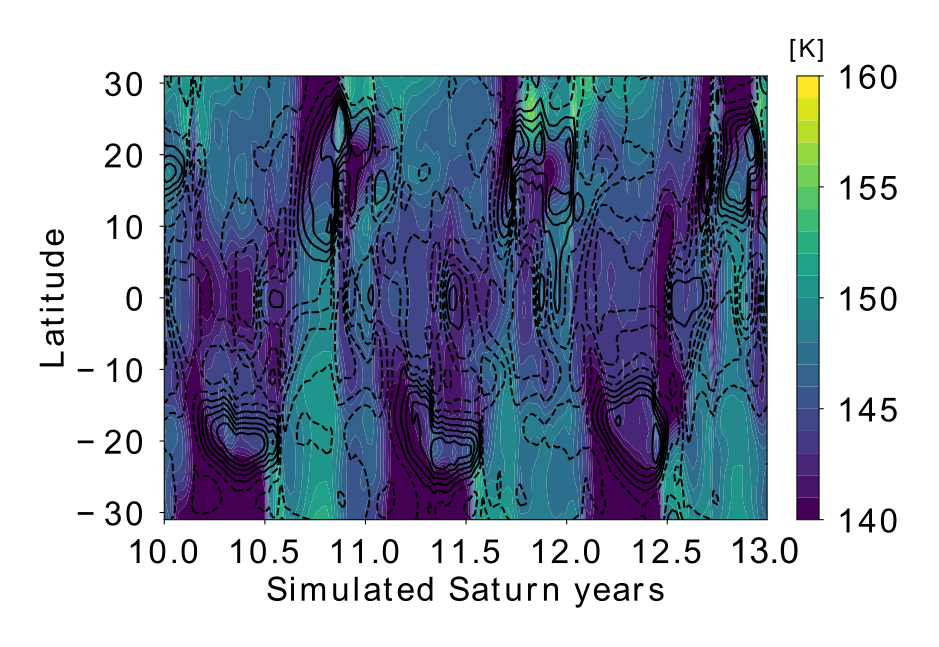}
    \caption{Time evolution of the zonal-mean temperature (color) and the zonal-mean zonal wind (contours spacing by 25 m s$^{-1}$ between -100 m s$^{-1}$ and 100 m s$^{-1}$) in Saturn's stratosphere (40 Pa) of our 13-years DYNAMICO-Saturn simulation.}
    \label{fig:temp_u_time_04mb}
\end{figure}

In our simulations, there is an interplay between the tropical eastward jets and temperature evolution, displayed in Figure \ref{fig:temp_u_time_04mb}.
In wintertime, at 20$^{\circ}$ latitude, the temperature starts by cooling rapidly (from 150K to 140K), which is likely due to the ring's shadow radiative effect. Then, when the eastward jet increases in strength, this low temperature region is significantly reduced, there is an increase of temperature from 140K to 145K. This is true only near the core of the eastward jet, at 20$^{\circ}$. Indeed, at 30$^{\circ}$ latitude, the temperature remains very cold (140K) throughout wintertime.
We will assess in more detail the dynamical impact of ring shadowing in the stratosphere in section \ref{oscillation/rings}.

%%%%%%%%
\subsection{Equatorial stratospheric planetary-scale waves}
\label{QBO-like_planetary_waves}

According to the terrestrial experience, equatorial oscillation results from both planetary-scale waves and mesoscale waves forcing. In our DYNAMICO-Saturn GCM simulation, the oscillation at the low latitudes can only result from resolved motions, therefore planetary-scale waves, because there is no sub-grid scale parameterization or bottom thermal forcing in our model. To study those waves, we performed a two-dimensional Fourier transform of the symmetric and antisymmetric components of temperature and zonal wind fields, following \cite{wheeler_1999} method, as in \cite{spiga_2018} section 3.3.1. We applied the two-dimensional Fourier analysis on a specific 1000-day-long run, with daily output frequency, after 270000 simulated Saturn days as in \cite{spiga_2018}. The spectral mapping in the wavenumber $s$ and frequency $\sigma$ space enables to evidence and characterize the planetary-scale waves driving the stratospheric equatorial oscillation. Symmetric component of temperature (about the equator) $T_S$ shows Rossby, inertia-gravity and Kelvin waves and antisymmetric component of temperature $T_A$ shows Yanai (Rossby-gravity) waves.

\begin{figure*}
    \centering
    \begin{subfigure}
        \centering
        \includegraphics[scale=0.3]{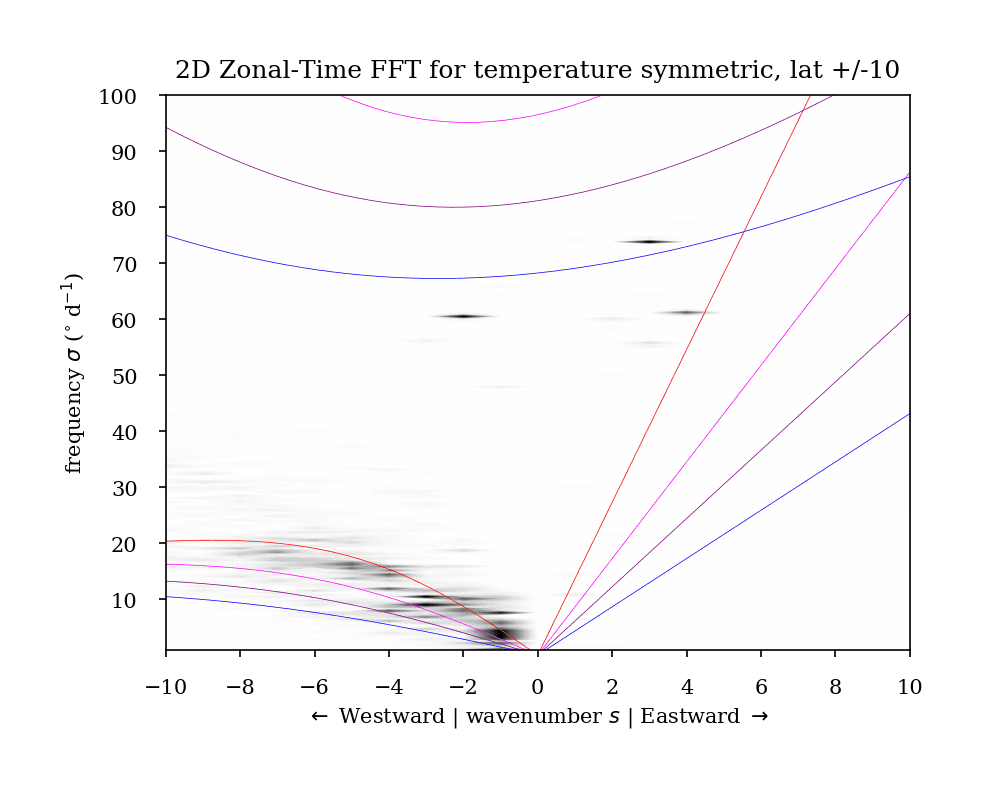}
    \end{subfigure}
    \begin{subfigure}
        \centering
        \includegraphics[scale=0.3]{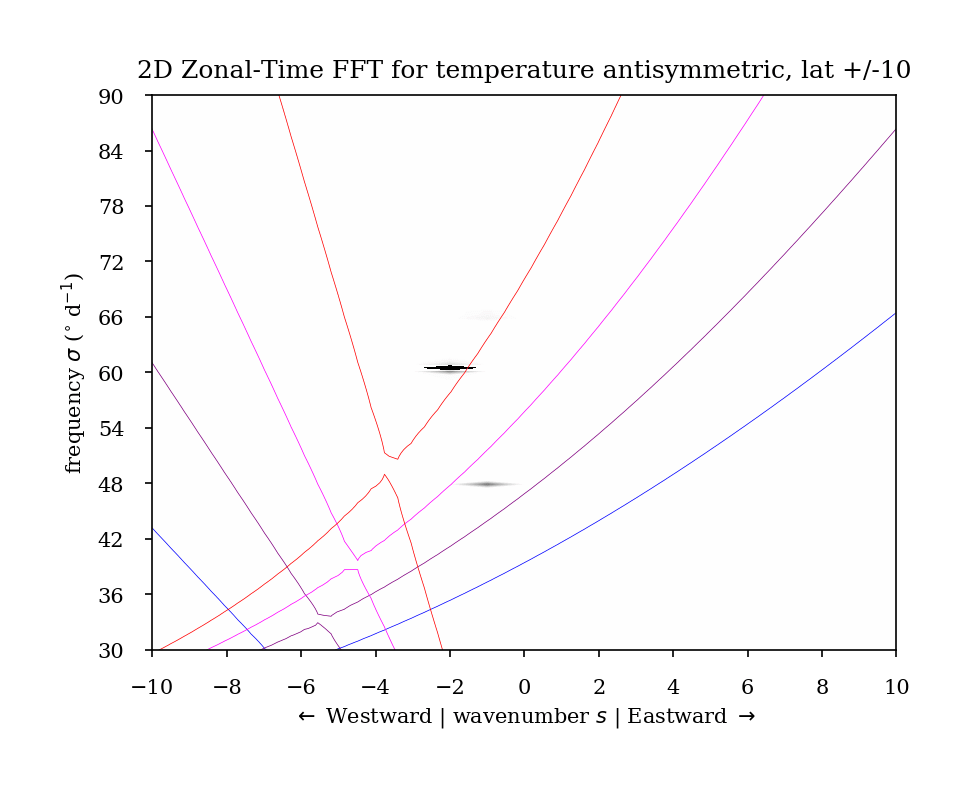}
    \end{subfigure}    
    \caption{Spectral analysis of the equatorial waves produced by our DYNAMICO-Saturn in dynamical steady-state at the 40th pressure-level (43 Pa). Grey shaded areas depict the waves identified by the spectral analysis. The colored curves correspond to the dispersion relation from the linear theory. Four values of equivalent depths are included into the linear theory: 5 km in blue, 10 km in purple, 20 km in magenta and 50 km in red.}
    \label{fig:dispersion_temperature_daily_z40}
\end{figure*}

The results of the two-dimensional Fourier transform for the stratospheric temperature and zonal wind fields at the 20-Pa pressure level are shown in Figure \ref{fig:dispersion_temperature_daily_z40} and in Table \ref{tab:dispersion_modes}. The spectral analysis demonstrates, both in the antisymmetric component of temperature and zonal wind fields, that the dominant wavenumber-2 is a westward-propagating Rossby-gravity wave with a period of 6 days and a frequency of 60$^{\circ}$ longitude per day. Westward-propagating Rossby and internal-gravity waves with wavenumber $s$ = -1 are the secondary prominent mode in temperature, with a period of 100 Saturn days. There are two other westward-propagating modes with wavenumbers $s$ = -2 and $s$ = -3, and respectively a period of 6 and 34 Saturn days. Finally, there are only two eastward-propagating Kelvin waves (identified both in the temperature and wind field) with $s$ = +3 and $s$ = +4, with a 5 Saturn days and 6 Saturn days periodicity. The zonal wind field indicates westward-propagating Rossby waves with $s$ = -1 to $s$ = -10, exhibiting long periods of several tens of Saturn days and frequencies of about fifteen degrees longitude per day.

\begin{table}
    \centering
    \begin{tabular}{cccc}
        \multicolumn{4}{c}{Dominant modes in $T_s$} \\ 
        \hline
        $s$ & $\sigma \, (^{\circ}$/d) & period (d) &  log(SP) \\ 
        \hline 
        -1 &      3.6 &      100 &      9.6 \\ 
        -2 &     60.5 &        6 &      9.4 \\ 
        -3 &     10.4 &       34 &      9.2 \\ 
        +3 &     73.8 &        5 &      8.9 \\ 
        -4 &      7.9 &       45 &      8.7 \\ 
        +4 &     61.2 &        6 &      8.7 \\ 
        -5 &     16.2 &       22 &      8.7 \\ 
        & \\
        \multicolumn{4}{c}{Dominant modes in $u_s$} \\ 
        \hline
        $s$ & $\sigma \, (^{\circ}$/d) & period (d) &  log(SP) \\ 
        \hline 
        -1 &      4.0 &       91 &     11.9 \\ 
        -2 &      7.9 &       45 &     11.9 \\ 
        -3 &     11.2 &       32 &     11.3 \\ 
        -7 &     13.7 &       26 &     11.3 \\ 
        +3 &     73.8 &        5 &     11.3 \\ 
        -4 &      8.3 &       43 &     11.2 \\ 
        -5 &     10.4 &       34 &     11.1 \\ 
        -10 &    17.3 &       21 &     11.0 \\ 
        -6 &     16.2 &       22 &     11.0 \\ 
        +4 &     61.2 &        6 &     11.0 \\ 
        & \\
        \multicolumn{4}{c}{Dominant modes in $T_A$, $u_A$, $v_S$} \\ 
        \hline
        $s$ & $\sigma \, (^{\circ}$/d) & period (d) &  log(SP) \\
        \hline 
        -2 &     60.5 &        6 &     11.3 \\ 
        & \\
    \end{tabular}
    \caption{Spectral modes detected by Fourier analysis and depicted in Figure \ref{fig:dispersion_temperature_daily_z40}. SP means spectral power and d Saturn days.}
    \label{tab:dispersion_modes}
\end{table}

It is worth noticing that there are numerous westward-propagating modes versus only two eastward-propagating mode, both in temperature and zonal wind components. The resolved QBO-like oscillation in our DYNAMICO-Saturn reference GCM simulation is thus characterized by an imbalance in eastward- and westward-wave forcing. This may be symptomatic of the absence of sub-grid scale waves parameterization, which contribute of almost 3/4 of the eastward momentum forcing in the equatorial oscillation of Earth. Indeed, the smaller the eastward-wave forcing is, the smaller the eastward momentum carried by waves to critical levels is. This lack of eastward momentum in the resolved dynamics could explain why, as noticed in section \ref{QBO-like_zonal_wind}, the eastward phase of the equatorial oscillation seems to be unstable in time compared to the westward phase. A plausible explanation is that the eastward momentum deposition of Kelvin wave is too low on its own to balance the westward phase induced by the Rossby-gravity, Rossby and inertia-gravity waves.

%%%%%%%%
\subsection{Eddy-to-mean interactions driving the equatorial stratospheric oscillation}
\label{QBO-like_TEM}

To determine the eddy-to-mean interactions within the flow, we use the Transformed Eulerian Mean (TEM) formalism. In terrestrial atmospheric studies, it is commonly employed to investigate momentum and heat transfers by wave-mean flow interactions \citep{andrews_1983}.

In all the following equations (\ref{eq:v_star} to \ref{eq:EP_flux}), the overline denotes the zonal-mean of each field, the prime symbol denotes the eddy component, i.e. departures from the zonal mean. 
By defining the transformed zonal-mean velocities as follows
\begin{equation}
\bar{v^{*}}=\bar{v}-\frac{\partial}{\partial P}\left(\frac{\bar{v'\theta'}}{\bar{\theta}_{P}}\right)
\label{eq:v_star}
\end{equation}
\begin{equation}
\bar{\omega^{*}}=\bar{\omega}+\frac{1}{a \cos{\phi}} \frac{\partial}{\partial \phi}\left(\frac{\cos{\phi}\bar{v'\theta'}}{\bar{\theta}_{P}}\right)
\label{eq:omega_star}
\end{equation}
\noindent we obtain the momentum equation in the Transformed Eulerian Mean formalism:
\begin{equation}
    % TEM: Momentum equation
    \frac{\partial \bar{u}}{\partial t} = \bar{X} - \underbrace{\left[ \frac{1}{a \cos{\phi}} \frac{\partial(\bar{u}\cos{(\phi}))}{\partial \phi} - f \right]\bar{v^{*}} - \bar{\omega^{*}}\frac{\partial\bar{u}}{\partial P}}_{I} + \underbrace{\frac{1}{a \cos{\phi}}\overrightarrow{\nabla}.\overrightarrow{\bar{F}}}_{II}
    \label{eq:acc_TEM}
\end{equation}
where $\bar{X}$ is the zonal-mean non-conservative friction and $\overrightarrow{F}=(F^{\phi},F^{P})$ is the Eliassen-Palm flux, which components are defined in equation \ref{eq:EP_flux} below.
\begin{equation}
\begin{array}{r c l}
    F^{\phi}&=&\rho_{0}a\cos{\phi}\left(\bar{v'\theta'}\frac{\frac{\partial\bar{u}}{\partial P}}{\frac{\partial\bar{\theta}}{\partial P}}-\bar{u'v'}\right) \\
    F^{P}&=&\rho_{0}a\cos{\phi}\left(\frac{\bar{v'\theta'}}{\frac{\partial\bar{\theta}}{\partial P}}\left(f-\frac{1}{a \cos{\phi}} \frac{\partial(\bar{u}\cos{\phi})}{\partial \phi}\right)-\bar{u'\omega'}\right)
\label{eq:EP_flux}
\end{array}    
\end{equation}
The Eliassen-Palm flux is defined as a wave momentum flux: it depicts both interactions between eddies and the mean flow (a momentum stress on the mean flow) and eddy propagation, as a wave activity. It allows one to link the meridional flux of heat~$\bar{v'{\theta}'}$ and zonal momentum~$\bar{u'v'}$ (due to waves forcing) to a horizontal~$F^{\phi}$ and vertical~$F^{P}$ flux of momentum on the mean flow (into Equation \ref{eq:acc_TEM}).
In equation \ref{eq:acc_TEM}, the term I is the residual-mean acceleration and the term II is the Eliassen-Palm flux divergence, i.e. the eddy forcing in momentum on the mean flow acceleration (named eddy-induced acceleration in the following). Studying the Eliassen-Palm flux, as well as its divergence, makes it possible to identify the regions of greatest interaction between the eddies and the mean flow, and to visually determine the propagation of eddy-induced momentum from one region of the atmosphere to another.

In the following analysis, we separate the lower stratosphere (2$\times$10$^{4}$ Pa to 6$\times$10$^{1}$ Pa) and the upper stratosphere (6$\times$10$^{1}$ Pa to 5$\times$10$^{-1}$ Pa) to clarify the analysis. For each stratospheric regions, the two components of the Eliassen-Palm vectors are scaled by their own averaged values which are distinct in the two stratospheric regions.

\begin{figure*}
    \centering
    \subfigure[][]{
        \label{subfig:EP_diagram_highstrato}
        \centering
        \includegraphics[scale=0.05]{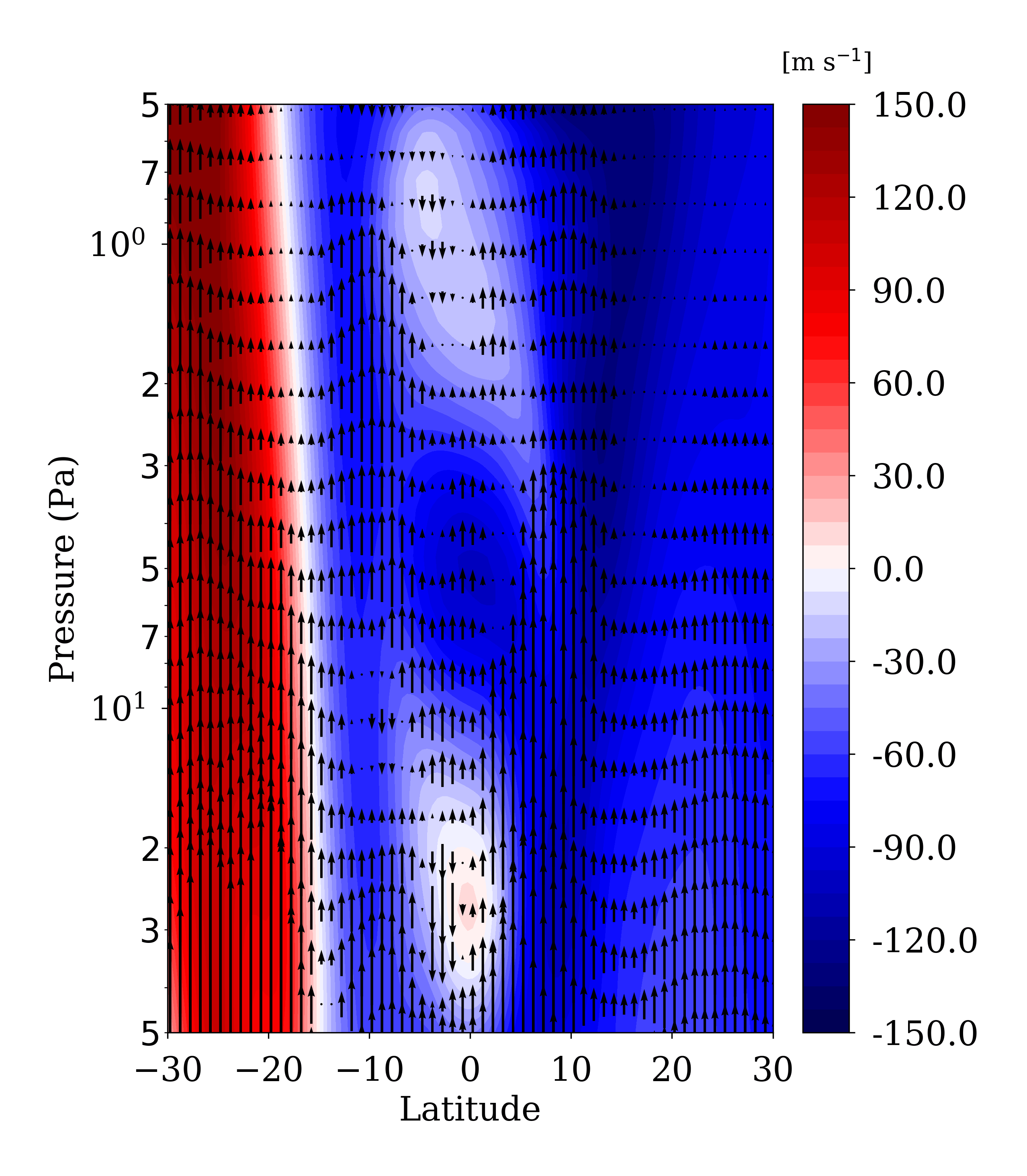}}
    \subfigure[][]{
        \label{subfig:EP_diagram_lowstrato}
        \centering
        \includegraphics[scale=0.05]{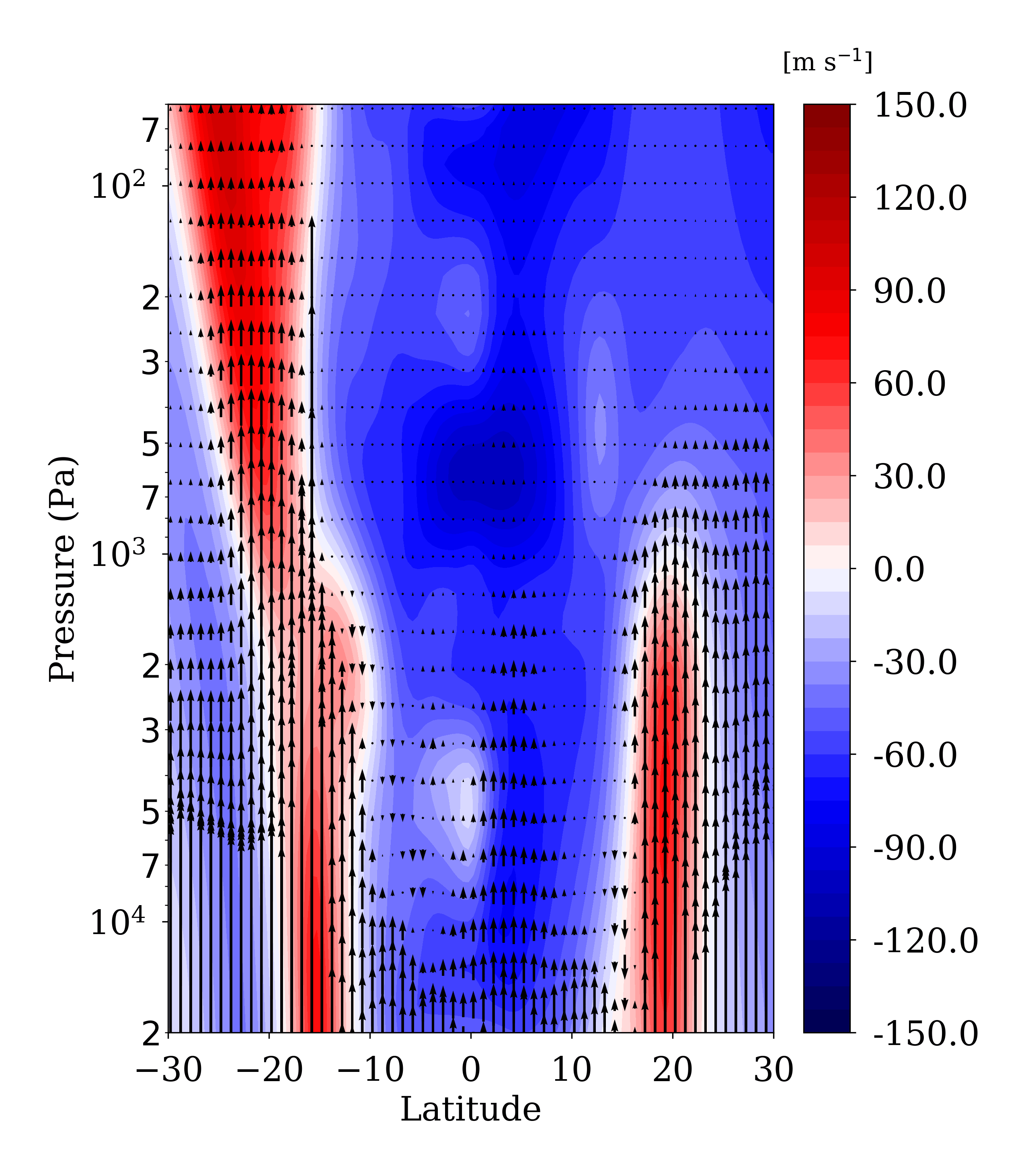}}    
    \caption{Altitude/latitude sections of the zonal mean zonal-wind (color) and the Eliassen-Palm flux in equatorial regions at Ls = 140$^{\circ}$ during the eighth simulated Saturn years, focused on high stratospheric levels (\ref{subfig:EP_diagram_highstrato}) and low stratospheric levels (\ref{subfig:EP_diagram_lowstrato}). Vectors scale is arbitrary for each section.}
    \label{fig:EPflux_diagram_zooms}
\end{figure*}

We ran an additional, specific 1000-day-long GCM simulations with a daily output frequency, restarted at a time of a strong reversal of wind direction from the westward to the eastward direction (at Ls = 140$^{\circ}$ during the eighth simulated Saturn years). Figure \ref{fig:EPflux_diagram_zooms} displays the time-average of the Eliassen-Palm flux diagram (vectors) superimposed on the zonal-mean zonal wind (color). The Eliassen-Palm flux essentially comes from the high troposphere in tropical regions, particularly the strong eastward jets located at $\pm$20$^{\circ}$ of latitude. In the southern hemisphere, the Eliassen-Palm flux vectors is significant between 2$\times$10$^{4}$ Pa and 6$\times$10$^{2}$ Pa in the lower stratosphere (Figure \ref{subfig:EP_diagram_lowstrato}) and between 5$\times$10$^{1}$ Pa and 1$\times$10$^{1}$ in the upper stratosphere (Figure \ref{subfig:EP_diagram_highstrato}). The eddy momentum transferred to the upper stratospheric layers mainly comes from the strong eastward jet correlated to the rings shadow (centered at 25$^{\circ}$S). Comparison between northern and southern tropical regions shows that the largest eddy momentum forcing is correlated to this strong eastward jet, even in the highest stratospheric layers, where the forcing is greatly reduced. Furthermore, there is an inversion of the Eliassen-Palm vectors' direction at the equator (downward) compared to the two tropical regions (upward). We deduce that there is a conceivable cell of eddy flux on both sides of the equatorial stacked jets, which suggests an eddy forcing of the equatorial oscillation.

\begin{figure*}
    \centering
    \includegraphics[scale=0.2,angle=90]{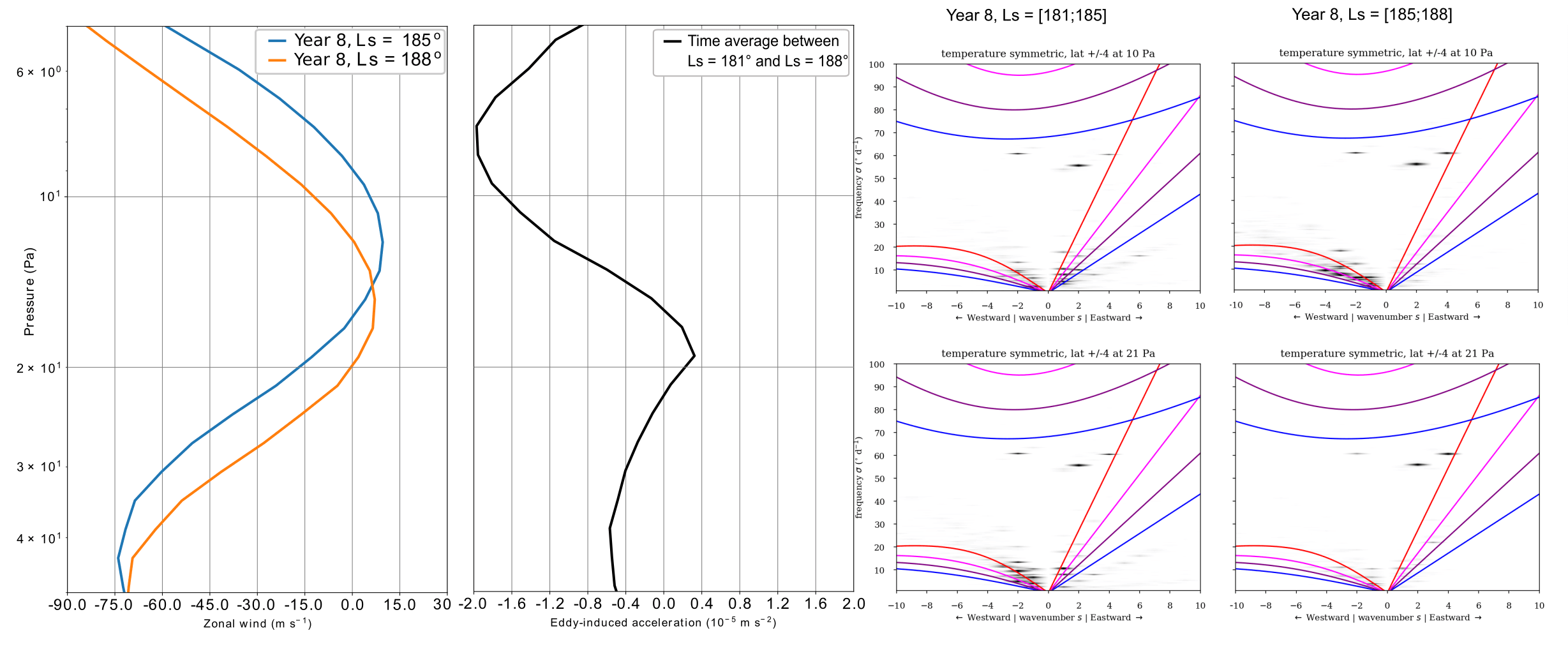}
    \caption{Left panel: Zonal-mean zonal wind profiles at Ls = 185$^{\circ}$ (blue line) and Ls = 188$^{\circ}$ (orange line). Middle panel: zonal-mean eddy-induced acceleration (term II of Equation \ref{eq:acc_TEM}, black line), average over 500 days around Ls = 185$^{\circ}$. Each profiles are average over $\pm$ 5 degrees of latitude. Right panel: two-dimensional Fourier transforms of the symmetric component of temperature field, following \cite{wheeler_1999} method, over two time intervals (Ls = 181-185$^{\circ}$ and Ls = 185-188$^{\circ}$) and at 21 and 10 Pa.}
    \label{fig:eddy-to-mean_for_downward_propagation}
\end{figure*}

Eddy-to-mean interactions are indeed involved in the downward propagation of the terrestrial equatorial oscillation. 
We investigate those interactions in the case of our simulated Saturn equatorial oscillation by choosing a representative case study, the downward propagation of the equatorial eastward zonal jets between Ls = 185$^{\circ}$ and Ls = 188$^{\circ}$ on simulated year 8.
For this test case, to diagnose the wave-induced forcing on the mean flow, we report the temporal evolution of zonal wind profiles, waves and the divergence of the Eliassen-Palm flux in Figure \ref{fig:eddy-to-mean_for_downward_propagation}. 
Additionally, to investigate the types of waves involved in this forcing, we perform two related spectral analysis, one in the interval Ls = 181-185$^{\circ}$ and the other in the interval Ls = 185-188$^{\circ}$, each one over 250 simulated Saturn days. 

At Ls = 185$^{\circ}$, zonal wind profile shows an intense westward jet (almost -75~m s$^{-1}$) between 50 and 15~Pa, a weak eastward jet between 15 and 9~Pa (only 10~m s$^{-1}$) and the lower part of a second intense westward jet (from 9~Pa to the top of the figure). 
At 21~Pa, between Ls=181$^{\circ}$ and 188$^{\circ}$, there is a small eastward eddy-induced acceleration, of about 0.3~m s$^{-2}$. Our spectral analysis shows that this is associated to a decrease of both Rossby and Kelvin waves activity at this pressure level. %From Ls = 185$^{\circ}$ to = 188$^{\circ}$, Rossby and Kelvin waves are absorbed at 21 Pa.
The eastward eddy-induced acceleration, correlated to the Kelvin waves activity at 21 Pa, forces the wind direction to eastward propagation and tends to ``push" the eastward jet down. So, at Ls = 188$^{\circ}$, this eastward jet is located from 20 to 12 Pa, close to the weak eastward eddy-induced acceleration.
At 10 Pa, the temporal average of the Eliassen-Palm flux divergence demonstrates a significant westward eddy-induced acceleration, associated to a great enhancement of Rossby waves and a vanishing of Kelvin waves activity in spectral analysis.
During the 500 Saturn days around the date Ls = 185$^{\circ}$, at 10~Pa, the westward waves transfer westward momentum to the mean flow, the large westward acceleration changes the wind direction and attenuates the eastward jet. As a result, the westward jet, located between 9~Pa to the top of the figure at Ls = 185$^{\circ}$, downs to 11 Pa at Ls = 188$^{\circ}$. At this pressure level, we observe the transition form the eastward phase to the westward phase of the equatorial oscillation. 
Eastward and westward accelerations occur just below eastward and westward jets, respectively: this may be due to the location of critical levels. Approaching a critical level from below, the wave vertical group velocity and vertical wavelength both decrease, facilitating the momentum exchange with the mean flow, although the maximum transfer of momentum can occur somewhere before encountering the critical level (e.g. through radiative damping). This could explain the downward propagation of the stacked jets causing the QBO-like oscillation. 

%%%%%%%%%%%%%%%%%%%%%%%%%%%%%%%%%%%%%%%%%%%%%%%%%%%%%%%%%%%%%%%%%%%%%%%%%%%%%%%%%%%%%%%%%%%%%%%%
%%%%%%%%%%%%%%%%%%%%%%%%%%%%%%%%%%%%%%%%%%%%%%%%%%%%%%%%%%%%%%%%%%%%%%%%%%%%%%%%%%%%%%%%%%%%%%%%

\section{Saturn seasonal cycle to synchronize its equatorial oscillation}
\label{oscillation/season}

%%%%%%%%
Idealized terrestrial QBO simulations show that the seasonal cycle could lock the equatorial oscillation period to an integer multiple of the annual cycle period \citep{rajendran_2016}. In our reference simulation, the QBO-like oscillation period is nearly the observed mean period, namely about half a Saturn year. This raises a new question: does Saturn's annual seasonal cycle synchronize the period of the equatorial oscillation? To address this question, we performed an alternate simulation (named hereinafter the ``no-season'' simulation) in which the seasonal cycle of the incoming solar radiation is neglected: the planet's position around the Sun is fixed at Ls=0$^{\circ}$. As a consequence, there is no ring shadowing in this simulation to avoid any singularities in the radiation at the equator. All other settings are similar to the DYNAMICO-Saturn reference simulation described previously. This ``no-season'' simulation starts with an initial state derived from the reference simulation after 7 simulated Saturn years. In this alternate simulation, the DYNAMICO-Saturn GCM is run for six simulated years: three years to reach a dynamical steady-state (a spin-up phase) and three additional years to compare the results with the eleventh to the thirteenth years of our reference simulation.

\begin{figure*}
    \centering
    \includegraphics[scale=0.15]{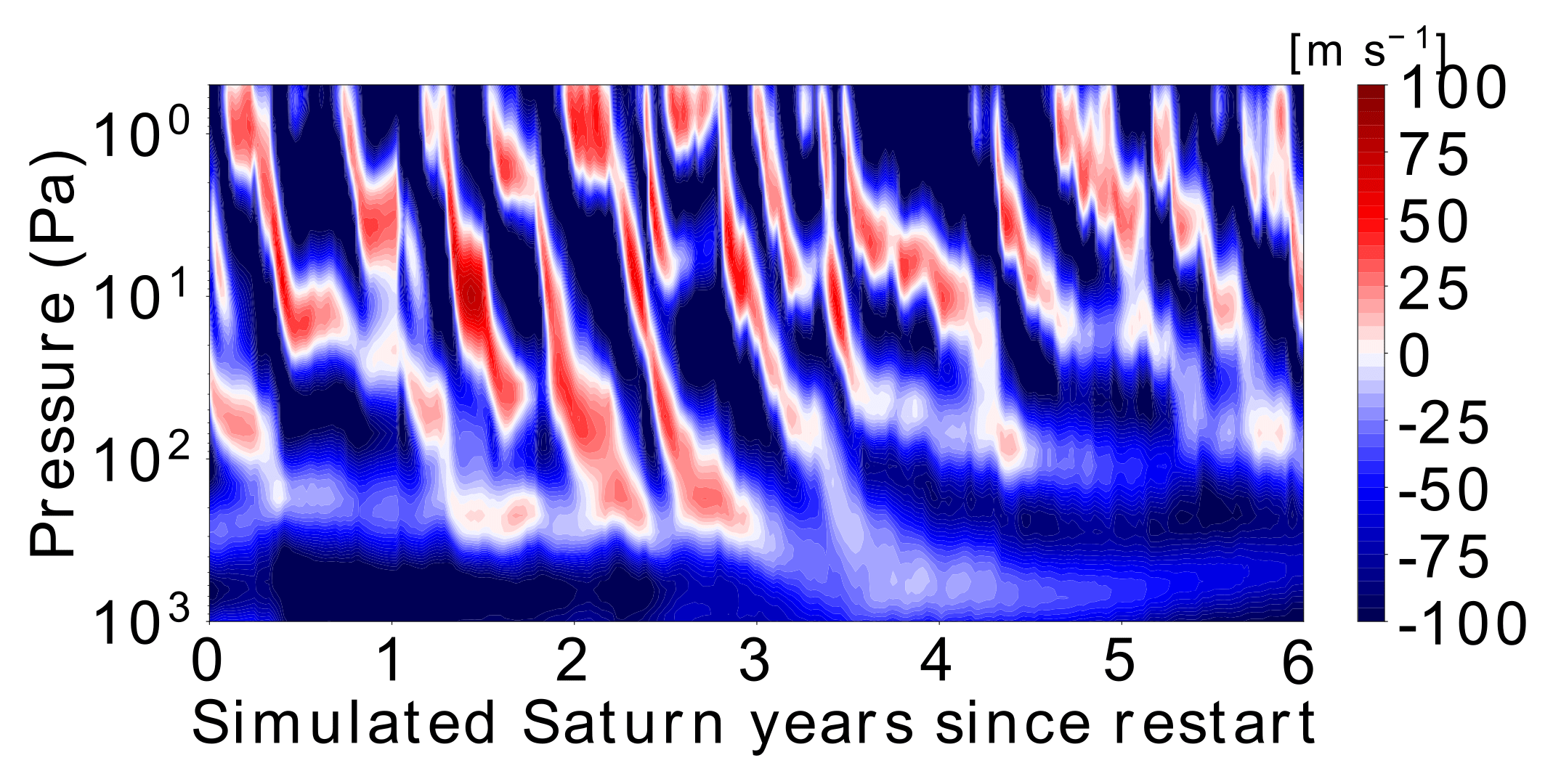}
    \caption{Altitude/time section at the equator of the zonal-mean zonal wind in Saturn's stratosphere for the no-season DYNAMICO-Saturn simulation.}
    \label{fig:u_y00_strato_noseason}
\end{figure*}

Figure \ref{fig:u_y00_strato_noseason} shows the time evolution of the zonal-mean zonal wind vertical structure at the equator for the six simulated Saturn years of the ``no-season'' simulation. Compared to the reference simulation (Figure \ref{fig:u_y00_strato}), two conclusions emerge from the ``no-season'' simulation: 
\begin{itemize}
    \item During the spin-up phase, the eastward wind intensity is higher and the eastward phase of the equatorial oscillation is more regular and longer duration than in the reference simulation. 
    \item after three simulated Saturn years, the oscillation begins to disappear in the lower stratosphere, toward 5$\times$10$^{1}$ Pa. 
\end{itemize}
There is an interconnection between the stratospheric jets (mean flow) and the eddies that force the equatorial oscillation. The presence of established jets, coming from the reference simulation, at the beginning of the ``no-season'' simulation influences the eddies activity, maintaining a seasonal signature for about three years. Jets and eddies are intimately coupled: once the seasonal cycle is removed, the equilibrium between jets and eddies is slightly disturbed, the effect of which is only noticeable after three simulated years.

After spin-up, the equatorial oscillation disappears in the lower stratosphere, between 10$^{3}$ and 5$\times$10$^{1}$ Pa, but it is maintained between 5$\times$10$^{1}$ Pa to the model top with a different periodicity than in the reference simulation. At any pressure level, from the fourth simulated Saturn year to the end of the ``no-season'' simulation, we notice only 3 or 4 stratospheric eastward jets. We obtain, in this ``no-season'' simulation, a periodicity of about 0.7 simulated Saturn year which is less consistent with observations than the reference simulation. This is consistent with one of the conclusions of \citet{rajendran_2016}: the annual cycle can regulate an equatorial oscillation by enhancing the eddies forcing induced by the seasonality of atmospheric waves.

We conclude from our ``no-season'' simulation that the seasonal cycle of Saturn is a key parameter to establish and regulate the stratospheric equatorial oscillation modeled by our DYNAMICO-Saturn GCM. Both eddy activity and residual-mean circulations are impacted by seasons in such a way that the periodicity of the equatorial oscillations is ``locked'' close to a semi-annual periodicity. How much which effect dominates over the other is left for a future study.

%%%%%%%%%%%%%%%%%%%%%%%%%%%%%%%%%%%%%%%%%%%%%%%%%%%%%%%%%%%%%%%%%%%%%%%%%%%%%%%%%%%%%%%%%%%%%%%%
%%%%%%%%%%%%%%%%%%%%%%%%%%%%%%%%%%%%%%%%%%%%%%%%%%%%%%%%%%%%%%%%%%%%%%%%%%%%%%%%%%%%%%%%%%%%%%%%

\section{Impact of the rings on Saturn's stratospheric dynamics}
\label{oscillation/rings}

%%%%%%%%
\subsection{Zonal wind in the tropical regions}
\label{Rings_zonal_wind}

To determine Saturn's rings contribution to the atmospheric dynamics in the tropical channel, we performed an alternate simulation named the ``no-ring'' simulation, in which the rings' shadowing of the incoming solar radiation is neglected, all other settings being equal with the DYNAMICO-Saturn reference simulation described previously. Contrary to the ``no-season'' simulation, a seasonal cycle is simulated in the ``no-ring'' simulation: it is different from the reference simulation since the shadow of the rings is removed. This ``no-ring'' simulation starts with an initial state derived from the reference simulation after 8 simulated Saturn years. The DYNAMICO-Saturn GCM is run for five simulated years: three years devoted to spin-up to reach a dynamical steady-state (see section \ref{oscillation/season}) and two additional years to compare the results with the eleventh and the twelfth years of our reference simulation.

\begin{figure*}
    \centering
    \subfigure[][]{
        \label{subfig:u_04mb_ring}
        \centering
        \includegraphics[scale=0.1]{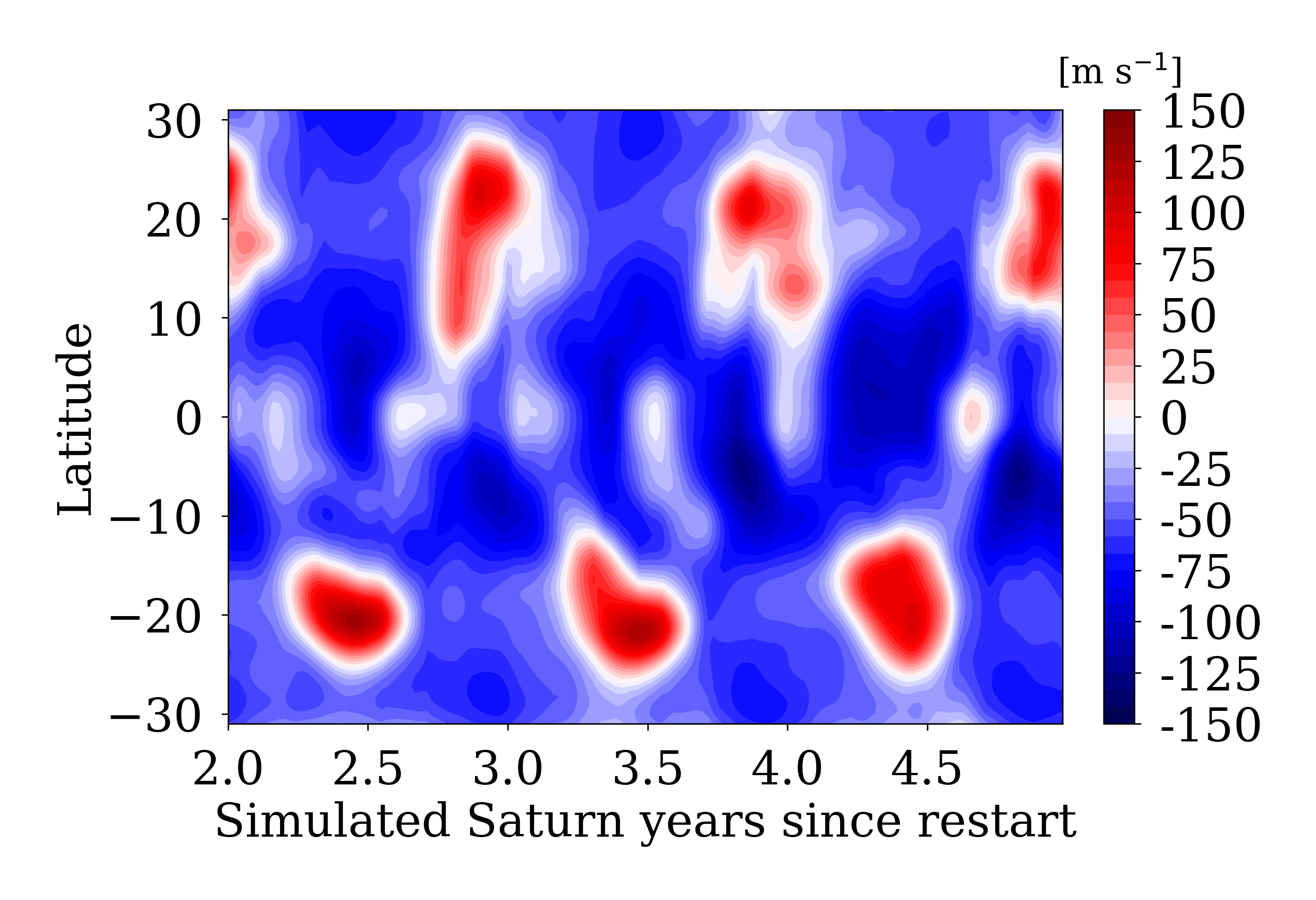}}
    \subfigure[][]{
        \label{subfig:u_04mb_noring}
        \centering
        \includegraphics[scale=0.1]{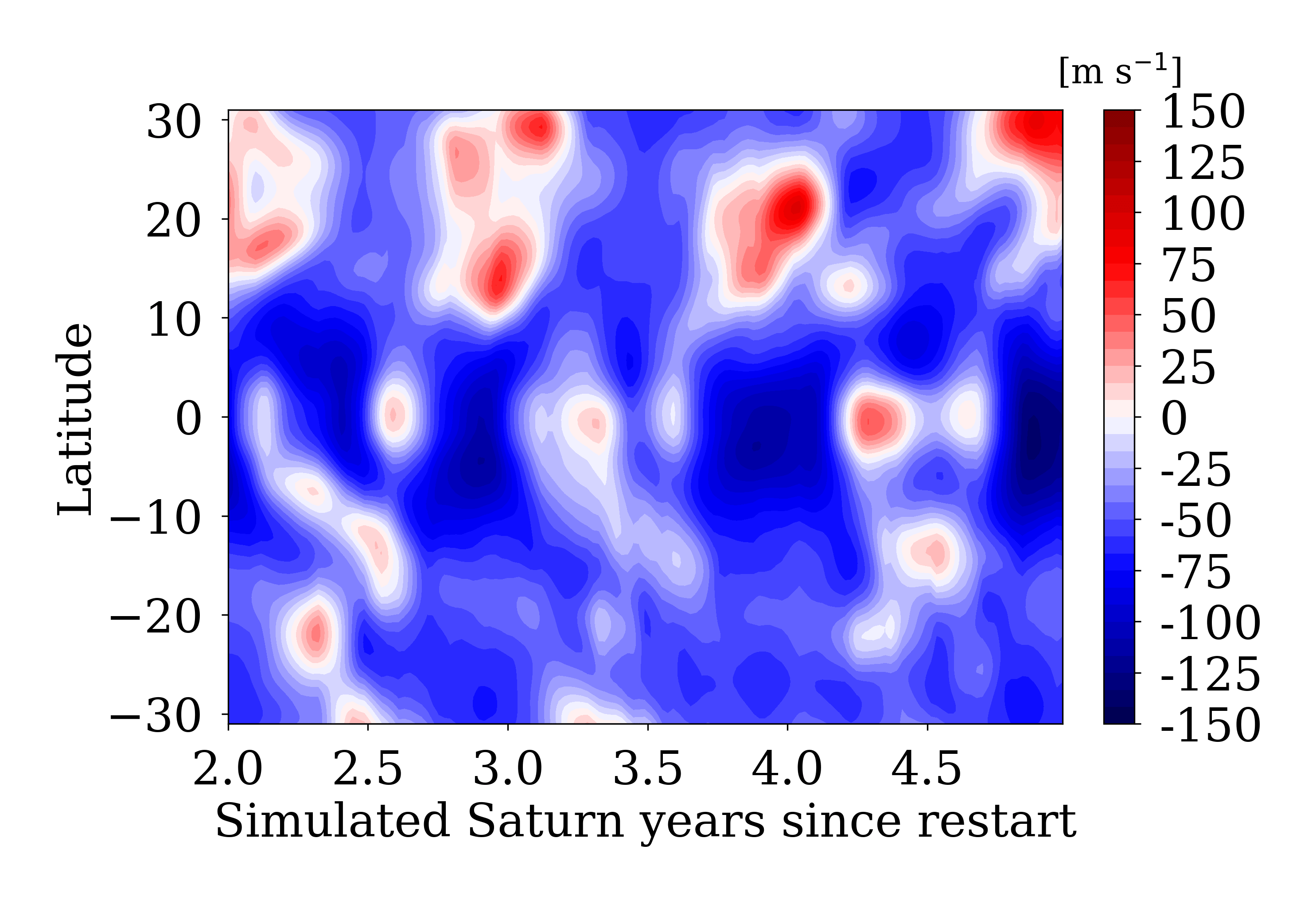}}
    \caption{Two-month running mean evolution of the zonal-mean zonal wind in Saturn's stratosphere (40 Pa) between 30$^{\circ}$N and 30$^{\circ}$S of our DYNAMICO-Saturn comparative simulations: \ref{subfig:u_04mb_ring} with rings'~shadow (reference simulation) and \ref{subfig:u_04mb_noring} without rings' shadow. The two first simulated years of the alternate simulation are not shown (model spin-up).}
    \label{fig:u_04mb_rings}
\end{figure*}

Figure \ref{fig:u_04mb_rings} compares the two Saturn GCM simulations. In the no-ring simulation compared to the reference simulation, we note a disturbance of the structure of the zonal-mean zonal wind at the tropics, as well as an enhancement of the zonal wind intensity at the equator. Eastward jets at the tropics in the no-ring simulation split in several weaker eastward jets in the northern hemisphere. In the southern tropics, eastward jets even disappear after 3.5 Saturn years of simulation. Hence, rings' shadowing (or the absence thereof) impacts equatorial and tropical dynamics. 

\begin{figure*}
    \centering
    \includegraphics[scale=0.15]{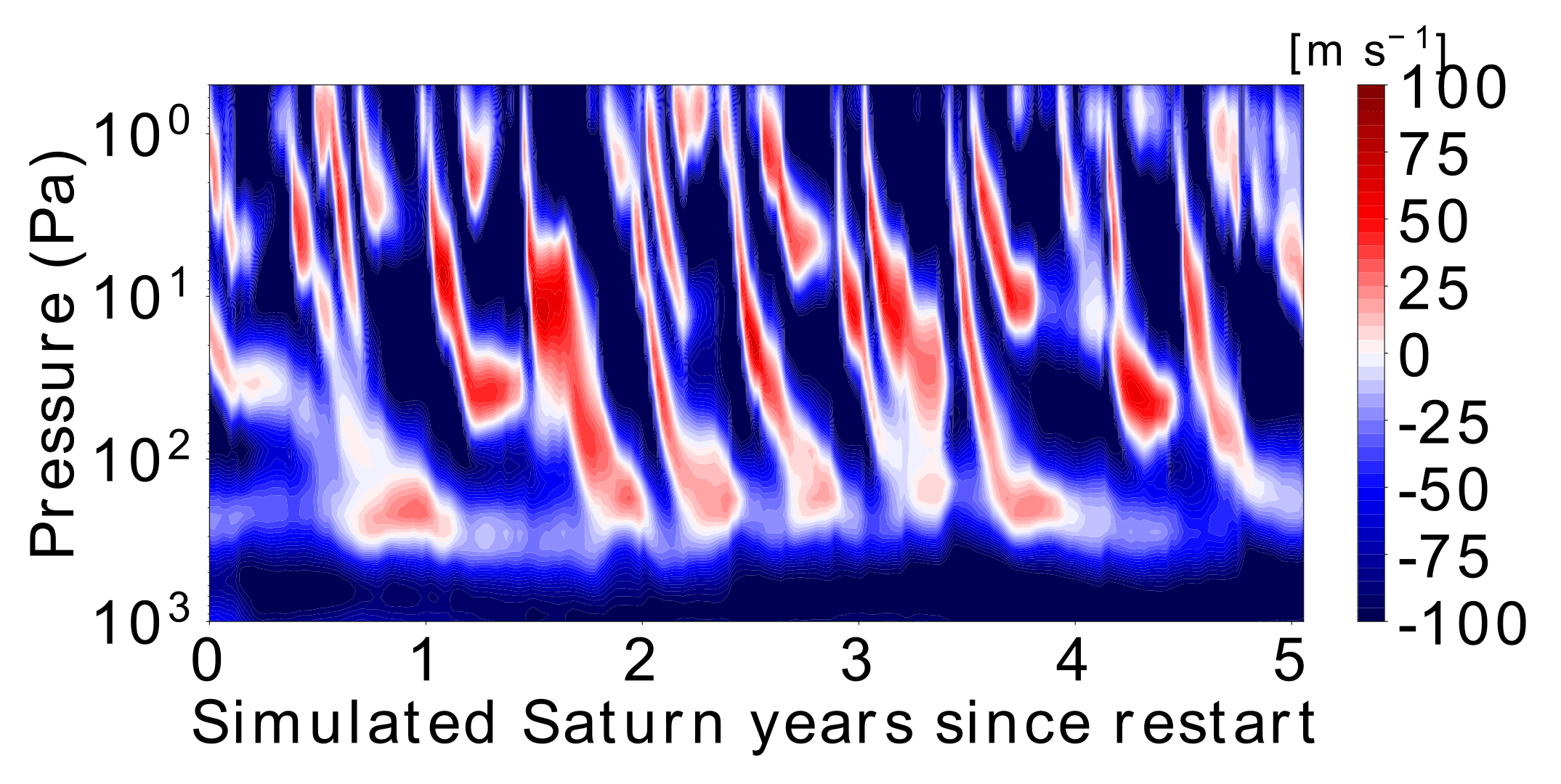}
    \caption{Altitude/time section at the equator of the zonal-mean zonal wind in Saturn's stratosphere for the no-ring DYNAMICO-Saturn simulation.}
    \label{fig:u_y00_strato_norings}
\end{figure*}
We present in Figure \ref{fig:u_y00_strato_norings} the vertical structure of the stratospheric equatorial oscillation resulting from the no-ring simulation. As in the reference simulation, the eastward phase of this QBO-like oscillation is irregular compared to the westward phase. However, eastward stratospheric jets in the no-ring simulation are larger in amplitude compared to the reference simulation (see Figure \ref{fig:u_y00_strato}), and the eastward phase of the equatorial oscillation lasts longer in the no-ring simulation than in the reference simulation. Eastward and westward phases are less contrasted. Neglecting the ring shadowing, we obtain an eastward phase of the simulated Saturn QBO-like oscillation enhanced by around 25 m s$^{-1}$ and a periodicity between 0.4 and 0.5 simulated Saturn years.
%%%%%%%%
\subsection{Temperature field in the tropical region}
\label{Rings_temperature}

\begin{figure*}
    \centering
    \includegraphics[scale=0.15]{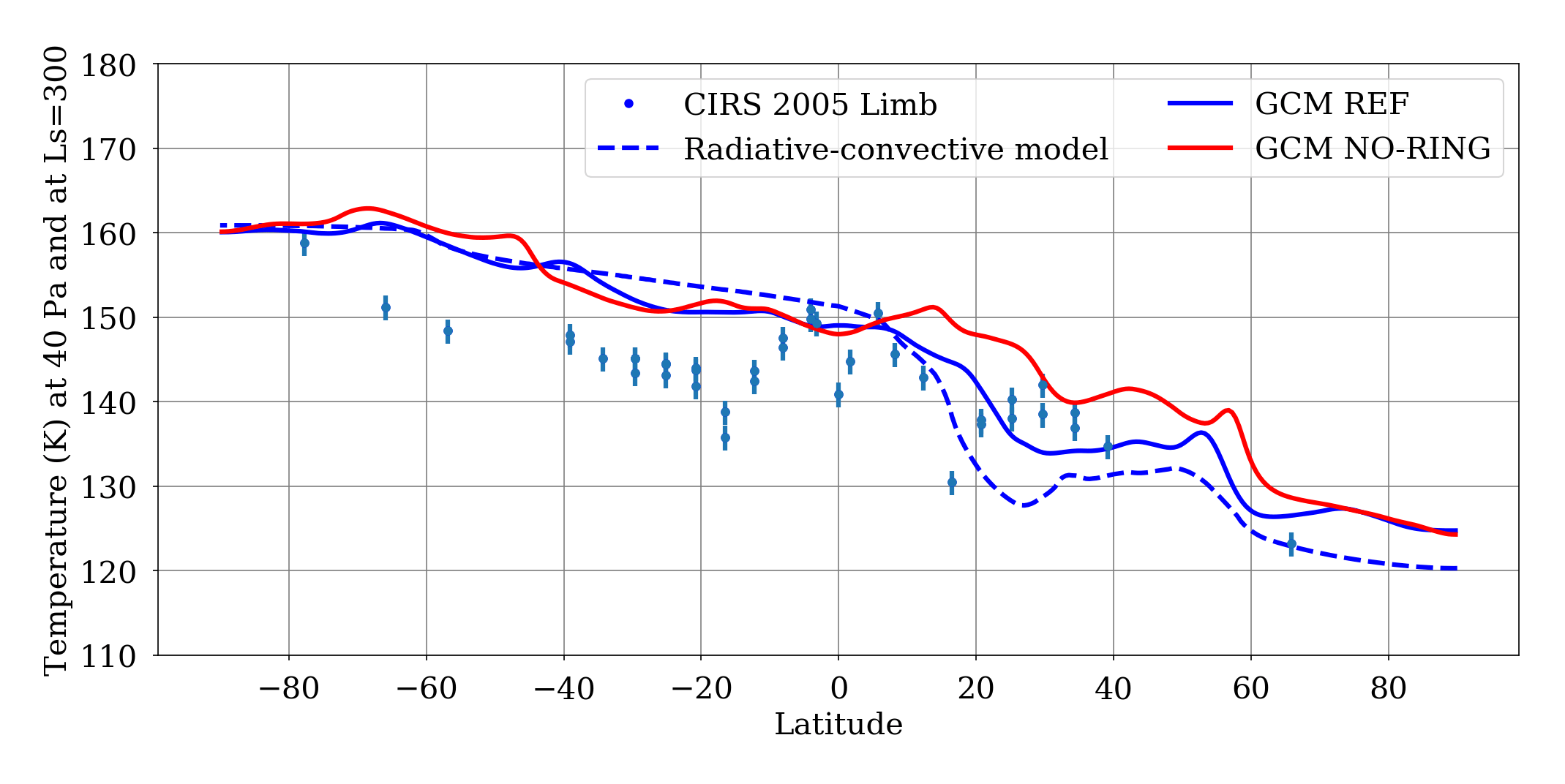}
    \caption{Temperature comparison between CIRS 2005 limb data (blue dots), reference simulation (blue line), NO-RING simulation (red line) and the radiative-convective model \emph{à la \cite{guerlet_2014}} (blue dashed line) at the 40 Pa pressure level. Errorbars for the CIRS data contain the error due to the measurement noise and also related to the uncertainty of the CH$_{4}$ abundance.}
    \label{fig:comparison_CIRSmap-rings-norings-radconv}
\end{figure*}

Figure \ref{fig:comparison_CIRSmap-rings-norings-radconv} compares at the 40-Pa pressure level
the reference DYNAMICO-Saturn simulation, the no-ring DYNAMICO-Saturn simulation, with radiative-convective 2D seasonal modeling \emph{à la} \citet{guerlet_2014}, and the 2005 CIRS limb observations \citep{guerlet_2009}.
This comparison is carried out at Ls = 300$^{\circ}$ in southern summer, when rings shadowing occurs at 20-25$^{\circ}$N. As mentioned previously, the radiative-convective model fails to reproduce the local temperature maximum under the rings' shadow (between 20 and 30$^{\circ}$N). It predicts instead a temperature $\sim$12 K colder than the CIRS observations. Besides, the reference GCM simulation presents higher temperatures at these latitudes, which are still $\sim$5 K too cold compared to measurements. At northern mid-latitudes in the winter hemisphere, the modeled temperature decreases from~20$^{\circ}$N to~40$^{\circ}$N, that we can associate with the signature of the rings shadow. Even if the dynamics of the DYNAMICO-Saturn GCM raises temperature underneath the ring, there remains a failure to replicate hot anomalies in the shadows. The no-ring simulation also exhibits warm temperature between 20$^{\circ}$N and 30$^{\circ}$N; however, these temperatures are higher than the reference simulation (probably because the tropical eastward jets are impacted by the lack of ring shadowing) and too warm compared to the observations for the whole northern hemisphere except between 30-35$^{\circ}$N. Hence Figure \ref{fig:comparison_CIRSmap-rings-norings-radconv} suggests a possible dynamical origin of the unexpected high temperature below Saturn's rings, but this dynamics is not fully addressed by DYNAMICO-Saturn \emph{with} rings shadow. We defer the detailed study of the seasonal circulation to a future work.

%%%%%%%%
\subsection{Eddy-to-mean interactions in the tropical region}
\label{Rings_TEM}

Ring shadowing impacts the tropical eastward jets. Those jets are forced by the residual-mean circulation and the eddy-to-mean interactions. To further characterize the impact of ring shadowing on those two contributions, we use the TEM formalism (Equation \ref{eq:acc_TEM}). 

\begin{figure*}
    \centering
    \begin{subfigure}
        \centering
        \includegraphics[scale=0.05]{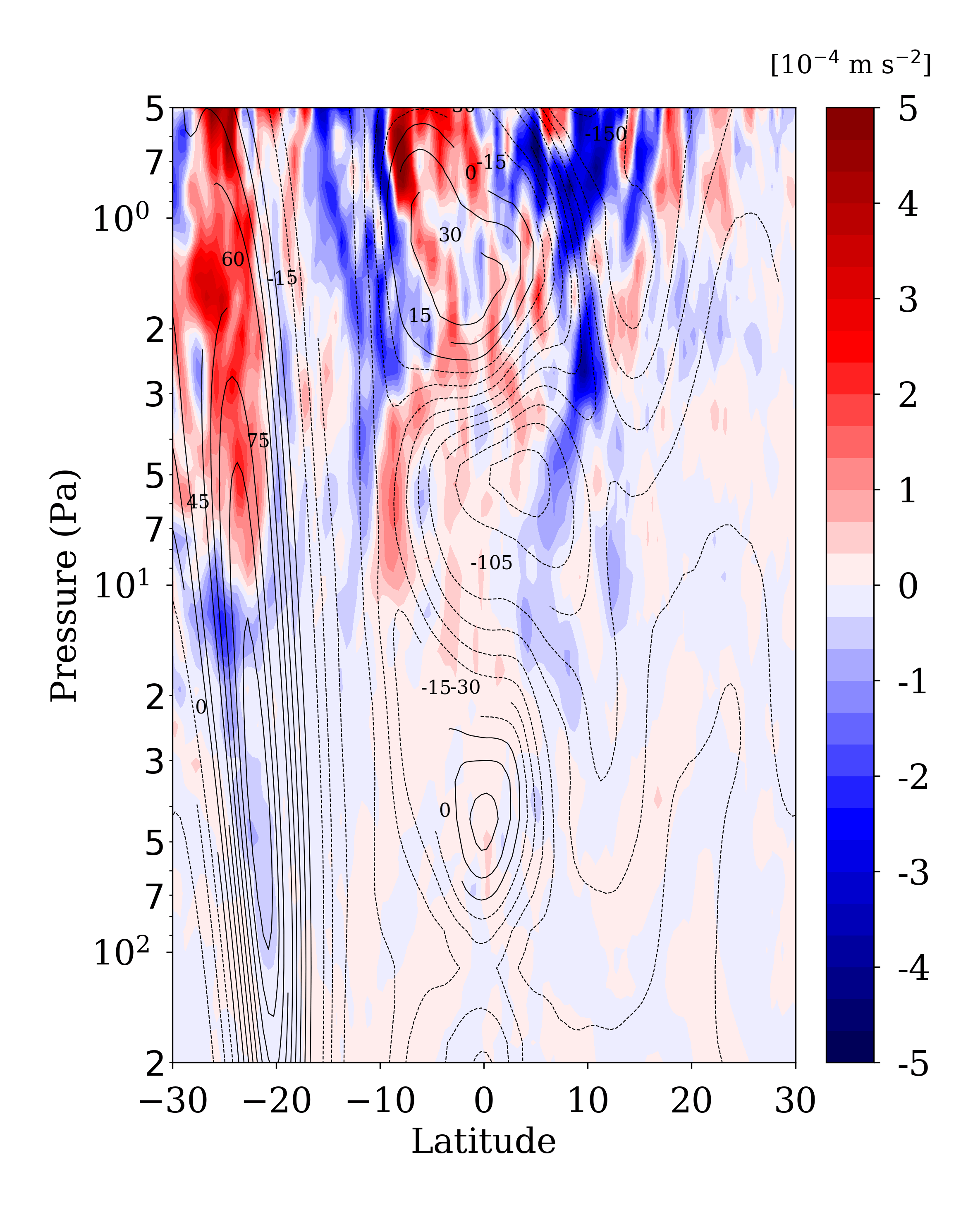}
    \end{subfigure}
    \begin{subfigure}
        \centering
        \includegraphics[scale=0.05]{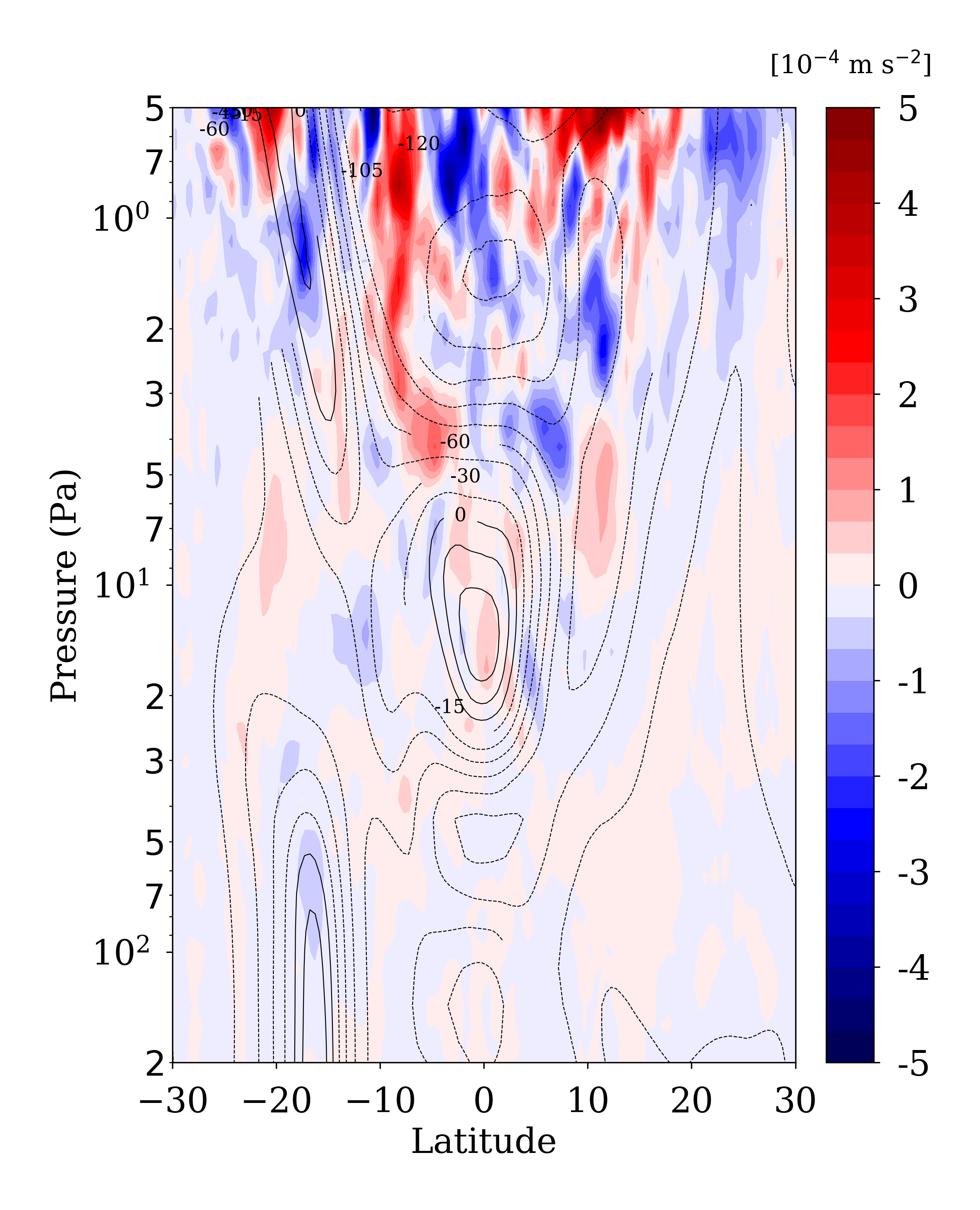}
    \end{subfigure}
    \centering
    \begin{subfigure}
        \centering
        \includegraphics[scale=0.05]{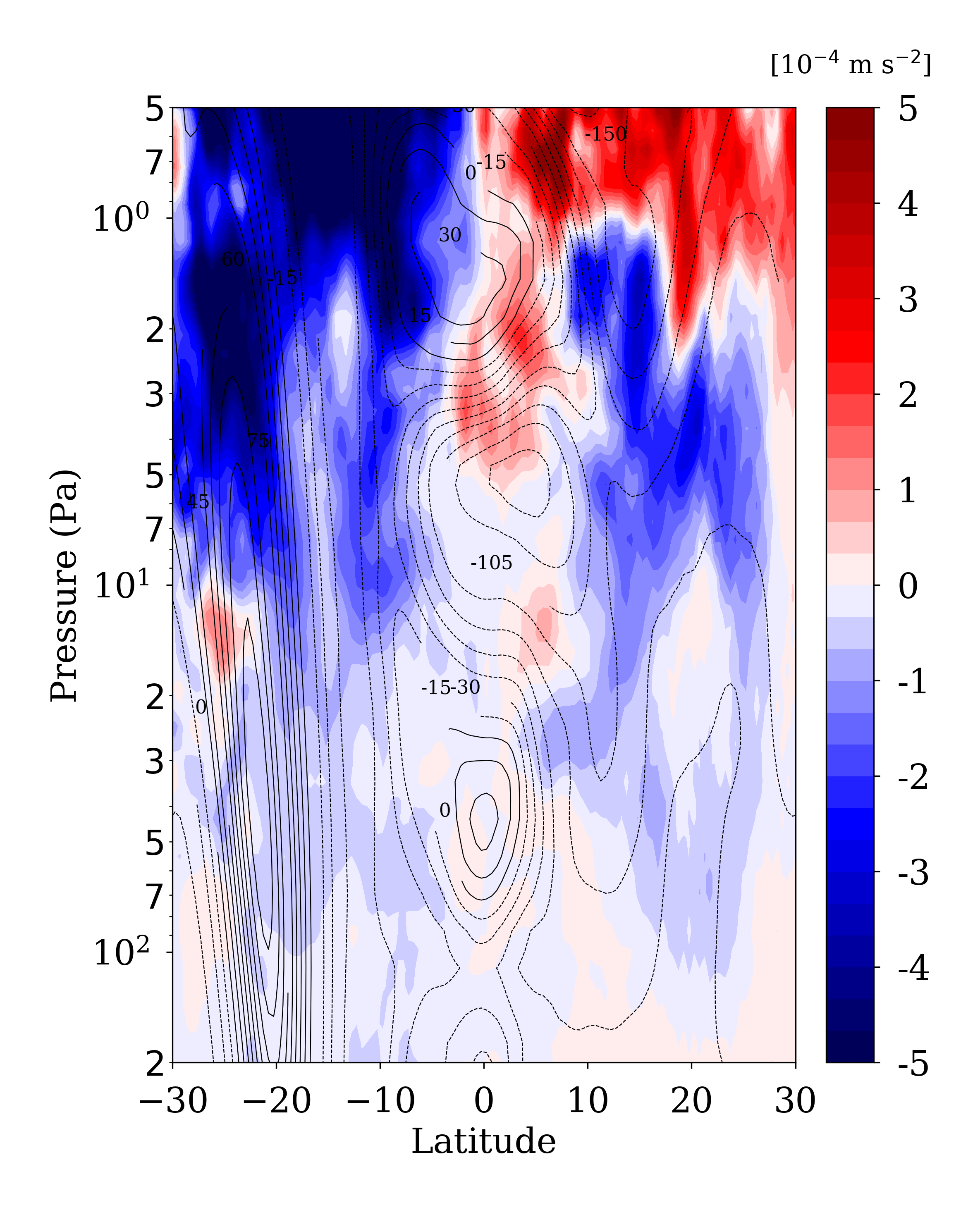}
    \end{subfigure}
    \begin{subfigure}
        \centering
        \includegraphics[scale=0.05]{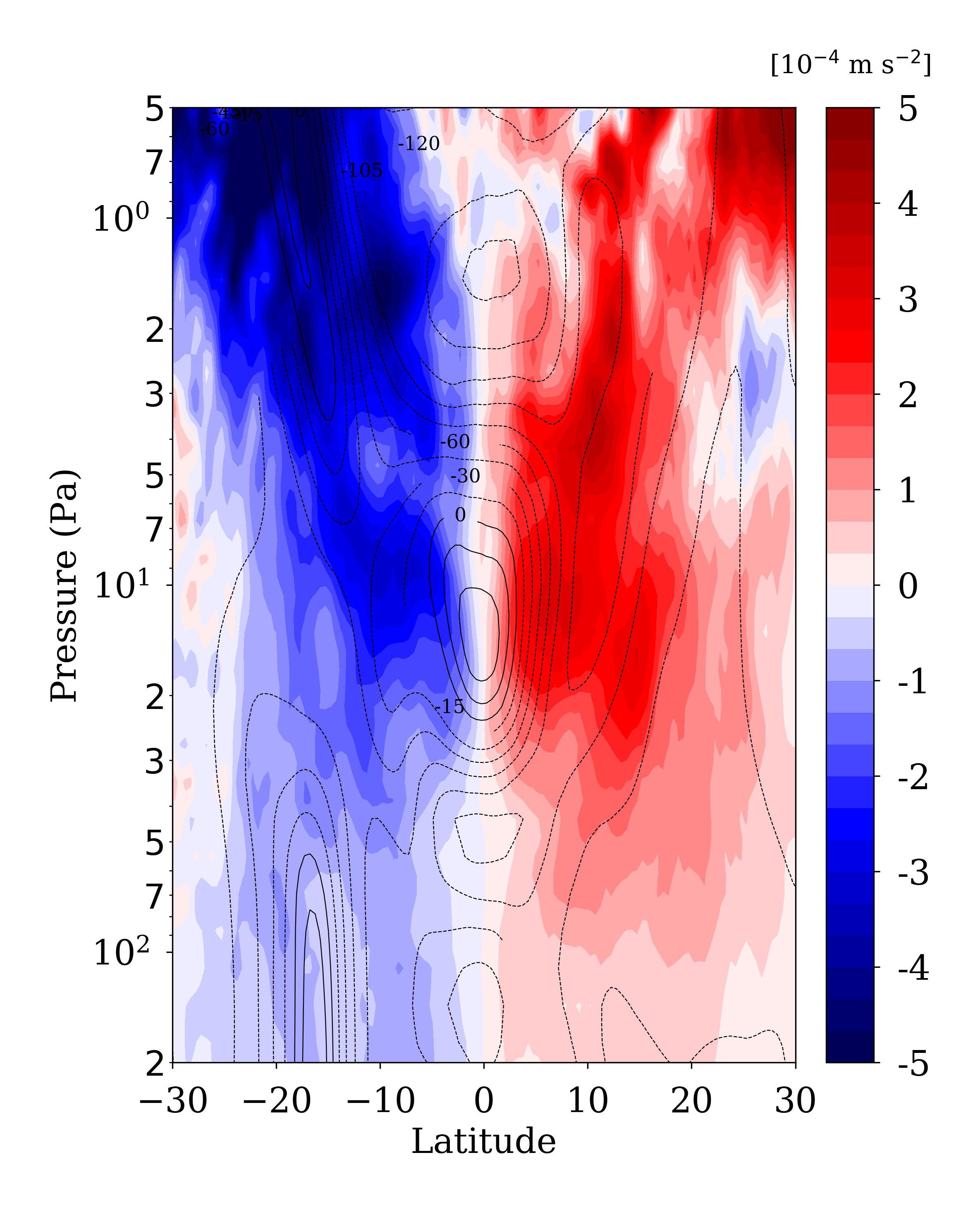}
    \end{subfigure}
    \caption{Comparative meridional sections of zonal-mean zonal wind (contours, solid lines depict eastward jets and dashed lines depict westward jets), eddy-induced acceleration (color, top rows) and mean-circulation-induced acceleration (color, bottom rows) in TEM formalism between rings- (left columns) and no-ring- (right columns) simulations at the 3.5 simulated Saturn year since restart. This dates correspond to the northern summer: rings' shadow takes place at around 20$^{\circ}$S and produces a strong eastward jet centered at 25$^{\circ}$S. }
    \label{fig:acc-edd-rmc_rings-norings_35}
\end{figure*}

In the two simulations, eddy-induced (term II of Equation \ref{eq:acc_TEM}) and mean-circulation-induced (term I of Equation \ref{eq:acc_TEM}) accelerations are of the same order of magnitude. It is also of the same order of magnitude between both simulations (Figure \ref{fig:acc-edd-rmc_rings-norings_35}). The presence of the tropical southern eastward jet in the reference simulation is associated with a significant eddy-induced acceleration that disappears completely when ring shadowing is not included (Figure \ref{fig:acc-edd-rmc_rings-norings_35} top rows). The equatorial behavior of eddy-induced acceleration is of opposite sign at the top of the stratosphere in the no-ring simulation compared to the reference simulation. 

The acceleration due to the mean circulation is shown at the bottom of Figure \ref{fig:acc-edd-rmc_rings-norings_35}. Without rings shadowing, acceleration induced by the mean circulation is mainly positive in the north and negative in the south.
With ring shadowing, this pattern is disturbed both in the southern hemisphere (where a reduced acceleration is seen at the location of the tropical eastward jet) and in the opposite hemisphere as well. Indeed, in the northern hemisphere, the acceleration is now mainly negative in the range 2 -- 100 Pa when ring shadowing is taken into account.
A similar behavior is observed during southern summer (not shown).

\begin{figure*}
    \centering
    \subfigure[][]{
        \label{subfig:accedd-TEM-u_ring}
        \centering
        \includegraphics[scale=0.1]{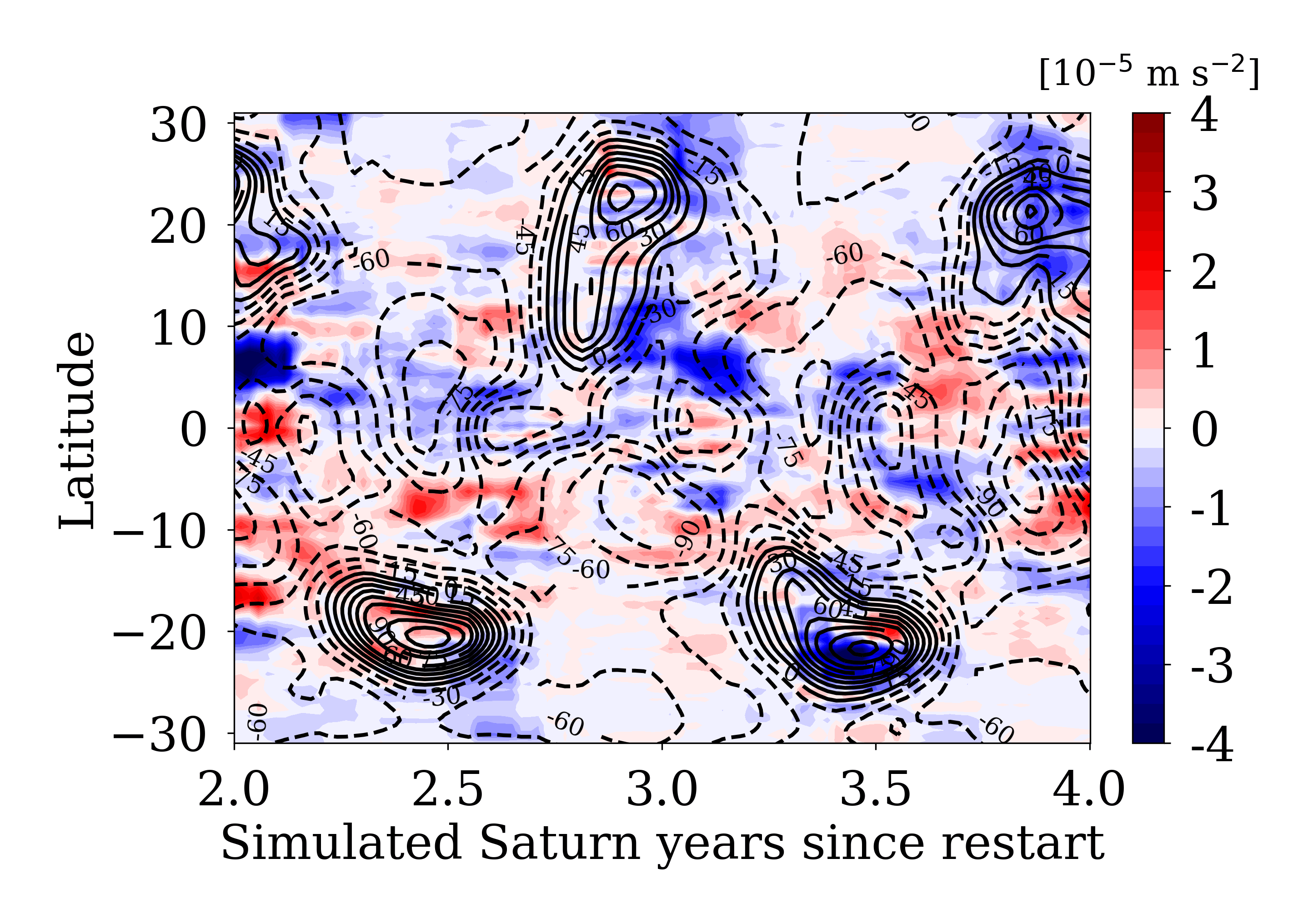}}
    \subfigure[][]{
        \label{subfig:accedd-TEM-u_noring}
        \centering
        \includegraphics[scale=0.1]{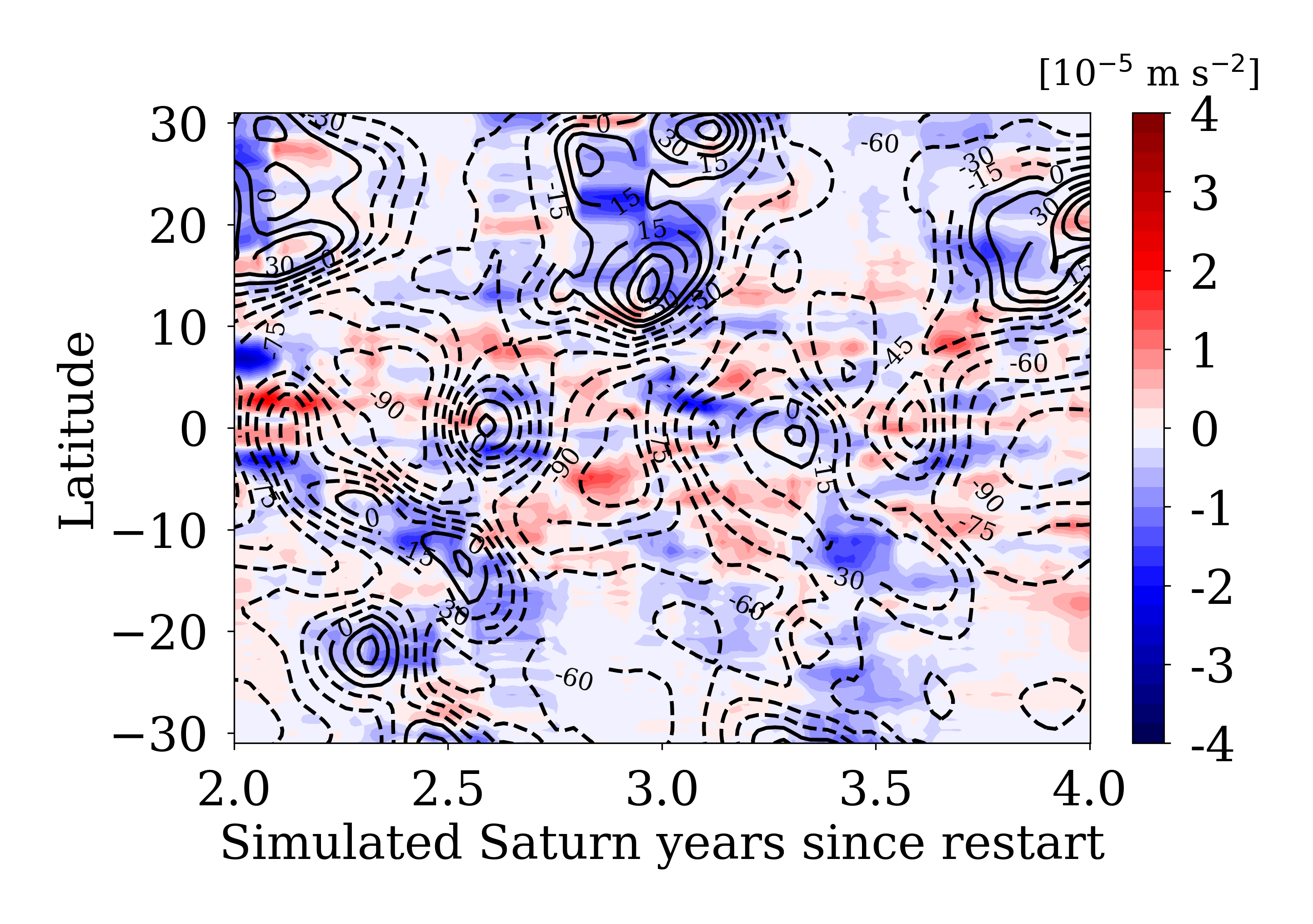}}
    \caption{Two-month running mean evolution of the zonal-mean eddy-induced acceleration (colors) and the zonal wind (contours) in Saturn's stratosphere (40 Pa) between 30$^{\circ}$N and 30$^{\circ}$S of our DYNAMICO-Saturn comparative simulations: \ref{subfig:accedd-TEM-u_ring} with rings'~shadow (reference simulation) and \ref{subfig:accedd-TEM-u_noring} without rings' shadow.}
    \label{fig:accedd-u_04mb_rings}
\end{figure*}

The amplification of the eddy-induced acceleration just below rings shadowing presents the same periodicity than the Saturn seasonal cycle. We map the time evolution of the acceleration due to eddies in the reference simulation (Figure \ref{subfig:accedd-TEM-u_ring}) and the no-ring simulation (Figure \ref{subfig:accedd-TEM-u_noring}). In the reference simulation, eddy-induced acceleration is enhanced at the tropics when eastward tropical jets are present. There is a strong deceleration due to eddies at the location of the tropical eastward jets (except for the jets located at 20$^{\circ}$S at 2.5 simulated Saturn year). The winter tropics, where the rings' shadow occurs and eastward jets are produced, are associated with an intense interaction between eddies and the mean circulation, which induces a significant deceleration due to eddy activity. Regarding the no-ring simulation, the eddy-induced acceleration is halved compared to the reference simulation at the location of the eastward jets. The seasonality of the eddy-to-mean interaction persists at the tropics without the rings shadowing, but the intensity of the eddy forcing in the mean flow is reduced. In this case, the whole tropical channel is driven by homogeneous absolute values of acceleration and deceleration (about 1.5$\times$10$^{-5}$m s$^{-2}$).

%%%%%%%%
To summarize the conclusions of this section~\ref{oscillation/rings}, rings' shadowing affects the stratospheric dynamics of Saturn. Rings shadowing impacts the thermal structure of the tropical regions with a seasonal periodicity. This causes seasonally-varying tropical eastward jets underneath the shadows, through two distinct dynamical interactions. First of all, there is a enhancement of the eddy-induced acceleration at the location of the shadow. Secondly, there is a reversal of sign of the acceleration (positive in the case without ring shadow to negative sign in the case with rings shadow) due to the residual-mean circulation in summer hemisphere (opposite of the "rings shadow" hemisphere). Moreover, the strong meridional gradients of wave activity disturb the eastward phase of the equatorial oscillation (and the overall periodicity of it) by producing low-intensity eastward jets.

A putative impact of the ring's shadow on Saturn's stratospheric dynamics was already suspected by previous authors, based on the observation of a local maximum in hydrocarbons and anomalously high temperatures under the ring's shadow in northern winter, disappearing at the spring equinox \citep{guerlet_2009, sylvestre_2015}.
Upward motion at summer mid-latitudes associated with a tropical adiabatic subsidence in the winter hemisphere could explain heat transport under the ring shadowing, as well as the observed hydrocarbon abundances asymmetries (\cite{guerlet_2009}, \cite{guerlet_2010}). 
With our study, we show that such interpretations of anomalies seen in observations invoking upward and downward motions may reflect a too simplistic view. Rather, eddy forcings might play a role as much as significant as the residual-mean circulation in shaping Saturn's stratospheric temperatures and dynamics.

%%%%%%%%%%%%%%%%%%%%%%%%%%%%%%%%%%%%%%%%%%%%%%%%%%%%%%%%%%%%%%%%%%%%%%%%%%%%%%%%%%%%%%%%%%%%%%%%
%%%%%%%%%%%%%%%%%%%%%%%%%%%%%%%%%%%%%%%%%%%%%%%%%%%%%%%%%%%%%%%%%%%%%%%%%%%%%%%%%%%%%%%%%%%%%%%%

\newpage
\section{Conclusions and perspectives}
\label{conclusions}

We use a troposphere-to-stratosphere Saturn GCM without any prescribed wave parameterization to study the equatorial stratospheric dynamics, especially the stratospheric equatorial oscillation.
Concerning the global atmospheric dynamics, spherical harmonic decomposition of the horizontal velocity shows that statistical properties are similar in the tropospheric and stratospheric levels. These properties differ when we extend the model top toward higher stratospheric levels, but still predict a stratosphere more energetic than the troposphere. However, the 32-level simulation shows energetic non-axisymmetric modes $m=1,2,3$ at large scale (i.e. small indices $n$) that vanish in our reference simulation. These modes likely results from energy accumulation at the model top which artificially enforces large-scale waves in the 32-level simulation of \citet{spiga_2018}. Our spectral analysis emphasizes a currently open question: does the statistical framework of 2D-turbulence with a $\beta$-effect apply to energy spectra of Saturn's atmosphere, and more generally to 3D planetary macroturbulence? If \cite{spiga_2018} simulation retrieves the well-known KK-law of quasi-2D turbulence, the $-5/3$ slope does not apply when we raise the model top in our reference simulation. These statistical predictions of our Saturn climate model have now to be validated by direct observations of the Saturn’s atmospheric flow.

Regarding the lower latitudes, our DYNAMICO-Saturn GCM reproduces an almost semi-annual equatorial oscillation with contrasted eastward and westward phases. This oscillation shows a similar behavior in temperature than the Cassini/CIRS retrievals, with alternatively local maxima and minima of temperature stacked on the vertical at the equator. The DYNAMICO-Saturn signal exhibits a two times smaller vertical characteristic size of the oscillation and underestimates by a factor of two the amplitude of the temperature anomalies, compared with CIRS observations. Regarding the zonal-mean zonal wind, we determined an irregular period of wind reversal and a downward propagation rate faster than observations. Spectral analysis at the equator demonstrated that this QBO-like oscillation is produced by planetary-scale waves. The equatorial oscillation in the DYNAMICO-Saturn GCM is mainly driven by strong westward-propagating waves, such as Rossby, Rossby-gravity and inertia-gravity waves, which deposit westward momentum in the stratosphere. There are only two eastward-propagating modes (Kelvin waves) in the spectral analysis with a weaker impact in the stratospheric dynamics than the westward-propagating waves. This lack of eastward waves prevents the eastward momentum deposition in the mean zonal wind and explains the erratic behavior of the eastward phase of the modeled QBO-like oscillation. Using the Transformed Eulerian Mean formalism to determine how the eddy-to-mean interactions drives the Saturn equatorial oscillation, we are able to conclude that the maximum of eddy forcing comes from the high-troposphere tropical regions. Moreover, we demonstrate the wave forcing origin of the vertically-stacked stratospheric eastward and westward jets that propagate downward with time to form the equatorial oscillation. At the equator, in the stratosphere, the eddy-induced eastward acceleration maximum is located just under the eastward jets and the eddy-induced westward acceleration maximum is located just under the westward jets at every step of the downward propagation of this Saturn QBO-like equatorial oscillation. A control experiment, using a perpetual equinox radiaitve forcing, demonstrate that the seasonal cycle of Saturn plays a significant role to establish and regulate the stratospheric equatorial oscillation modeled by our DYNAMICO-Saturn GCM. Both eddy activity and residual-mean circulations are impacted by seasons in such a way that the periodicity of the equatorial oscillations is ``locked'' close to a semi-annual periodicity.

For future improvements of the modeling of the QBO-like oscillation in Saturn’s equatorial stratosphere, we will draw inspiration from Earth’s atmospheric modeling. Adequate vertical resolution is needed to obtain a more realistic Quasi-Biennial Oscillation in Earth models (\cite{richter_2014}, \cite{hamilton_2001}). We plan to refine the vertical resolution in the stratosphere both to lead to a downward propagation rate of the oscillating zonal wind consistent with the observations, and a large-enough amplitude of Kelvin and Rossby-gravity waves to enhance the westward and eastward forcing of the Saturn equatorial oscillation phases. Furthermore, the Earth's QBO eastward phase is primarily induced by gravity waves, triggered by tropospheric convection (around 70\% of the total eastward forcing) and Kelvin waves for the remaining 30\% \citep{baldwin_2001}. To overcome the lack of eastward momentum in our Saturn stratospheric modeling, we plan to add a stochastic gravity wave drag parameterization \citep{lott_2012} in our DYNAMICO-Saturn GCM. This is expected to produce a more realistic wave spectrum (with equivalent eastward- and westward-propagating waves), which would strongly impact the simulation of the equatorial oscillation and the downward propagation of winds.

%%%%%%%%
Another main result of this study is the impact of ring shadowing on the stratospheric dynamics. In our DYNAMICO-Saturn reference simulation we obtained strong tropical eastward jets, which are seasonally periodic, and correlated with Saturn rings' shadow. With an additional simulation not including rings' shadowing, we show that the tropical eastward jets are caused by rings' shadowing in the stratosphere. Without rings' shadowing, the tropical eastward jets disappear after 3 simulated Saturn years. Comparisons between temperature predicted by our dynamical GCM simulations, computed with radiative-convective equilibrium, and measured from Cassini/CIRS observations also suggests a dynamical impact of rings' shadowing in the stratosphere. The transformed Eulerian mean formalism shows that the dynamical impact of ring shadowing on tropical eastward jets is strong both on the eddy-induced acceleration and the residual-mean-induced acceleration. In presence of rings' shadowing, eddy-induced acceleration is increased in the tropical channel and there is a reversal of the residual-mean-induced acceleration in the opposite hemisphere to the rings. Wave activity is correlated with rings' shadowing: the seasonality of the eddies is enhanced by rings' shadowing, which contributes to the annual periodicity of the strong eastward tropical jets. 
Dynamics in the equatorial region, in particular the QBO-like oscillation, is also influenced by rings' shadowing. 
%%%%%%%%%%

Stratospheric winds have never been measured on Saturn and were determined using the thermal wind balance from the CIRS retrievals of stratospheric temperatures. Obviously, the tropical eastward jets could be a particularity of our DYNAMICO-Saturn GCM. Because of the small vertical wind shear associated with these tropical jets, there is no specific temperature signature associated to it. In other words, we cannot invalidate their existence with the available observations of the temperature field. Future measurements of winds in Saturn's stratosphere would help to validate the predictive scenario drawn by our model.

%%%%%%%%
Future studies using the DYNAMICO-Saturn GCM could adopt a wider analysis scope and study the global meridional circulation in the stratosphere. Results presented here are focused on the tropical regions, while the GCM extent is global. For instance, we could further characterize heat transport under the ring shadows, as well as the observed asymmetries of hydrocarbon abundances (\cite{guerlet_2009}, \cite{guerlet_2010}). More generally, a complete study on the possible Brewer-Dobson-like circulation in the stratosphere of Saturn, and its impact on the hydrocarbons distribution, is warranted. A subsequent coupling of DYNAMICO-Saturn with photochemical models would then allow to refine this picture.

%%%%%%
Sudden stratospheric warming on Earth's high and mid-latitude regions are associated with downward wind propagation anomalies of the Quasi-Biennial Oscillation \citep{lu_2008}.
Is it possible that the extremely warm stratospheric disturbance in the aftermath of the Great White Storm of 2010-2011 \citep{fletcher_2011,fletcher_2012,fouchet_2016}
is due to a disruption of the downward propagation of the equatorial oscillation on Saturn?
Conversely, Cassini observations showed that the occurrence of this huge storm disturbed the equatorial oscillations \citep{fletcher_2017}.
Global stratospheric dynamical simulations by DYNAMICO-Saturn could help to address these questions in future studies.

\section*{Acknowledgments}
The authors acknowledge the exceptional computing support from Grand Equipement National de Calcul Intensif (GENCI) and Centre Informatique National de l'Enseignement Supérieur (CINES). All the simulations presented here were carried out on the Occigen cluster hosted at CINES. Bardet, Spiga, Guerlet acknowledge funding from Agence National de la Recherche (ANR), project EMERGIANT ANR-17-CE31-0007.
We also acknowledge funding by CNES as support of interpretation of Cassini/CIRS data analysis.

This paper is dedicated to the memory of Adam Showman who reviewed our paper with his perfect balance of expertise and open-mindedness that our entire field will sorely miss.

\newpage
\bibliography{bib_these}

\end{document}